\documentclass[preprint,12pt,3p]{elsarticle}
\usepackage[T1]{fontenc}
\usepackage{times}
\usepackage{graphics,graphicx,subcaption}
\usepackage{enumerate,epsfig}
\usepackage{multirow,afterpage,rotating}
\usepackage{amsfonts}
\usepackage{amsmath,amssymb}
\usepackage{caption}
\usepackage{soul}
\usepackage{url}
\usepackage{inputenc}
\usepackage[misc,geometry]{ifsym}
\usepackage[colorlinks=true]{hyperref}
\usepackage{dcolumn}
\usepackage{wrapfig,dsfont}
\usepackage{mathtools}
\usepackage{color}
\usepackage{tikz}

\DeclareCaptionFont{red}{\color{red}}

\newcommand{\bluesolidline}{\raisebox{2pt}{\tikz{\draw[-,blue,solid,line width = 0.9pt](0,0) -- (5mm,0);}}}
\newcommand{\cyansolidline}{\raisebox{2pt}{\tikz{\draw[-,cyan,solid,line width = 0.9pt](0,0) -- (5mm,0);}}}
\newcommand{\blacksolidline}{\raisebox{2pt}{\tikz{\draw[-,black,solid,line width = 0.9pt](0,0) -- (5mm,0);}}}
\newcommand{\redsolidline}{\raisebox{2pt}{\tikz{\draw[-,red,solid,line width = 0.9pt](0,0) -- (5mm,0);}}}
\newcommand{\greensolidline}{\raisebox{2pt}{\tikz{\draw[-,green,solid,line width = 0.9pt](0,0) -- (5mm,0);}}}
\newcommand{\magentasolidline}{\raisebox{2pt}{\tikz{\draw[-,magenta,solid,line width = 0.9pt](0,0) -- (5mm,0);}}}
\newcommand{\yellowsolidline}{\raisebox{2pt}{\tikz{\draw[-,yellow,solid,line width = 0.9pt](0,0) -- (5mm,0);}}}

\newcommand{\reddashedline}{\raisebox{2pt}{\tikz{\draw[-,red,dashed,line width = 0.9pt](0,0) -- (5mm,0);}}}
\newcommand{\greendashedline}{\raisebox{2pt}{\tikz{\draw[-,green,dashed,line width = 0.9pt](0,0) -- (5mm,0);}}}
\newcommand{\bluedashedline}{\raisebox{2pt}{\tikz{\draw[-,blue,dashed,line width = 0.9pt](0,0) -- (5mm,0);}}}

\newcommand{\solidcircle}[1][black,fill=red]{\tikz{\draw[#1] (0,0) circle (.5ex);}}

\definecolor{skyblue}{rgb}{0,0.5,0.5}
\definecolor{henna}{rgb}{0.5,0.5,0}
\definecolor{purple}{rgb}{0.5,0,0.5}
\definecolor{darkpink}{rgb}{1,0,1}

\biboptions{numbers,sort&compress}

\hypersetup{urlcolor=blue, citecolor=red}

\newlength{\defbaselineskip}
\newcommand{\setlinespacing}[1]%
           {\setlength{\baselineskip}{#1 \defbaselineskip}}

\begin{document}

\begin{frontmatter}

\title{Estimation of energy consumption of electric vehicles using Deep Convolutional Neural Network to reduce driver's range anxiety}

\author[label1]{Shatrughan Modi\corref{cor1}}
\address[label1]{Computer Science \& Engineering Department, \\
Thapar Institute of Engineering \& Technology, Patiala - 147004, India}

\cortext[cor1]{Corresponding author}

\ead{shatrughanmodi@gmail.com}

\author[label1]{Jhilik Bhattacharya}
\ead{jhilik@thapar.edu}

\author[label2]{Prasenjit Basak}
\address[label2]{Electrical \& Instrumentation Engineering Department, \\
Thapar Institute of Engineering \& Technology, Patiala - 147004, India}
\ead{prasenjit@thapar.edu}

\journal{ISA Transactions, Elsevier}

\begin{abstract}
    The goal of this work is to reduce driver's range anxiety by estimating the real-time energy consumption of electric vehicles using deep convolutional neural network. The real-time estimate can be used to accurately predict the remaining range for the vehicle and hence, can reduce driver's range anxiety. In contrast to existing techniques, the non-linearity and complexity induced by the combination of influencing factors make the problem more suitable for a deep learning approach. The proposed approach requires three parameters namely, vehicle speed, tractive effort and road elevation. Multiple experiments with different variants are performed to explore the impact of number of layers and input feature descriptors. The comparison of proposed approach and five of the existing techniques show that the proposed model performed consistently better than existing techniques with lowest error.
\end{abstract}

\begin{keyword}
Electric Vehicle \sep Deep Convolutional Neural Network \sep Energy Consumption Estimation.
\end{keyword}

\end{frontmatter}

\section{Introduction}\label{Sec:Introduction}
The demand of Electric Vehicles (EVs) is increasing at a very rapid rate and they have great potential to overcome the problems faced by transportation sector like depletion of fossil fuels and increasing pollution. A comprehensive survey of 162 EV drivers was conducted by Hubner et al. \cite{hubner2012use}, which highlights the four main barriers for EVs, namely (i) high purchasing price, (ii) limited driving range, (iii) limited availability of public infrastructure for charging and (iv) recharging time for the vehicle. Amongst these barriers, the major challenge faced by a driver sums down to the limited driving range of EVs. As reported by Zhang et al. \cite{ZHANG2018527}, 37.8\% EV owners out of 2193 in Japan feels that limited driving range is the major concern for EVs. The driver is anxious about the ability of the vehicle to reach the destination with the amount of charge present in the battery. Hence, it is very important to predict the driving range of an EV in real time. So, to achieve that the first step is to estimate the energy consumption of the EVs by studying the battery behaviour in different conditions.

Energy consumption of EVs depends on various factors like vehicle characteristics, vehicle speed, road elevation, and acceleration etc. In real life, these factors vary a lot and hence make the estimation of energy consumption a complex problem. A number of analytical models were proposed in the literature to solve this problem, for instance, Genikomsakis et al. \cite{GENIKOMSAKIS201798} developed a simulation model, for energy estimation and route planning, of Nissan Leaf and compared its results for 9 typically used drive cycles with Future Automotive Systems Technology Simulator (FASTSim). Halmeaho et al. \cite{7902061} developed four simulation models for a city electric bus and validated the models with the data collected from a bus prototype. The models were different in the type of methods used to simulate the behaviour of the motor. One model used the efficiency map and other models try to represent the losses of the motor as resistive losses. The validation results show that the model using efficiency map of the motor had an error of -3.4\% to 5\% whereas the models based on resistive loads gave an error of -0.4\% to 11.9\%. Gao et al. \cite{4168023} discussed various modeling and simulation tools and techniques for electric and hybrid vehicles.

The simulation-based models are hard to generalize as they require calibration according to the specific vehicle and require internal vehicle parameters, like motor efficiency curve, battery internal resistance etc., from the manufacturer, which are not readily available and sometimes hard to obtain. In contrast to simulation-based models, a number of data-driven models (mostly regression models) were also proposed for energy consumption estimation of EV. Ferreira et al. \cite{ferreira2012data} considered factors (like battery's State of Charge (SOC), speed, weather information, road type, driver profile) and collected data from an EV named Pure Mobility Buddy 09. They have proposed a data mining approach which internally uses regression to predict the driving range for an EV. The approach is computationally costly as for each destination, calculations need to be performed again and again. A statistical model based on physical model of an EV was proposed by Yuan et al. \cite{YUAN20171955} and regression analysis was used to obtain the driving condition independent energy consumption characteristics. An instantaneous energy consumption model considering the vehicle speed and acceleration was developed by Fiori et al. \cite{FIORI2016257}. The model was able to capture the regenerative energy efficiency as the function of deceleration of the vehicle. C.D. Cauwer et al. \cite{de2017data} developed a cascade NN-MLR model for energy consumption prediction. The Neural Network (NN) in the model was used to predict the speed profile based on road characteristics, weather conditions and traffic and then Multiple Linear Regression (MLR) was used to predict the energy consumption based on these parameters.

Similar to these, a number of researchers studied the factors influencing energy consumption and proposed different regression-based approaches for estimating energy consumption. For instance, Liu et al. \cite{LIU2018324} studied the impact of temperature and auxiliary loads and proposed three models, calibrated using ordinary least square and multilevel mixed-effects linear regression, for estimating energy consumption. From the study, they inferred that the temperature range of $21.8 - 25.2^\circ$C is the most economical in terms of energy efficiency and with proper usage of auxiliary loads vehicle energy consumption can be reduced significantly. Wang et al. \cite{WANG2015710} assessed the energy consumption of EVs in real-world driving conditions and concluded that small driving range and severe driving conditions make EVs a better choice than conventional vehicles due to less energy consumption. Yang et al. \cite{YANG201441} studied the impact of roads with different elevation on energy consumption of EVs and concluded that energy consumption increases sharply with increase in speed and road slope. Liu et al. \cite{LIU201774} also studied the effect of road elevation/grade using 12 grade ranges and developed eight regression models (4 linear and 4 logarithmic) for energy consumption estimation and demonstrated that EVs are more energy efficient than conventional vehicles in areas where road gradient changes frequently, due to regenerative braking. The comparison of proposed models shows that the logarithmic models performed better than linear models. Galvin \cite{GALVIN2017234} had found that the energy efficiency of EVs get compromised significantly when acceleration changes at a high rate regardless of the speed. A multivariate energy estimation model for EVs have been developed and validated the results with eight EVs. Wu et al. \cite{5524052} proposed analytical methodologies to estimate the power and energy consumption by Plug-In EVs (PEVs) considering the travel patterns from the 2009 National Household Travel Survey (NHTS) database. They argued that energy consumption of an EV depends upon time and location of driving, like the vehicles in the rural area consumed more energy than in urban areas. A cyber-physical system based approach was developed by Lv et al. \cite{8400581} considering vehicle characteristics and the different driving styles to optimally control the EV for best performance in terms of energy. The experimental results confirm that vehicles can perform better with respect to energy consumption in conservative, moderate and aggressive driving style with optimized control strategy. Fetene et al. \cite{FETENE20171} collected and combined data from four different sources namely, GPS driving patterns of 741 drivers over two years, road type, weather conditions and driver characteristics. A model to estimate Energy Consumption Rate (ECR) of an EV was proposed and the impact of various factors on ECR was studied and it has been found that more energy gets consumed in winters as compared to summers. Other than these regression-based techniques, Alvarez et al. \cite{6861542} trained Artificial Neural Networks (ANN) for estimating the energy consumption of EVs. For this, the neural networks were given input of vehicle speed, acceleration, and jerk. Felipe et al. \cite{7313117} extended \cite{6861542} by adding route information to the input of the neural network. The ANN developed in \cite{7313117, 6861542}, gave only one output of total energy consumed for the trip at the end of the trip. So, these models can not be used in real time to guide the driver about remaining energy in the battery.

Number of research gaps or problems have been found from the comprehensive literature review. Two types of techniques have been discussed in the literature review, one are analytical / simulation models and other are regression models. From the literature, it can be concluded that the analytical / simulation models lack the applicability in real world as they require internal vehicle data from the manufacturer (like battery's internal resistance, SOC curve of the battery, motor's efficiency curve etc.), which is not readily available. Also, it is very difficult to generalize the simulation models as they require vehicle specific calibration. Similarly, statistical regression techniques rely heavily on the availability of real-world data and vary in the extent to which they can be linked to underlying physical principles. Although, the existing ANN based models \cite{7313117, 6861542} show promising results with high prediction power and robustness, but the existing ANN based models instead of providing real-time output during the trip, provide only one result at the end of the trip i.e. total energy consumed for the trip. Hence, can not be used to provide real time guidance to the driver about remaining driving range of the vehicle, route to be taken which can help the driver reach his destination etc.

To overcome the problems discussed in the literature review, in this paper a Deep Convolutional Neural Network (D-CNN) based methodology has been developed. To the best of author's knowledge, D-CNN based approach for power/energy estimation of EV is being developed for the first time. One of the challenges was the requirement of internal vehicle data from the manufacturers for calibration of simulation models, which is vary hard to obtain as the manufacturer do not share the data in public domain. The proposed methodology requires only three parameters namely, vehicle speed, road elevation and tractive effort. Also, the required input parameters can be easily obtained or calculated, for instance, vehicle speed and road elevation can be easily obtained using Global Positioning System (GPS) and Geographic Information System (GIS), respectively. Similarly, tractive effort can be calculated easily using equation \eqref{Eq:TractiveEffort}, as discussed in Section \ref{Sec:DataSets}, which requires very basic parameters like linear acceleration (can be easily calculated from speed), vehicle weight (readily available) etc. There are some ANN based approaches \cite{7313117, 6861542} which provide very promising results and do not require internal vehicle data from the manufacturer but they do not provide real-time output and hence are not useful in the real-world as they can not be used to guide the driver in real-time. In contrast to this, the proposed deep learning based solution provide real-time energy consumption as output and can be used to provide real-time guidance to the driver about remaining energy in the battery and hence, remaining driving range of the vehicle. Also, the deep learning architectures can learn more complex patterns than shallow networks, as existing ANN based models have only one hidden layer. Recent advances in computing power and fast learning algorithms have made training deep learning architectures feasible. Due to this, deep learning architectures have gained a lot of interest in the automotive sector also and have been successfully applied in numerous problems like image classification, object detection, traffic flow prediction etc \cite{Liu:2019:DNN:3309769.3231741, MASOOD2018, WU2018166, GENG2018895, SONG2018381}. Also, the nonlinearity and complexity induced by the combination of all the influencing parameters make the problem of energy consumption estimation more suitable for a deep learning approach, in contrast to other regression techniques. This motivates the authors to focus on the deep learning based models to solve the problem of estimation of energy consumption of EV. The current work, by providing experimental results, proves that a deep learning architecture is suitable for the problem at hand. Considering the success of the current experiments, the task of thoroughly evaluating different kinds of deep learning networks will next be considered and is not in the scope of this work. So, following are the main contributions of this work:

\begin{enumerate}[i)]
    \item A D-CNN based methodology has been developed which requires only three external parameters, namely, Road Grade, Tractive Effort and Vehicle Speed and can accurately estimate the energy consumption.
    \item The proposed model can be easily trained for other vehicles either using the real world driving data which is subject to availability or using the simulated data from the simulated model, as done in the current case.
    \item The effect of different input features descriptors on models performance has also been studied by training the D-CNN models with three different feature descriptors namely, Gramian Angular Field (GAF), Covariance and Eigen Vectors.
    \item The effect of different number of hidden layers has been explored by training the models with different number of layers.
\end{enumerate}

The rest of the article has been organized as follows. Section \ref{Sec:DataSets} describe the data set used for training, validating and testing the proposed approach. In Section \ref{Sec:ProposedApproach}, the architecture of the proposed methodology has been discussed. Section \ref{Sec:Results} discuss the experimental results obtained from the proposed approach while the comparative analysis is being done in Section \ref{Sec:ComparativeAnalysis}. Finally, Section \ref{Sec:Conclusion} provide the future directions and concludes the paper.

\section{Datasets}\label{Sec:DataSets}
Data from two different sources for an EV namely, Nissan Leaf 2013, was used for training, validating and testing the proposed methodology.

One dataset was obtained from Downloadable Dynamometer Database \cite{D3ANL} generated at the Argonne National Laboratory (ANL) of Advanced Powertrain Research Facility (APRF), under the funding and guidance of the U.S. Department of Energy.  It contains data from several dynamometer tests conducted on various EVs at road grade of 0\% for several drive cycles.

As this dataset was quite small and not enough for training, validating and testing the proposed methodology, another dataset was generated in this work using a simulation model of Nissan Leaf 2013. The simulation model was developed in FASTSim \cite{brooker2015fastsim} using vehicle specific parameters of Nissan Leaf 2013 \cite{factNissan2013, burress2012benchmarking}, shown in Table \ref{Tab:VehicleParameters}. Similar to this, simulated models for other EVs can also be developed based on the availability of manufacturer data like motor efficiency curve, battery internal resistance etc. Using the simulated model of Nissan Leaf 2013 the data was generated for 80 standard drive cycles (like Supplemental Federal Test Procedures (SFTP), Urban Dynamometer Driving Schedule (UDDS) and New European Driving Cycle (NEDC)), which have been widely used by other researchers \cite{GENIKOMSAKIS201798, YUAN20171955} also, and 30 road grade profiles by varying the road grade from -20\% to 20\%. It is to be noted that for checking the robustness of the proposed approach a number of other custom generated road grade profiles or drive cycles can also be used.

Henceforth, the dataset generated through the simulation model and dataset obtained from the Downloadable Dynamometer Database will be referred as $DS-I$ and $DS-II$, respectively. Training and validation was done using dataset $DS-I$, while $DS-II$ was used for testing the proposed approach. The datasets $DS-I$ and $DS-II$ both contain data recorded at 10 Hz frequency i.e., 10 readings for every sec. The dataset $DS-I$ contain various parameters like vehicle speed, battery power supplied, battery's state of charge, environmental temperature, tractive effort, road elevation and auxiliary loads. In the current work, for training, validating and testing the proposed methodology four parameters were selected, namely Vehicle Speed ($v_{sp}$), Tractive Effort ($t_{eff}$), Elevation of the road ($r_{el}$) and Power Supplied by battery ($p_{batt}$) at environmental temperature of $25^\circ$C and constant auxiliary load of 150W. The effect of auxiliary load on energy consumption is additive in nature, so does not increase the complexity of the problem. Also, the environmental temperature does not affect the energy consumption of EV much, unless there is huge change in climatic temperature. So, this dataset, although recorded at $25^\circ$C, is valid for wide range of temperature. The three parameters $v_{sp}$, $r_{el}$ and $p_{batt}$ are straight forward but $t_{eff}$ refers to the driving force required by the vehicle to move forward which is a combination of multiple components and can be calculated using the following equation, provided in \cite{GENIKOMSAKIS201798}:

\begin{equation}\label{Eq:TractiveEffort}
t_{eff} = f_{ad} + f_{rr} + f_{hc} + f_{la} + f_{wa}
\end{equation}

where $f_{ad}$ represent the opposing force due to aerodynamic drag, $f_{rr}$ is the opposing rolling resistance force, $f_{hc}$ is the gravitational force component which acts while hill climbing, $f_{la}$ is the opposing force due to linear acceleration and $f_{wa}$ is the inertial force due to rotating parts of the vehicle. So, tractive effort $t_{eff}$ contains the combine effect of all these forces, which in turn depend upon number of vehicle characteristics, like frontal area of the vehicle, vehicle's aerodynamic drag coefficient, vehicle's mass and rolling resistance coefficient etc. Due to this, tractive effort $t_{eff}$ along with road elevation $r_{el}$ and vehicle's speed $v_{sp}$ are the ideal candidate to be considered for input and instantaneous power supplied by battery $p_{batt}$ for output.


\begin{table}[h!]
\centering
\caption{Parameters of Nissan Leaf 2013 used for developing simulation model in FASTSim \cite{factNissan2013, burress2012benchmarking}}
\label{Tab:VehicleParameters}
\begin{tabular}{|c|c|c|}
\hline
Component                     & Parameter                           & Value                           \\ \hline
\multirow{3}{*}{Motor}        & Type                                & Permanent Magnet AC Synchronous \\ \cline{2-3}
                              & Max. Power (kW)                     & 80                              \\ \cline{2-3}
                              & Max. Torque (Nm)                    & 253                             \\ \hline
\multirow{2}{*}{Transmission} & Type                                & Single Speed                    \\ \cline{2-3}
                              & Final Drive Ratio                   & 7.9                             \\ \hline
\multirow{7}{*}{Battery}      & Type                                & Lithium Ion                     \\ \cline{2-3}
                              & Number of Cells                     & 192                             \\ \cline{2-3}
                              & Cell Configuration                  & 2 Parallel, 96 Series           \\ \cline{2-3}
                              & Nominal Cell Voltage (V)            & 3.7                             \\ \cline{2-3}
                              & Nominal System Voltage (V)          & 364.8                           \\ \cline{2-3}
                              & Rated Pack Capacity (Ah)            & 66.2                            \\ \cline{2-3}
                              & Rated Pack Energy (kWh)             & 24                              \\ \hline
\multirow{8}{*}{Vehicle}      & Front \& Rear Track (m)             & 1.53                            \\ \cline{2-3}
                              & Vehicle Weight (kg)                 & 1498                            \\ \cline{2-3}
                              & Drive Train                         & Front Wheel Drive               \\ \cline{2-3}
                              & Aerodynamic Drag Coefficient        & 0.29                            \\ \cline{2-3}
                              & Frontal Area ($m^2$)                & 2.27                            \\ \cline{2-3}
                              & Wheelbase (m)                       & 2.7                             \\ \cline{2-3}
                              & Weight Distribution Front/Rear (\%) & 58/42                           \\ \cline{2-3}
                              & Wheel Radius (m)                    & 0.3162                          \\ \hline
\end{tabular}
\end{table}

\section{Proposed Methodology}\label{Sec:ProposedApproach}
Energy consumption of an EV depends upon number of factors like road elevation, vehicle speed and vehicle acceleration etc. These factors have a non-linear relation among them as they vary a lot in real world. So, to accurately estimate their non-linear relation a deep learning based methodology has been developed.

Deep learning architectures are capable of learning high dimensional non-linear functions using a sequence of semi-affine non-linear transformations. The deep architectures can be represented as a graph of nodes and edges. Each edge has a weight which signifies the relative importance of the link and each node applies an activation function to the weighted sum of incoming connections. A number of activation functions are available like sigmoid function, tanh etc. A particular deep learning architecture, namely, Convolutional Neural Network (CNN), has been used in this work for estimation of energy/power consumption of EV.

The CNN has a unique learning ability from images due to its two unique characteristics, namely, pooling mechanism and locally connected layers. The pooling mechanism significantly reduces the number of parameters required for training the network while preserving the important features. In locally connected layers, the output neurons of the layers are connected to their local input neurons only instead of all the input neurons, as in fully connected layers. This helps CNN extracting the critical local features from the images effectively because every layer tries to extract different feature for the prediction problem.

\begin{figure*}[h]
      \centering
      \includegraphics[width=\linewidth]{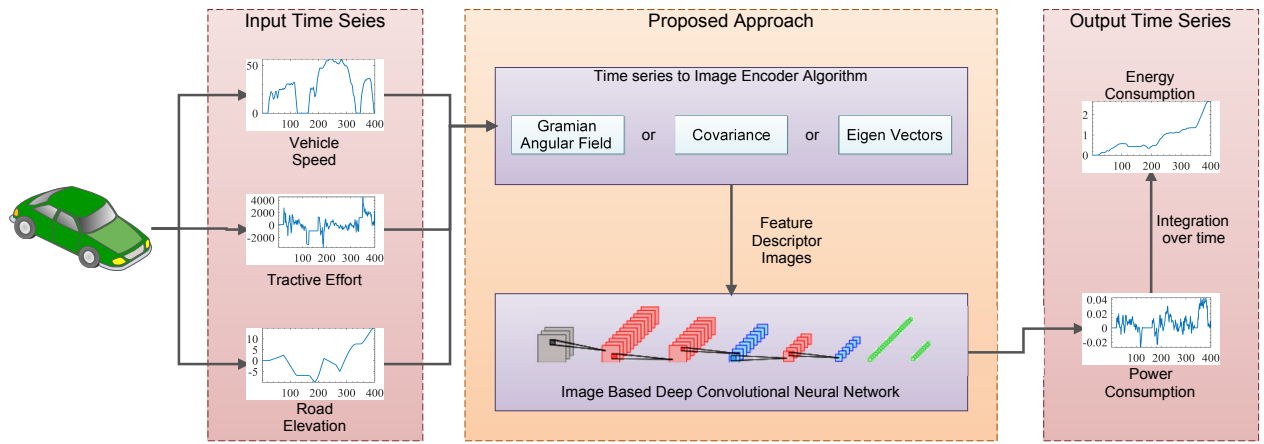}
      \caption{Architecture of the proposed methodology}\label{Fig:ProposedArchitecture}
      \vspace{5mm}
\end{figure*}

Considering the above-mentioned characteristics, image based CNN was chosen to be used for estimating the energy consumption of EV. Figure \ref{Fig:ProposedArchitecture} represents the complete architecture of the proposed methodology. There are mainly two modules namely, Time Series to Image Encoder and Image based Deep Convolutional Neural Network.

\subsection{Time Series to Image Encoder}\label{Subsec:TimeSeriesImageEncoder}
While there are recurrent neural networks for time-series classification, some researchers have also considered the transformation of time-series into a 2D signal thus taking advantage of a CNN based classification or regression. CNN models have proved their performance for recognizing patterns from images. Hence, to take advantage of the success of CNN models in learning features from images, literature was explored for existing algorithms to convert time series data into images. Also, different features in an image representation of a time-series data, not present in its 1D form,  elevates the performance of the task. A number of approaches have been proposed by researchers for encoding time series data to images, for instance, Yang et al. \cite{yang2015deep} have proposed a method, to encode time series data to images for human activity recognition, in which the multiple time series were concatenated as rows of image i.e. each time series correspond to the particular row in the image. This method is not suitable for the current problem, as only three parameters, namely, road elevation, the speed of the vehicle and tractive effort have been considered for the input. So, images with only three rows are not appropriate for training the CNN models. Wang et al. in 2015 \cite{wang2015encoding} proposed Gramian Angular Field (GAF) and Markov Transition Field (MTF) as two approaches to encode time series data to images for classification. It has been observed that there is a lot of information loss using MTF, as in this the time series need to be binned to a number of quantile bins. Hence, for our work,  it is very hard to even roughly recover the original signal after applying MTF whereas in GAF the information loss is comparatively lower, i.e. it is possible to approximately reconstruct the original signal.

Hence, in this work, GAF has been used as one of the approaches to convert time series data to images. Due to some information loss in GAF, the covariance and eigenvector methods were also considered for conversion. The covariance descriptor reflects the correlation information, hence accommodating the power consumption changes due to instant acceleration. Also, the covariance matrix being symmetric becomes computationally effective. The eigenvectors of a covariance matrix give a set of orthonormal vectors which indicate the directions in which the data varies the most (principal components). In general, CNN uses augmentation techniques (such as Principal Component Analysis and whitening) to reduce overfitting. Hence, motivating the use of eigenvectors as feature input.

In order to convert time series data into images, the selected time series namely, Vehicle Speed ($v_{sp}$), Tractive Effort ($t_{eff}$), Elevation of the road ($r_{el}$) and Power Supplied by battery ($p_{batt}$), from dataset $DS-I$ and $DS-II$ were partitioned into $m$ small time series each of 10 sec duration, such that dataset $DS-I$ contain approximately 3.5 lacs while $DS-II$ contains approximately 3500 partitions. Out of these, $70\%$ of the partitioned time series were randomly selected from $DS-I$ (say $DS-I_{tr}$) and were used for training and rest $30\%$ (say $DS-I_{val}$) were used for validating the CNN models. As discussed previously in section \ref{Sec:DataSets}, out of the four time series, the first three were used as input to the CNN model and the fourth one was taken as output. So, the partitions of input time series only were converted into images using three preprocessing techniques namely, GAF, Covariance and Eigenvectors, and generate three different sets of images as output. The output sets can be represented, in general, using equation \ref{Eq:inputX}.

\begin{equation}\label{Eq:inputX}
\mathds{X} = \{M^i\ | \ M^i \in \mathbb{R}^{100\times100\times3} \ and \ i = 1,2,...,m\}
\end{equation}

where $\mathds{X}$ is the output set obtained after using particular preprocessing technique, $M^i$'s are the images obtained from corresponding $i^{th}$ partition of input time series of $v_{sp}$, $t_{eff}$ and $r_{el}$, as shown in Figure \ref{Fig:DataPreprocessing}. In this figure, the input signals were the $i^{th}$ partition of time series $v_{sp}$, $t_{eff}$ and $r_{el}$ (highlighted in red) and denoted by $v_{sp}^i\_In$, $t_{eff}^i\_In$ and $r_{el}^i\_In$. The preprocessing algorithm can be any of the three methods, namely GAF, Covariance and Eigen Vectors. For each input, the preprocessing algorithms gave the corresponding output of size $100 \times 100$ , denoted by $v_{sp}^i\_Out$, $t_{eff}^i\_Out$ and $r_{el}^i\_Out$. The output matrices generated were then concatenated to obtain the corresponding image $M^i$ which was then fed to CNN models as input. The three preprocessing methods have been discussed as follows:

\begin{figure}[h!]
  \centering
  \includegraphics[width=\linewidth]{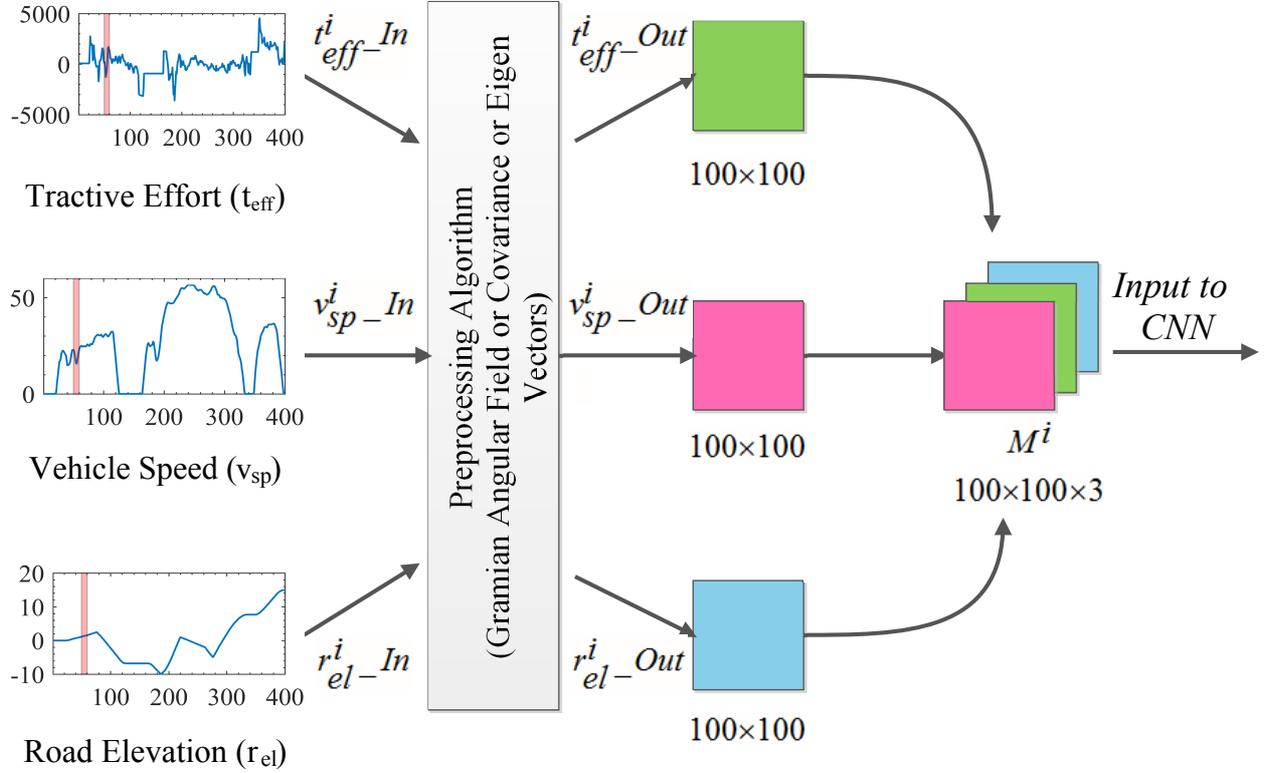}
  \caption{Preprocessing of Time Series Data}
  \label{Fig:DataPreprocessing}
\end{figure}

\begin{enumerate}[i)]
  \item \emph{Gramian Angular Field (GAF):} In Gramian Angular Field method, proposed in \cite{wang2015encoding}, initially all the $m$ partitioned time series (say $x^i$'s) of Vehicle Speed ($v_{sp}$), Tractive Effort ($t_{eff}$) and Road Elevation ($r_{el}$) were normalized into the range of [-1,1], using Equation \eqref{Eq:GAFNormalize}, to generate three new sets $\hat{V}$, $\hat{T}$ and $\hat{R}$ each containing the corresponding normalized partitions $\hat{x^i} \in \mathbb{R}^{100\times1}$ of $v_{sp}$, $t_{eff}$ and $r_{el}$ respectively.

        \begin{equation}\label{Eq:GAFNormalize}
        \hat{x^i} = \frac{(x^i - \max(x)) + (x^i - \min(x))}{\max(x) - \min(x)}
        \end{equation}

        In this equation, $\max(x)$ and $\min(x)$ represents the maximum and minimum values of time series $x$. Then, the sets $\hat{V}$, $\hat{T}$ and $\hat{R}$ were further transformed to three new sets $V$, $T$ and $R$ of GAF matrices $G^i \in \mathbb{R}^{100\times100}$ obtained from corresponding normalized partitions $\hat{x^i}$, by using Equation \eqref{Eq:GAF}.

        \begin{equation}\label{Eq:GAF}
        G^i = \hat{x^i} \cdot \hat{x^i}^T - \sqrt{I - \hat{x^i}^2} \cdot \sqrt{I - \hat{x^i}^2}^T
        \end{equation}

        where $I$ represents the 1D array $[1,1,...,1]^T$ of length $100$. The sets $V$, $T$ and $R$ were then used to generate the input set $\mathds{X}$ of images $M^i$s. The $j^{th}$ element of $\mathds{X}$, i.e. $M^j \in \mathbb{R}^{100\times100\times3}$, was obtained by concatenating $j^{th}$ GAF matrices $G^j \in \mathbb{R}^{100\times100}$ from $V$, $T$ and $R$ after normalizing them to the range of [0,1], i.e., $G^j$ from $V$, $T$ and $R$ after normalizing became the first, second and third layer, respectively, of $M^j$. It can be observed that there is some information loss, as explained in \cite{gamboa2017deep}, due to the negative term (second term with square roots) in Equation \eqref{Eq:GAF} which can effect the estimation accuracy of the proposed models.

  \item \emph{Covariance:} The loss in information in above method motivated to use Covariance matrix as feature input. The first step in this method was to normalize the partitions $x^i$'s of $v_{sp}$, $t_{eff}$ and $r_{el}$ and obtain three new normalized sets $\hat{V}$, $\hat{T}$ and $\hat{R}$. After normalization, three sets $V$, $T$ and $R$ were generated each containing the covariance matrices $\hat{C^i} \in \mathbb{R}^{100\times100}$ of $\hat{V}$, $\hat{T}$ and $\hat{R}$ respectively. The sets $V$, $T$ and $R$ were then used to create the set $\mathds{X}$ by concatenating the corresponding $\hat{C^i}$'s from $V$, $T$ and $R$.

  \item \emph{Eigen Vectors:} In this method, the covariance matrices $C^i \in \mathbb{R}^{100\times100}$ of each partition of $v_{sp}$, $t_{eff}$ and $r_{el}$ was calculated but without normalization. It generated three sets $V^c$, $T^c$ and $R^c$ each containing the covariance matrices $C^i$s of $v_{sp}$, $t_{eff}$ and $r_{el}$. Then eigen vector matrices $E^i$s from these covariance matrices $C^i$s were calculated and then these eigen vectors were normalized to the range of [0,1]. So three new sets $V$, $T$ and $R$ were generated each containing the normalized eigen vector matrices $\hat{E^i} \in \mathbb{R}^{100\times100}$. Then, the set $\mathds{X}$ was generated by concatenating the corresponding $\hat{E^i}$s from $V$, $T$ and $R$.
\end{enumerate}

\subsection{Image based Deep Convolutional Neural Network}\label{SubSec:D-CNN}
\begin{figure}[h!]
  \centering
  \begin{subfigure}[t]{.9\linewidth}
  \centering
  \includegraphics[width=0.9\linewidth]{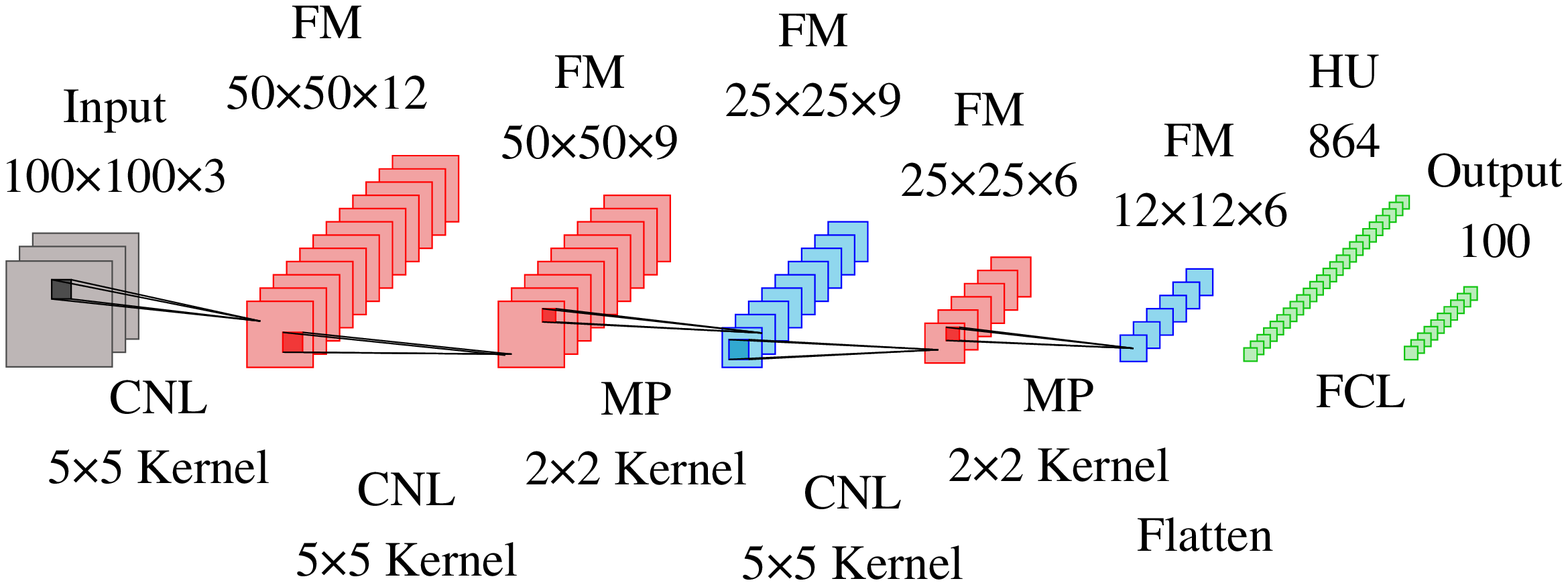}
  \caption{CNN Model with 7 layers ($CNN^7$)}\label{Fig:SmallNetwork}
  \end{subfigure}

  \begin{subfigure}[t]{.9\linewidth}
  \centering
  \includegraphics[width=0.9\linewidth]{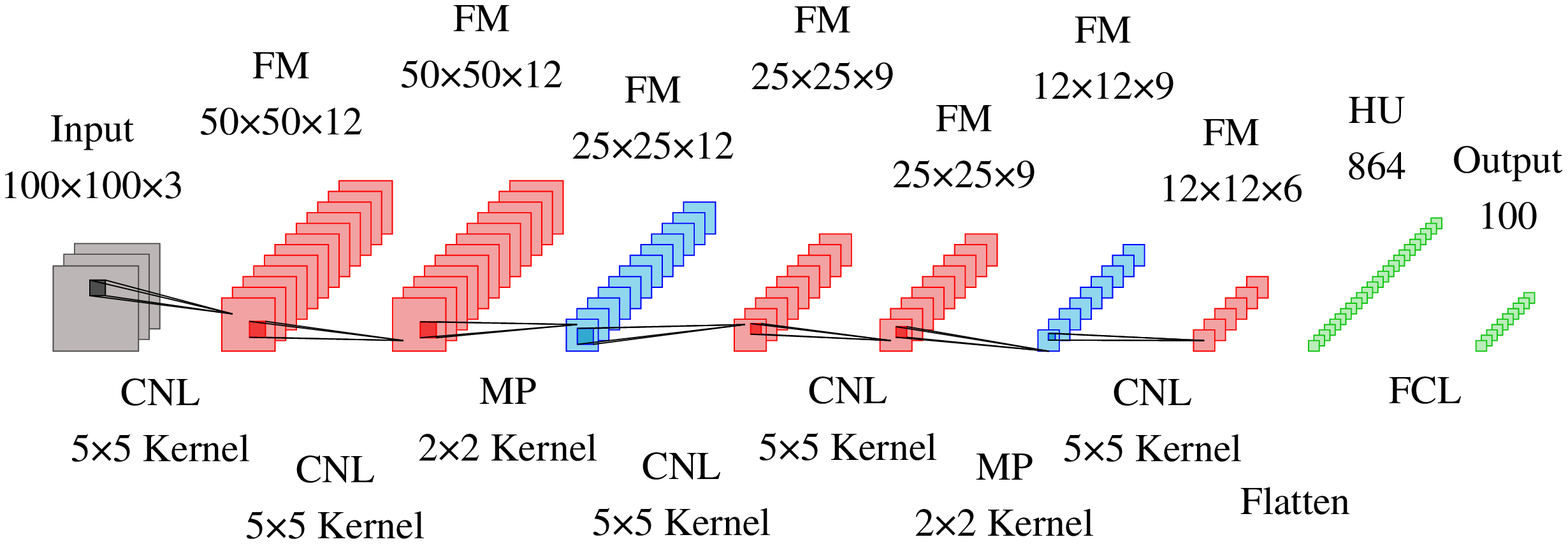}
  \caption{CNN Model with 9 layers ($CNN^9$)}\label{Fig:BigNetwork}
  \end{subfigure}

  \caption{Two CNN Models with different architecture (Convolution with Non-Linearity (CNL), Fully Connected Layer (FCL), Max Pooling (MP), Feature Maps (FM), Hidden Units (HU))}\label{Fig:CNNModels}
\end{figure}

CNN architectures have gained a lot of popularity in the field of pattern recognition. AlexNet \cite{NIPS2012_4824} is one of the most popular and vastly used architecture proposed in the field of pattern recognition. It has also been considered as a base reference for researchers applying deep learning in new domain \cite{7519080}. Considering the above,  initially the authors chose to start with CNN architecture considering AlexNet architecture as the base reference. AlexNet architecture has multiple convolution, pooling and fully connected layers stacked together. So in this work, experiments with multiple CNN architectures, having different number of layers, were performed. Later in the experiments, it has been observed that increasing the layers further after a particular number of layers (in this case 7) did not enhance the performance for the current data. So results for two architectures, as shown in Figure \ref{Fig:CNNModels}, are presented in this work. Let's call the CNN architecture with seven layers, shown in Figure \ref{Fig:SmallNetwork}, and architecture with nine layers, shown in Figure \ref{Fig:BigNetwork}, as $CNN^7$ and $CNN^9$, respectively.

$CNN^7$ architecture takes an image of size $100\times100\times3$, obtained from Time Series to Image Encoder module, as input and convolves it with $5\times5$ kernels. The kernels have depth of 3 as the input image has three channels. During the convolution operation, padding of two rows and two columns have been used along with stride of 2 positions. In first convolution layer, 12 such $5\times5$ kernels were used which gave output of $50\times50\times12$ feature maps (can be calculated using the equation $Output_{size} = ((Input_{size} - Kernel_{size} + 2 \times Padding)/Stride) + 1$). Here, it can be observed that number of channels have been increased in the multiple of 4 i.e. from 3 to 12 and size of the image has been reduced to $1/4^{th}$ i.e. from $100\times100$ to $50\times50$. So, the first convolution layer produces same number of features as the size of the input image i.e. $100\times100\times3$ becomes $50\times50\times12$. The number of kernels for the first layer is chosen as 12 as compared to standard sizes of 48 etc in AlexNet for two reasons. First data considered is time series data as opposed to more complicated image data. Second a larger number of kernels will require a bigger training set in order to converge. The output from the convolution layer was then passed through a non-linear (Tanh) activation layer which maps it using a function $tanh(x) = \frac{1-e^{-2x}}{1+e^{-2x}}$. After the first convolution with non-linearity (CNL) layer, there is a pattern of layers (i.e. one CNL layer then a max pooling layer), which has been repeated twice. This pattern has been used to decrease the dimension and number of feature maps and only keep the important features. For instance, the second CNL layer reduce the number of feature maps from 12 to 9 and then, a max pooling layer has been used which finds the maximum feature map over local neighborhood and reduce the size of feature maps from $50\times50$ to $25\times25$. After the repeated pattern of layers, a flatten layer has been used which change the shape of feature maps from 3D to 1D because the next layer which is a fully connected layer (FCL) take a 1D vector as input. So, the FCL maps the output of previous flatten layer to the desired output of length $100$. Similar to $CNN^7$ architecture, the $CNN^9$ architecture has been developed by increasing the number of layers. In $CNN^9$ architecture, the main difference is the number of layers and hence the dimension and number of feature maps decrease slowly. The main reason for this was to keep the important features as long as possible so that more accurate output can be obtained but it has been observed that there is no accuracy gain by increasing the number of layers further after a particular number of layers.


A set of images $\mathds{X}$, obtained from Time Series to Image Encoder module, was taken as input to the CNN models, and the models generated an output set $\mathds{Y}$. The output set $\mathds{Y}$, defined in Equation \eqref{Eq:outputY}, was the set of 1D arrays $O^i$'s each of length $100$, corresponding to the instantaneous power supplied by the battery. Each 1D array $O^i$ represent the $i^{th}$ partition of time series $p_{batt}$, normalized into the range of [0,1].

\begin{equation}\label{Eq:outputY}
\mathds{Y} = \{O^i \ | \ O^i \in \mathbb{R}^{100\times1} \ and \ i = 1,2,...,m\}
\end{equation}

\section{Results and Discussion}\label{Sec:Results}
A number of CNN models with different number of layers were trained with dataset preprocessed with three methods, namely GAF, Covariance and Eigen Vectors. Henceforth, CNN models with $n$ number of layers trained with GAF, Covariance and Eigen Vector features are denoted as $CNN_{gaf}^n$, $CNN_{cov}^n$ and $CNN_{eig}^n$, respectively. In all of these models, 70\% of the dataset $DS-I$ (mentioned in Section \ref{Sec:DataSets}) was used for training with 2000 epochs. The remaining 30\% of the dataset was used for validation. Furthermore, initially CNN architectures were considered as black boxes and the only performance indicators were the accuracy achieved, error etc but recently Shwartz-Ziv and Tishby \cite{DBLP:journals/corr/Shwartz-ZivT17} have presented an interesting approach to visualize the behaviour of internal hidden layers of Deep Neural Networks (DNN) in information plane using mutual information of layers. They have shown that the visualization of internal behaviour of DNN architecture can provide the insight about how well the model is training, how many epochs are actually required for fitting (called the drift phase), whether the particular architecture able to find the fitting solution etc.

\subsection{Mutual Information of Layers}\label{SubSec:MutualInformation}
The mutual information represents the amount of relevant information contained by a random variable $X$ about another random variable $Y$. The mutual information of any two random variables, $X$ and $Y$, with joint distribution $p(x,y)$, can be defined as:
\begin{equation}\label{Eq:MutualInformation}
    I(X;Y) = \sum_{x \in X, y \in Y} p(x,y) \log\bigg(\frac{p(x,y)}{p(x)p(y)}\bigg)
\end{equation}
where $p(x), p(y)$ represents the marginal distribution of the variables $X$ and $Y$ respectively. The mutual information obtained using above equation range from [0,$\infty$). So for comparison purpose $I(X;Y)$ was normalized to the range of [0,1] by using equation \eqref{Eq:NormalizedMutualInformation} as follows:
\begin{equation}\label{Eq:NormalizedMutualInformation}
    NMI(X;Y) = \frac{I(X;Y)}{\sqrt{H(X)H(Y)}}
\end{equation}
where $H(X)$ and $H(Y)$ represent the entropy of random variables $X$ and $Y$. A number of other normalizations are also possible based on the observation that $I(X;Y) \leq min(H(X),H(Y))$ using arithmetic or geometric mean of $H(X)$ and $H(Y)$. The geometric mean was used due to the analogy with the normalized inner product in Hilbert Space. As $H(X) = I(X;X)$, it can be observed that $NMI(X;X) = 1$ as desired.

\subsection{Training and Validation of the Models}
While training the CNN models, the kernels of each convolutional layer of CNN models were initialized with random numbers generated from a uniform distribution, defined in the range of $[-stdv, stdv)$ where $stdv = 1/\sqrt{kw \times kh \times numInPl}$. Here $kw$, $kh$ and $numInPl$ represent the kernel width, kernel height and number of input planes of the particular convolutional layer, respectively. The models were trained to learn the kernels for maximum 2000 epochs using Stochastic Gradient Descent (SGD) with initial learning rate and batch size set to 0.01 and 64, respectively. The learning rate was set to gradually decrease as the training progresses at a constant rate. The objective was to minimize the Mean Square Error (MSE) between the predicted and actual power consumption. Experiments were conducted with different number of layers such as 5, 6, 7, 8, 9, 10 and 11. It was found that by increasing the number of layers, the number of epochs to converge reduced, for instance, the CNN models with 5, 7, 9 and 11 layers when trained using dataset preprocessed with covariance method converged at approximately 420, 380, 330 and 290, respectively. Although by increasing the layers the models converge early but, it is also a well known fact that the architectures with more layers require more training data to achieve the same level of accuracy as the architecture with less number of layers. For comparison purpose, the results for two CNN models $CNN^7$ and $CNN^9$ are shown in this paper.


\begin{figure*}[h!]
  \centering
  \begin{subfigure}[t]{0.32\linewidth}
    \centering
    \includegraphics[width=\linewidth]{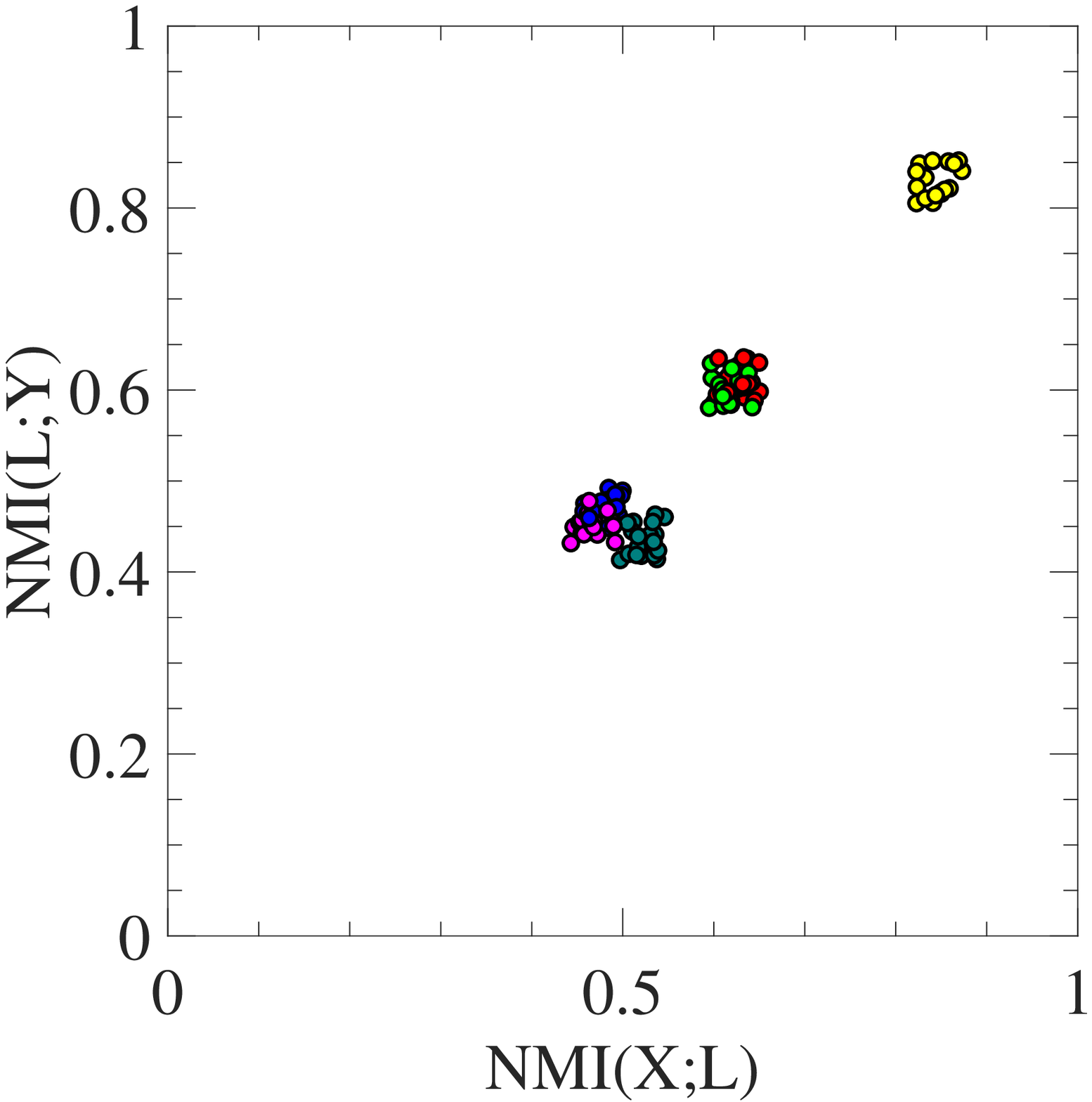}
    \caption{$CNN_{cov}^7$ at initial state}
  \end{subfigure}
  \hfill
  \begin{subfigure}[t]{0.32\linewidth}
    \centering
    \includegraphics[width=\linewidth]{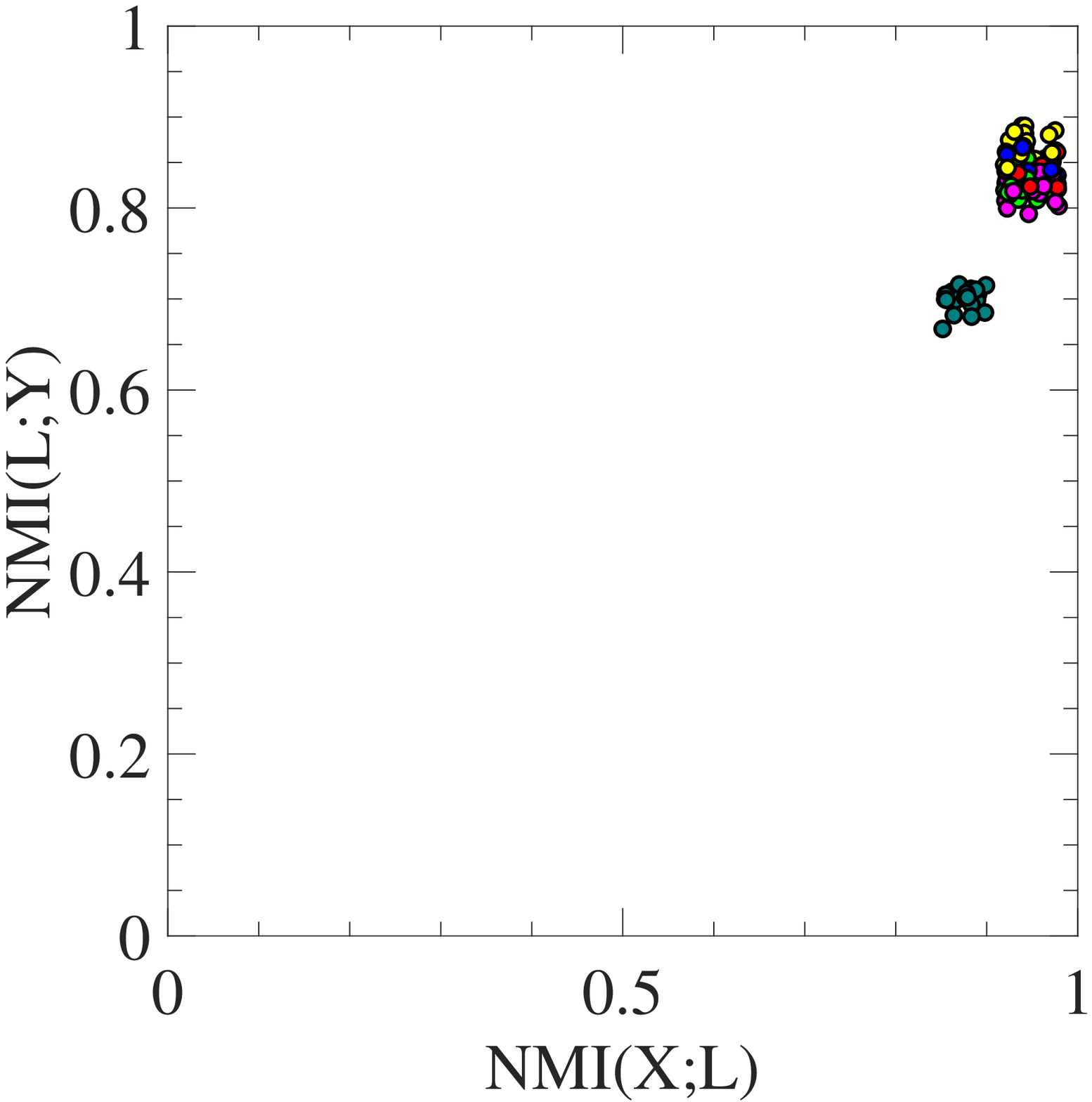}
    \caption{$CNN_{cov}^7$ after 300 epochs}
  \end{subfigure}
  \hfill
  \begin{subfigure}[t]{0.32\linewidth}
    \centering
    \includegraphics[width=\linewidth]{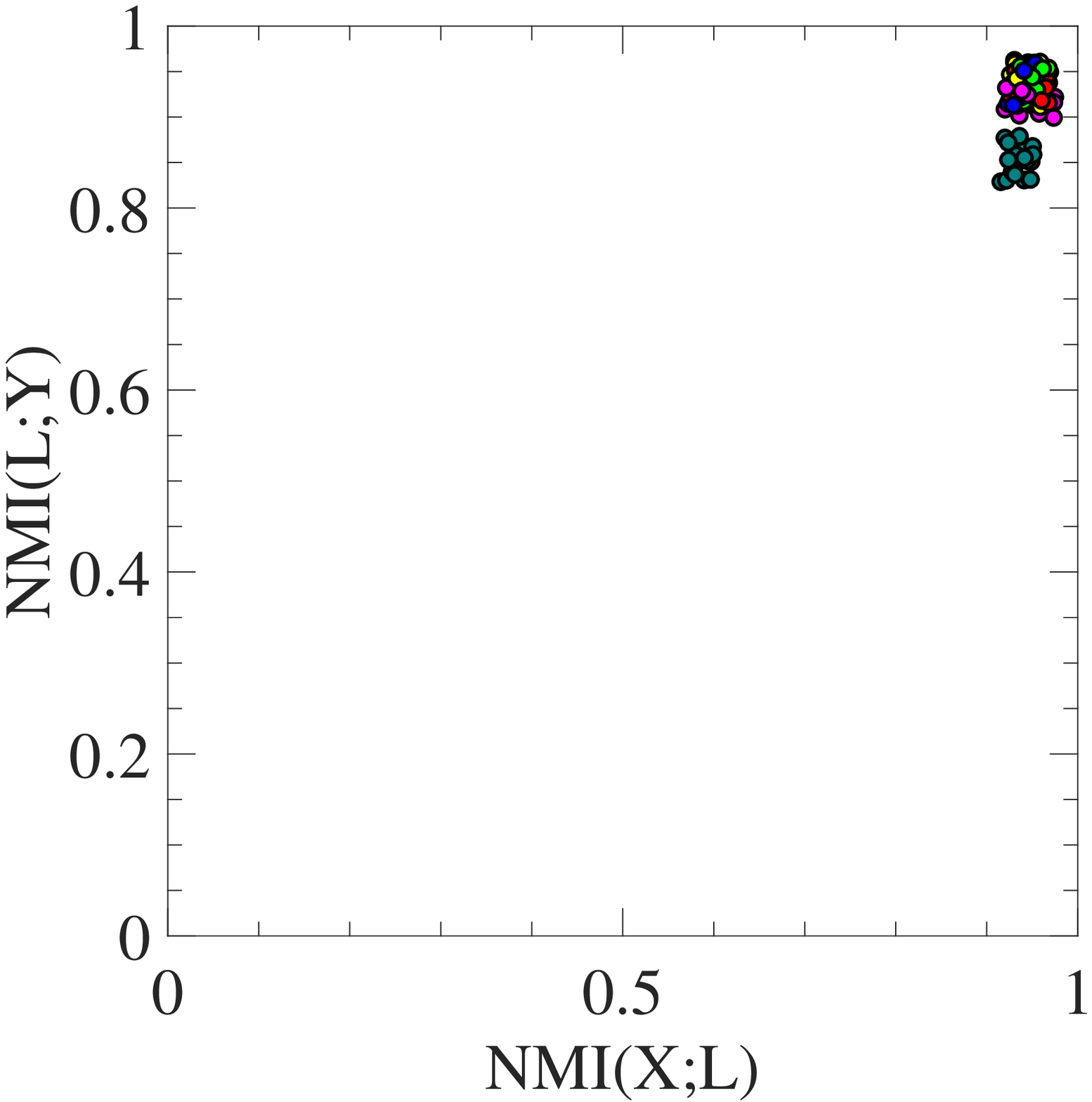}
    \caption{$CNN_{cov}^7$ after 2000 epochs}
  \end{subfigure}

  \begin{subfigure}[t]{0.32\linewidth}
    \centering
    {\includegraphics[width=\linewidth]{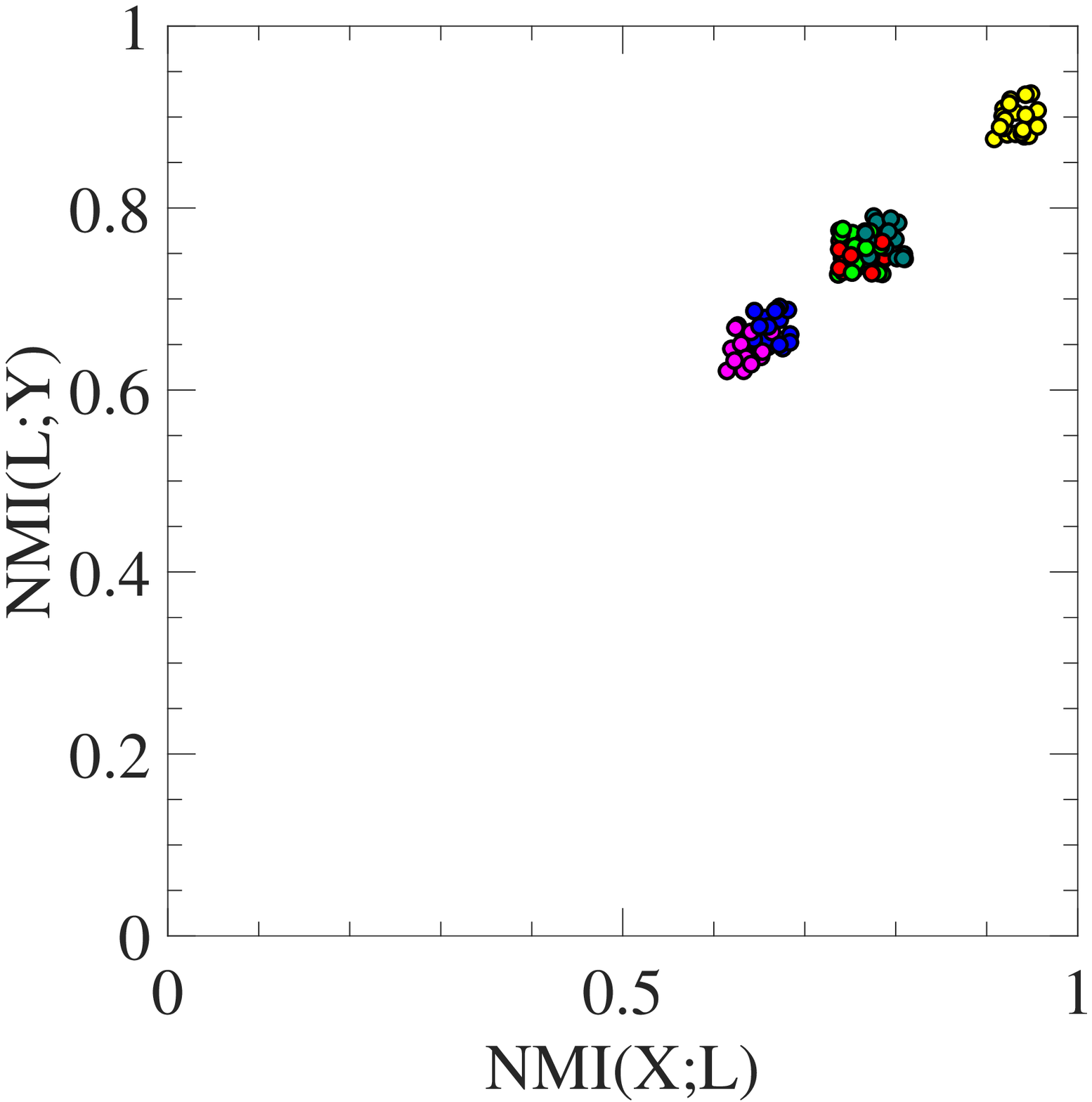}}
    \caption{$CNN_{gaf}^7$ at initial state}
  \end{subfigure}
  \hfill
  \begin{subfigure}[t]{0.32\linewidth}
    \centering
    \includegraphics[width=\linewidth]{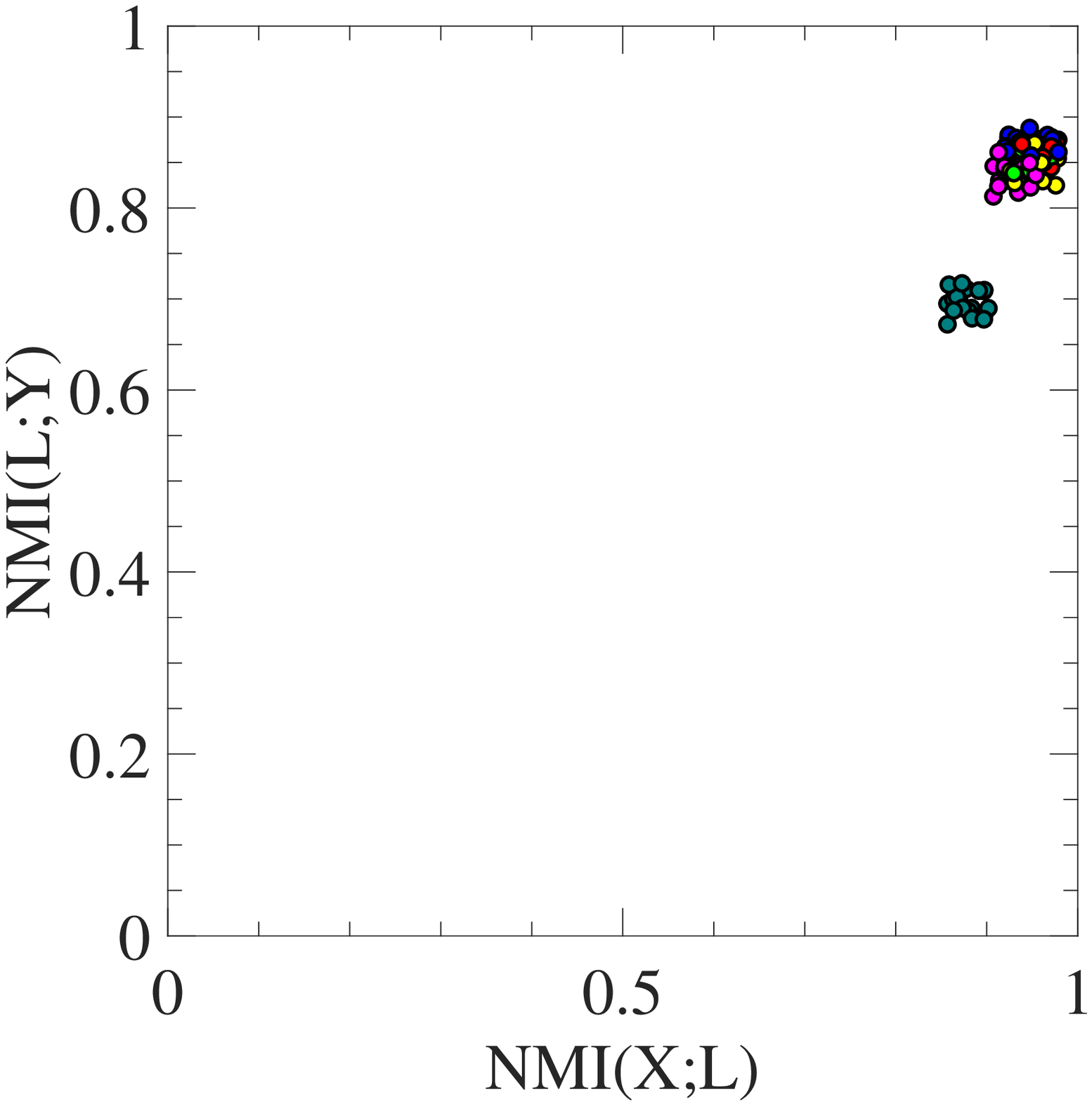}
    \caption{$CNN_{gaf}^7$ after 300 epochs}
  \end{subfigure}
  \hfill
  \begin{subfigure}[t]{0.32\linewidth}
    \centering
    \includegraphics[width=\linewidth]{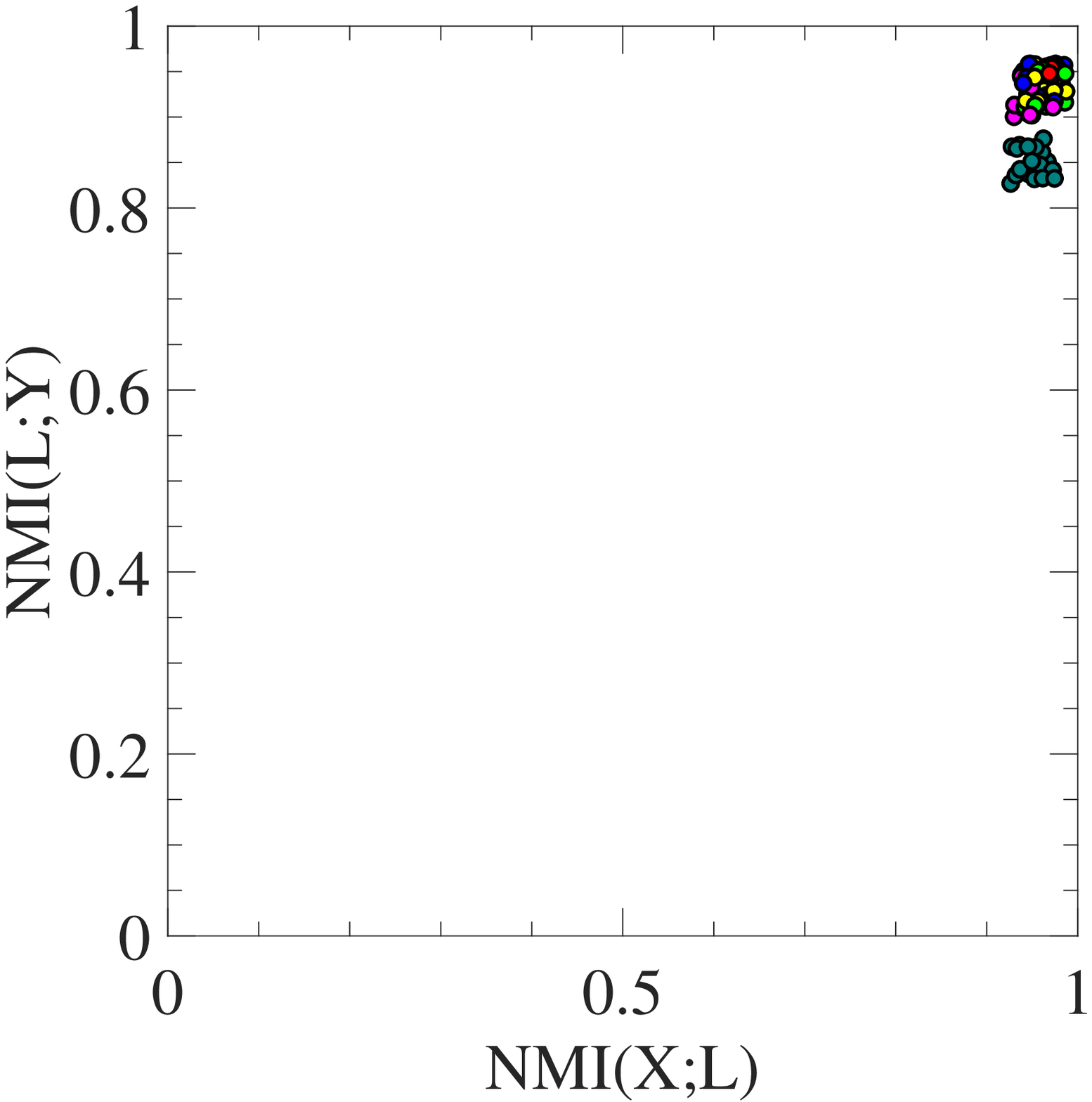}
    \caption{$CNN_{gaf}^7$ after 2000 epochs}
  \end{subfigure}

  \begin{subfigure}[t]{0.32\linewidth}
    \centering
    \includegraphics[width=\linewidth]{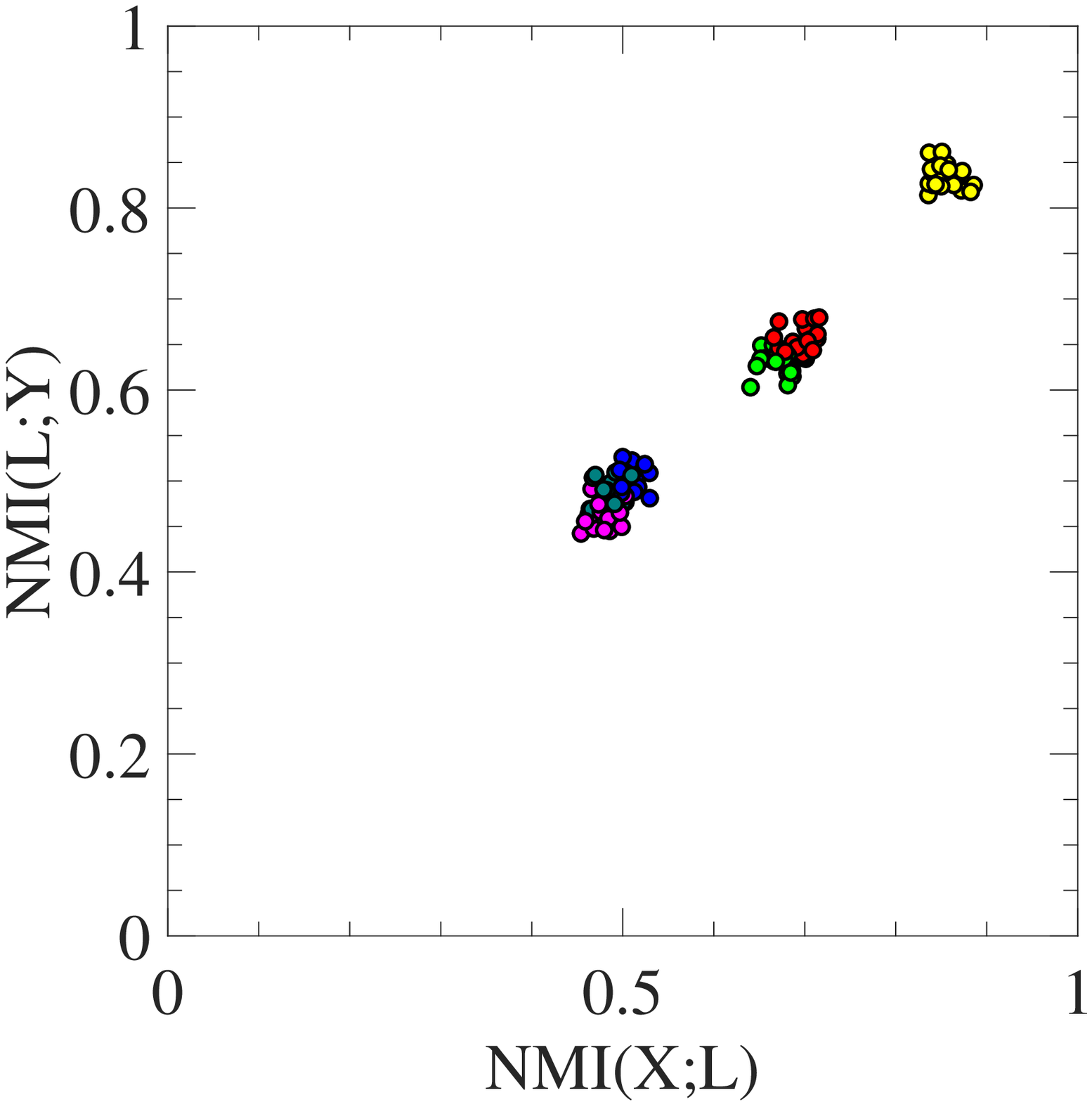}
    \caption{$CNN_{eig}^7$ at initial state}
  \end{subfigure}
  \hfill
  \begin{subfigure}[t]{0.32\linewidth}
    \centering
    \includegraphics[width=\linewidth]{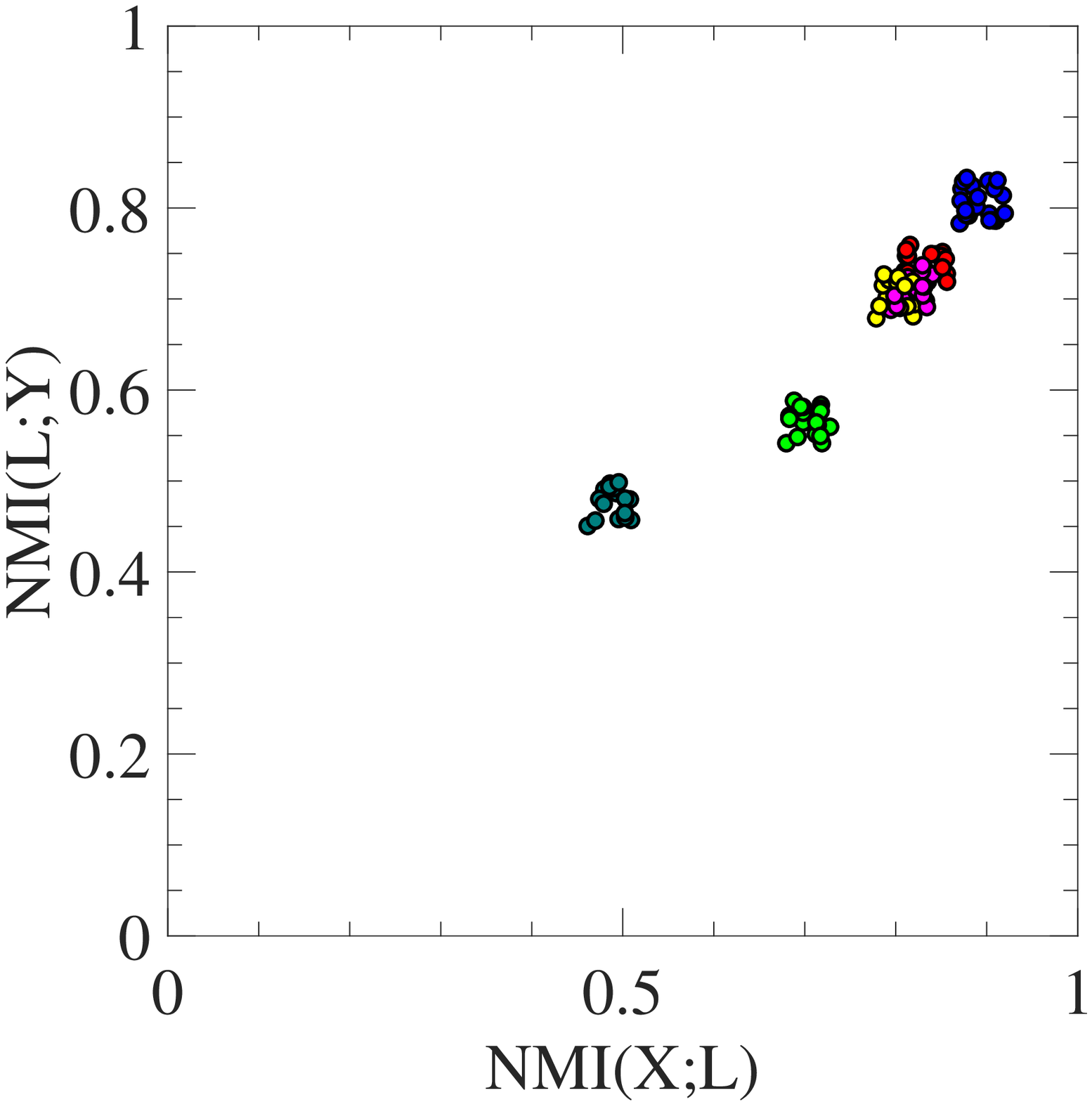}
    \caption{$CNN_{eig}^7$ after 300 epochs}
  \end{subfigure}
  \hfill
  \begin{subfigure}[t]{0.32\linewidth}
    \centering
    \includegraphics[width=\linewidth]{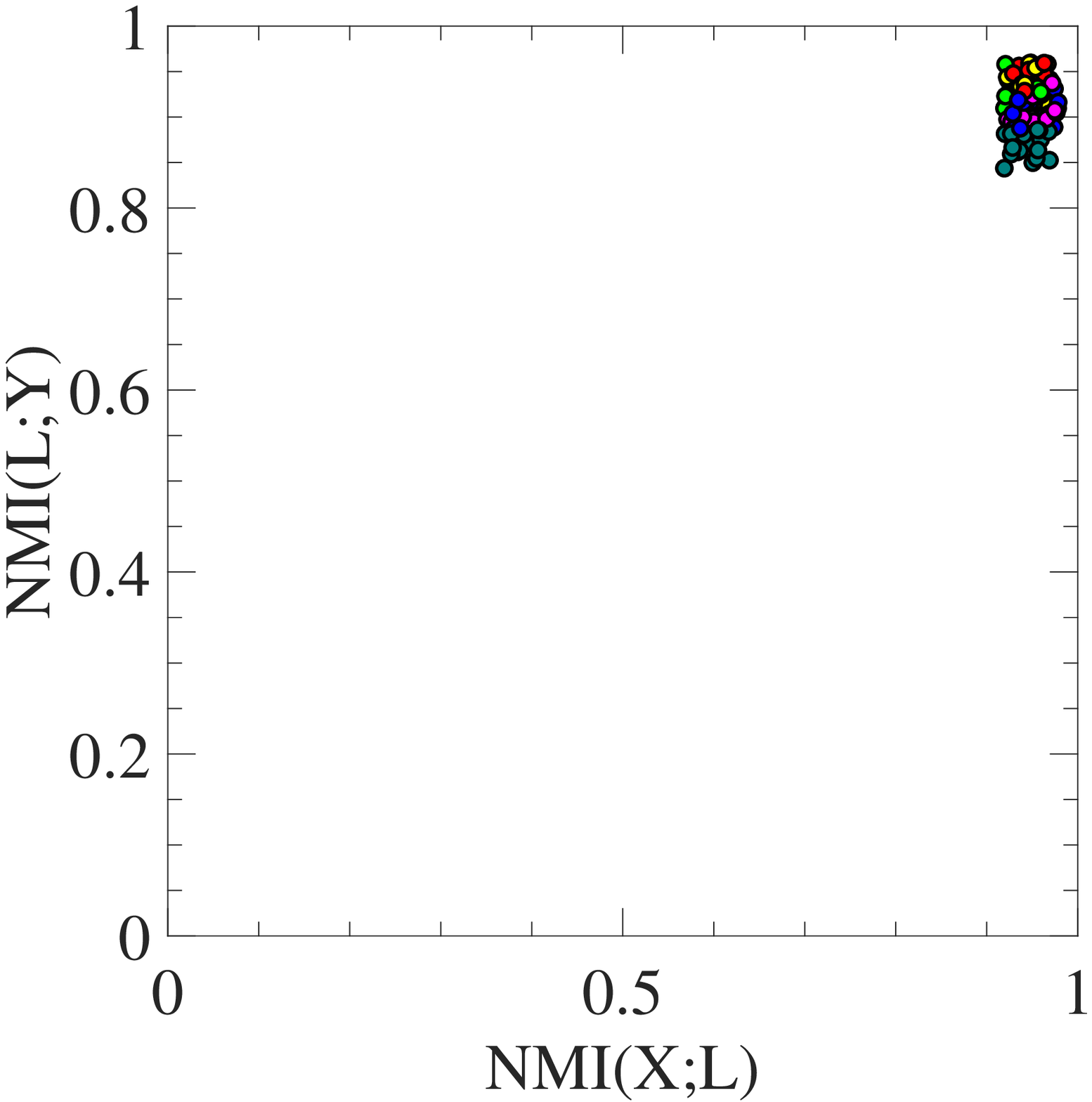}
    \caption{$CNN_{eig}^7$ after 2000 epochs}
  \end{subfigure}
  \caption{Normalized Mutual Information (NMI) for 20 randomly initialized $CNN_{cov}^7$, $CNN_{gaf}^7$ and $CNN_{eig}^7$ models. Legend: NMI between input/output and output of (\protect\solidcircle[black, fill=yellow]) Convolutional Layer 1, (\protect\solidcircle[black, fill=red]) Convolutional Layer 2, (\protect\solidcircle[black, fill=green]) Max Pooling Layer 1, (\protect\solidcircle[black, fill=blue]) Convolutional Layer 3, (\protect\solidcircle[black, fill=darkpink]) Max Pooling Layer 2, (\protect\solidcircle[black, fill=skyblue]) Fully Connected Layer}\label{Fig:Small-NMI}
\end{figure*}

\begin{figure*}[h!]
  \centering
  \begin{subfigure}{0.32\textwidth}
    \centering
    \includegraphics[width=\textwidth]{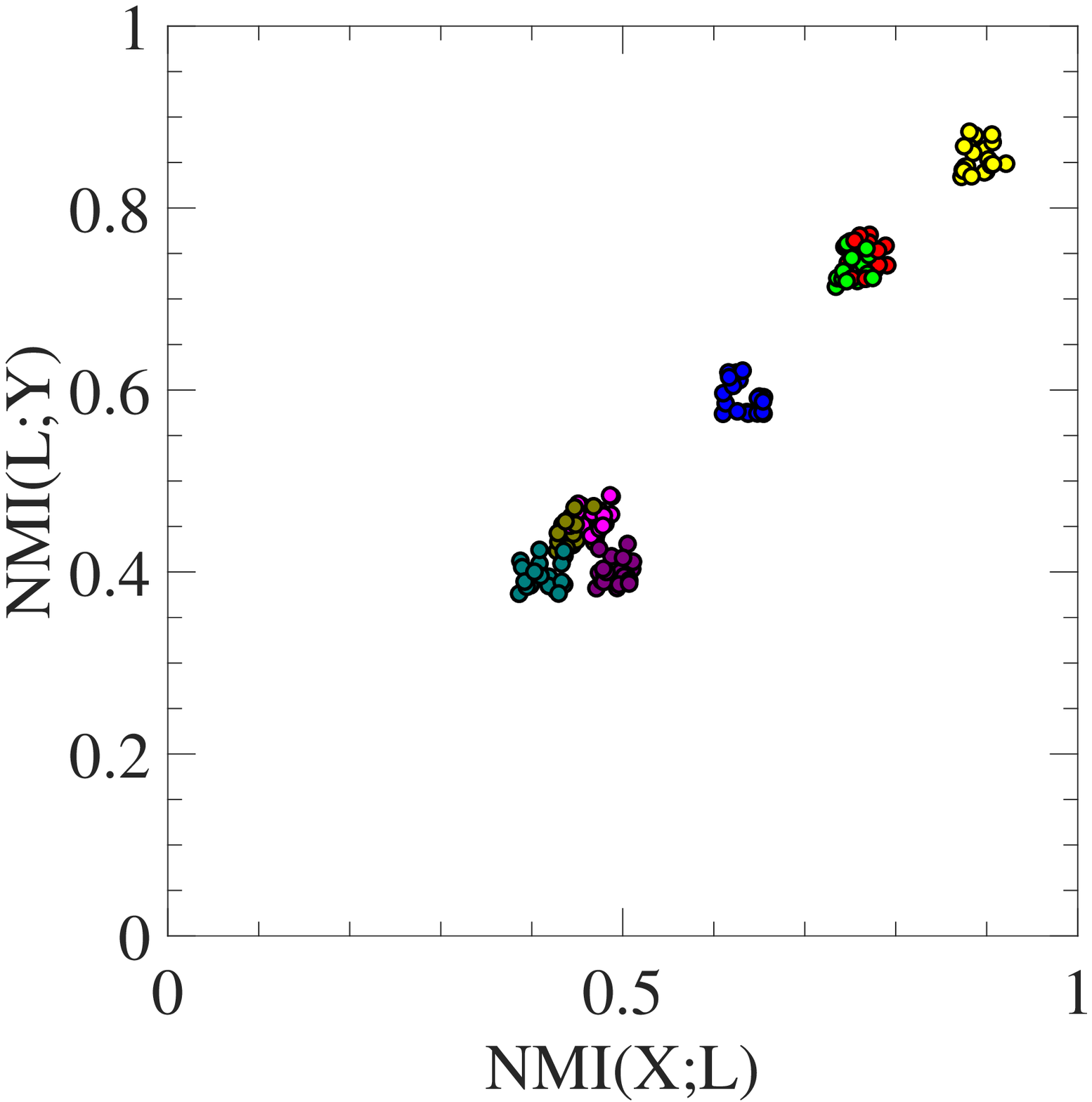}
    \caption{$CNN_{cov}^9$ at initial state}
  \end{subfigure}
  \hfill
  \begin{subfigure}{0.32\textwidth}
    \centering
    \includegraphics[width=\textwidth]{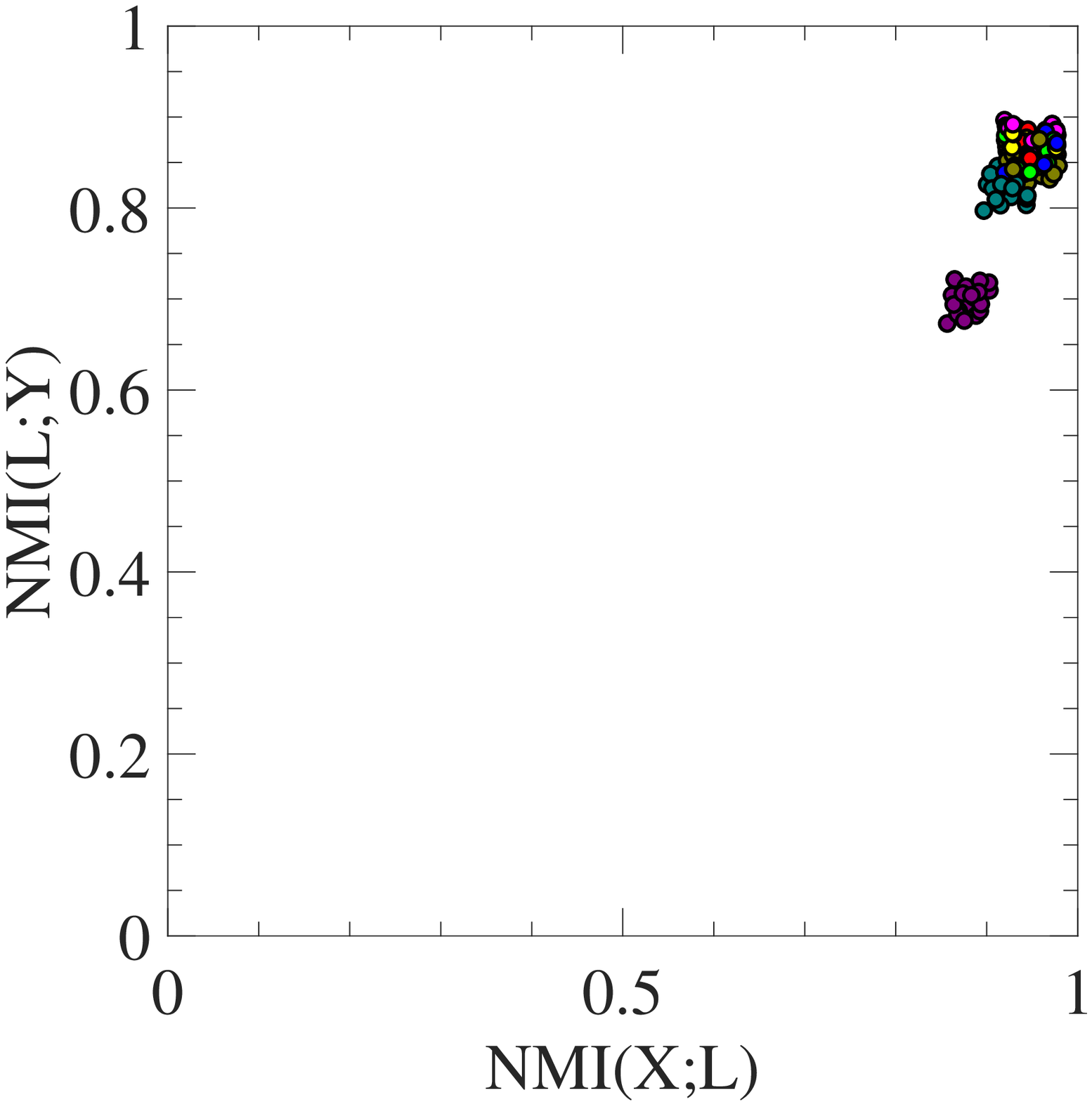}
    \caption{$CNN_{cov}^9$ after 300 epochs}
  \end{subfigure}
  \hfill
  \begin{subfigure}{0.32\textwidth}
    \centering
    \includegraphics[width=\textwidth]{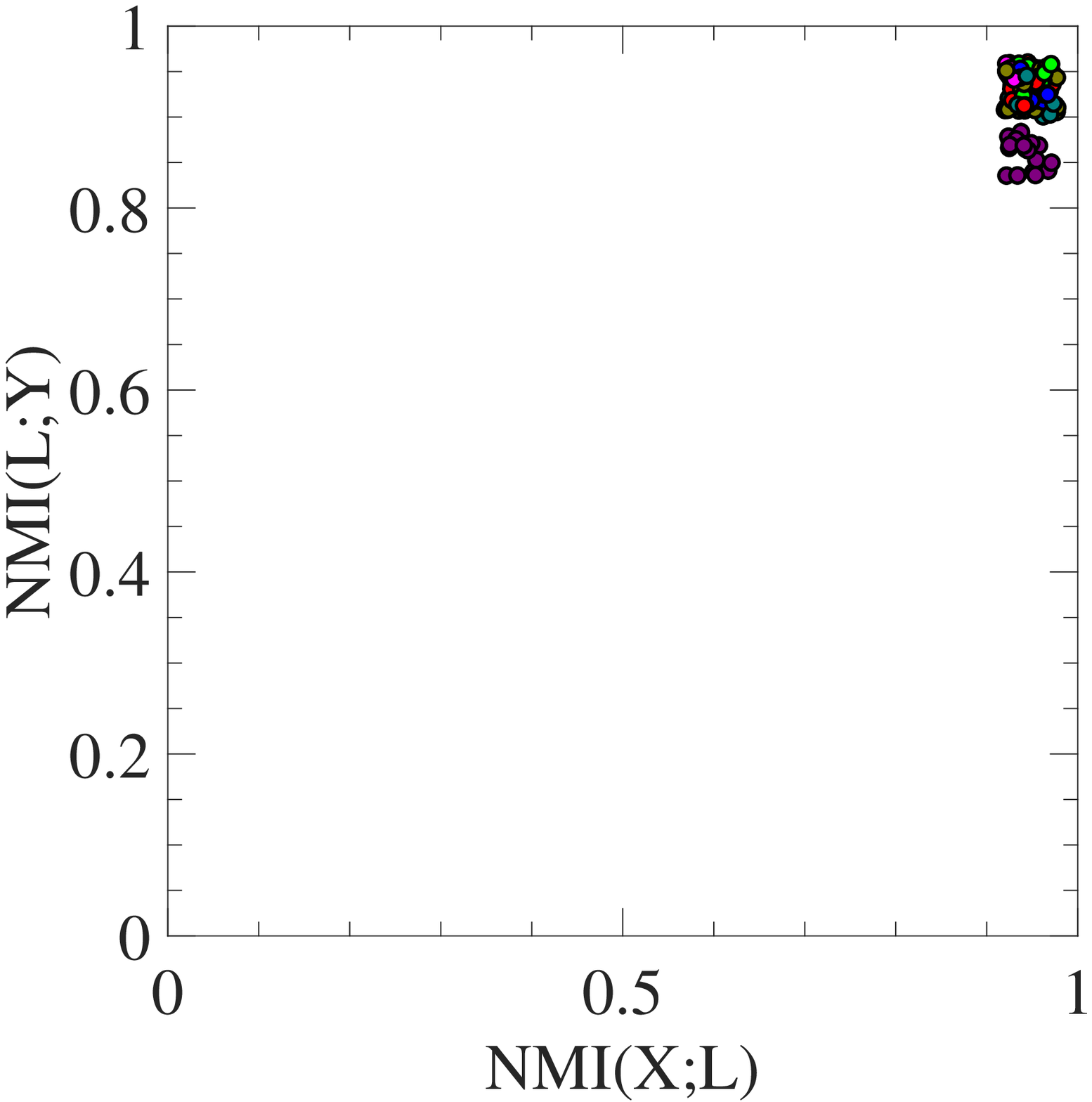}
    \caption{$CNN_{cov}^9$ after 2000 epochs}
  \end{subfigure}

  \begin{subfigure}{0.32\textwidth}
    \centering
    \includegraphics[width=\textwidth]{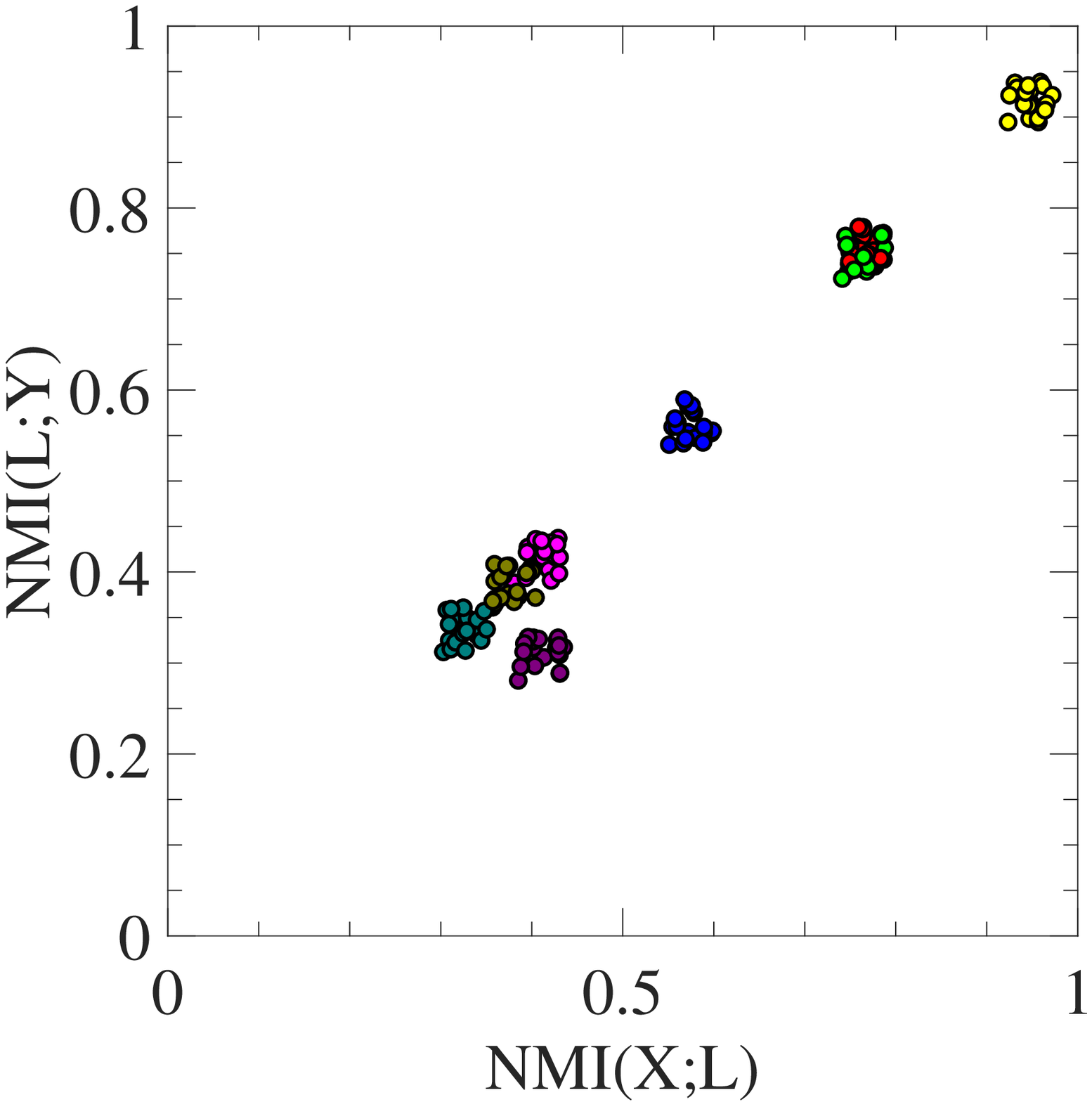}
    \caption{$CNN_{gaf}^9$ at initial state}
  \end{subfigure}
  \hfill
  \begin{subfigure}{0.32\textwidth}
    \centering
    \includegraphics[width=\textwidth]{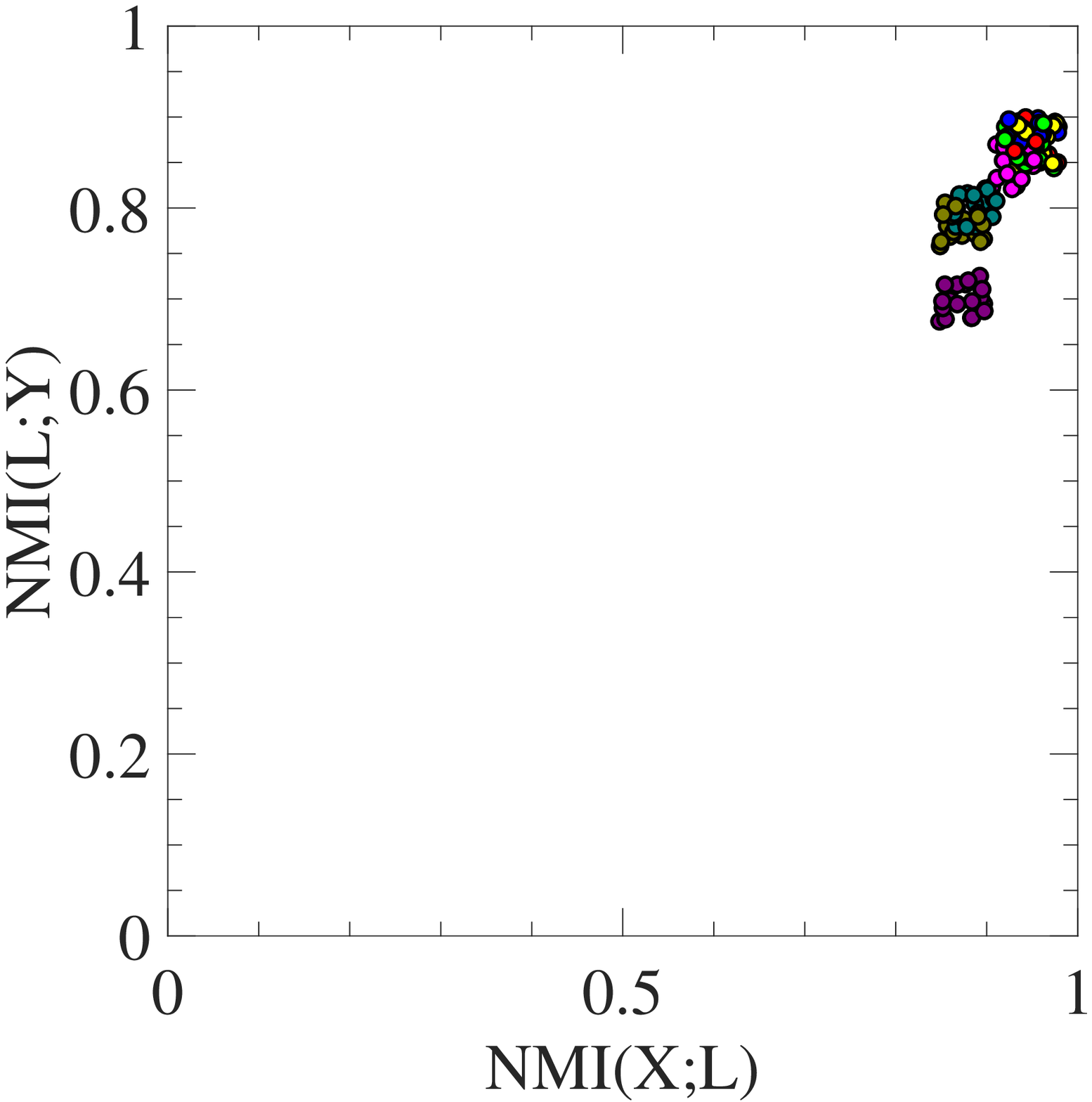}
    \caption{$CNN_{gaf}^9$ after 300 epochs}
  \end{subfigure}
  \hfill
  \begin{subfigure}{0.32\textwidth}
    \centering
    \includegraphics[width=\textwidth]{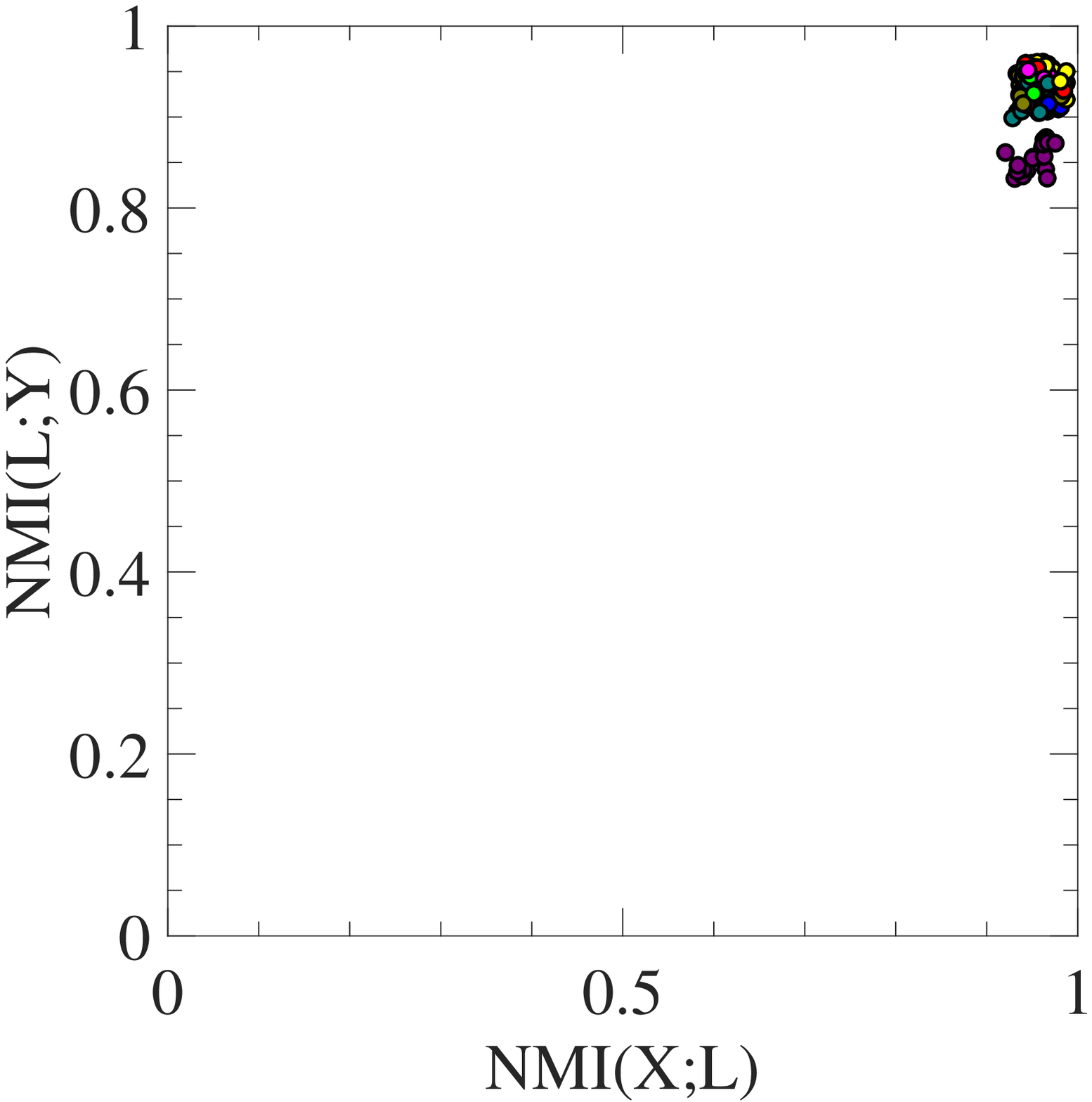}
    \caption{$CNN_{gaf}^9$ after 2000 epochs}
  \end{subfigure}

  \begin{subfigure}{0.32\textwidth}
    \centering
    \includegraphics[width=\textwidth]{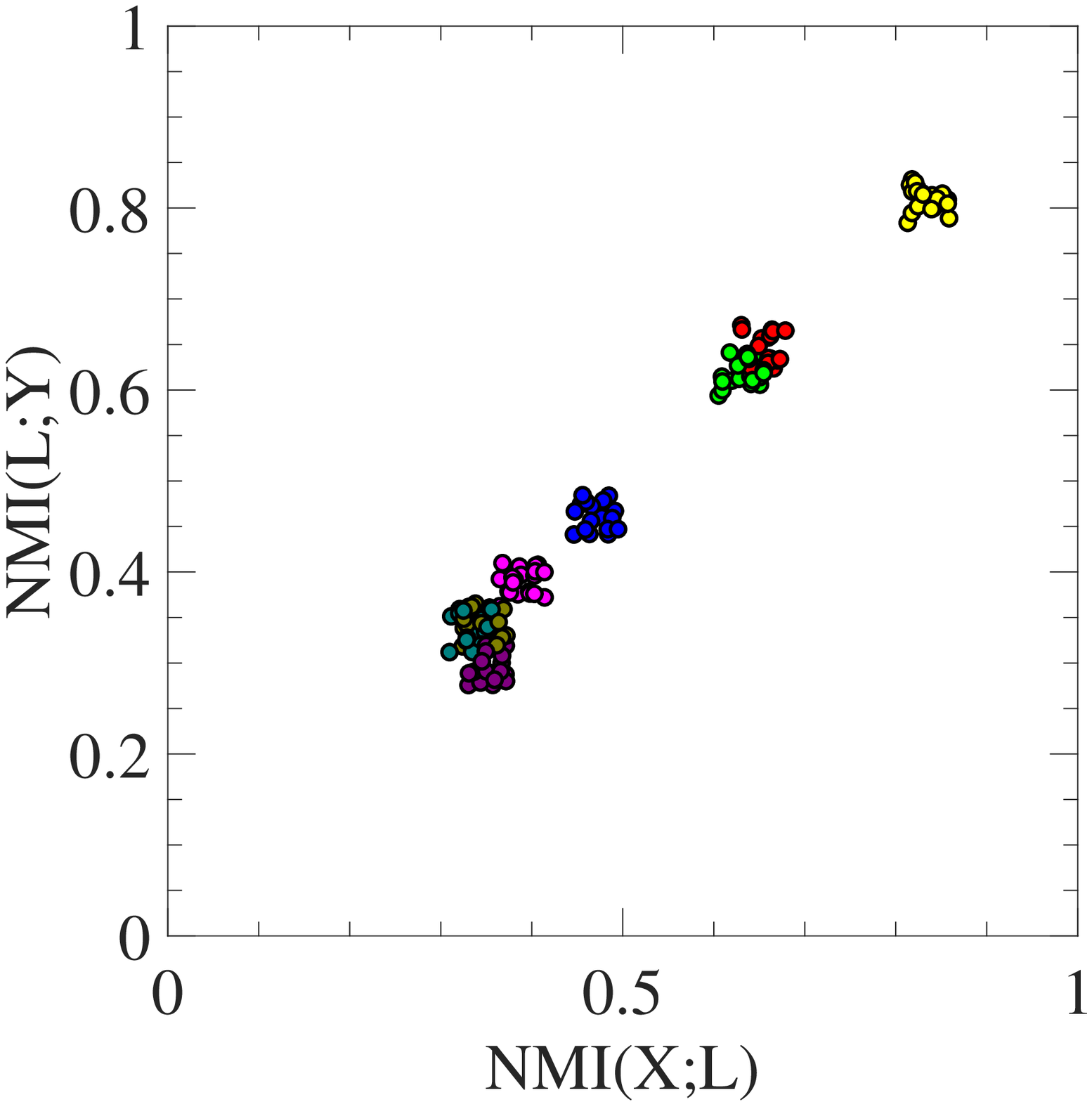}
    \caption{$CNN_{eig}^9$ at initial state}
  \end{subfigure}
  \hfill
  \begin{subfigure}{0.32\textwidth}
    \centering
    \includegraphics[width=\textwidth]{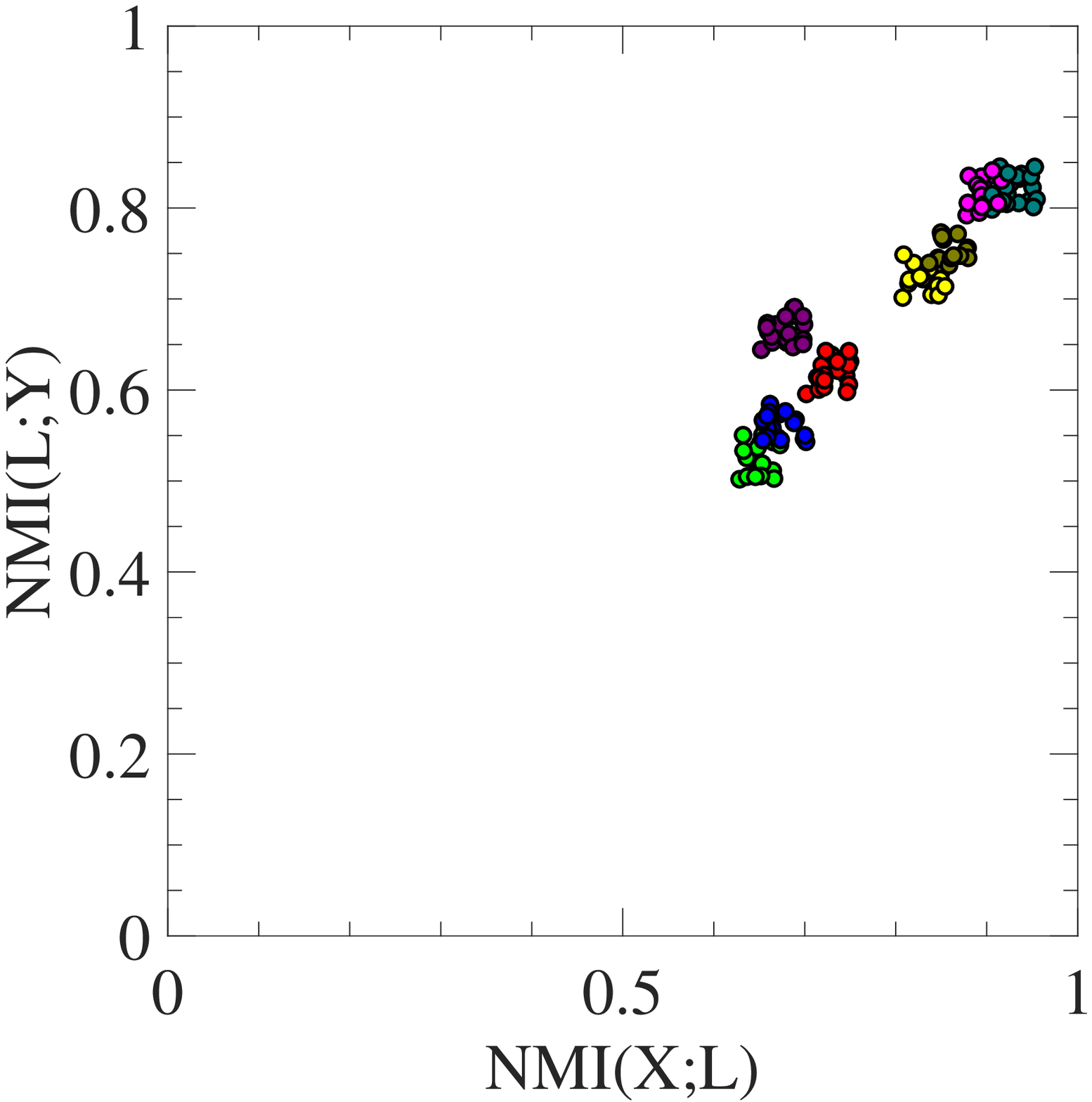}
    \caption{$CNN_{eig}^9$ after 300 epochs}
  \end{subfigure}
  \hfill
  \begin{subfigure}{0.32\textwidth}
    \centering
    \includegraphics[width=\textwidth]{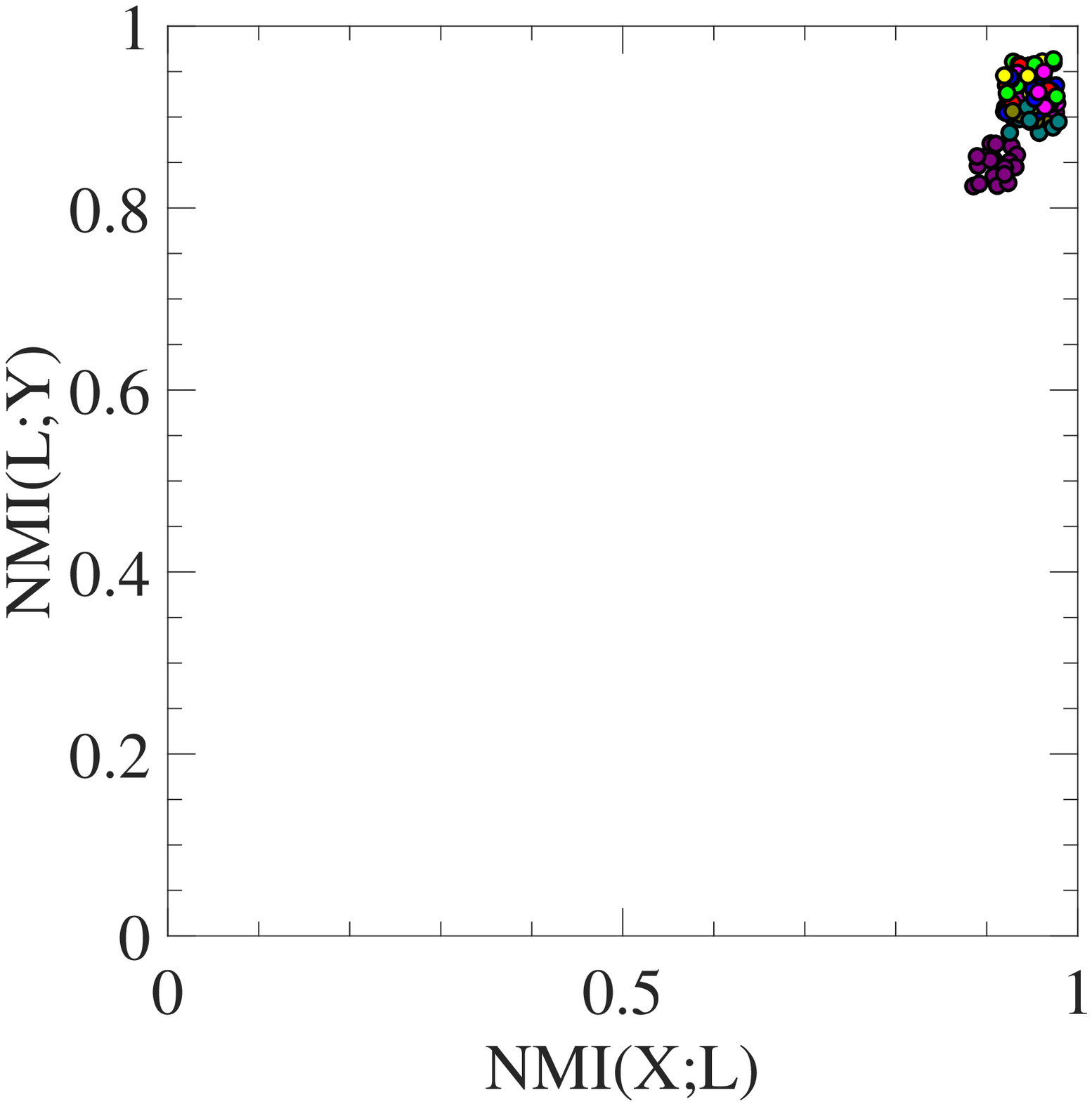}
    \caption{$CNN_{eig}^9$ after 2000 epochs}
  \end{subfigure}
  \caption{Normalized Mutual Information (NMI) for 20 randomly initialized $CNN_{cov}^9$, $CNN_{gaf}^9$ and $CNN_{eig}^9$ models. Legend: NMI between input/output and output of (\protect\solidcircle[black, fill=yellow]) Convolutional Layer 1, (\protect\solidcircle[black, fill=red]) Convolutional Layer 2, (\protect\solidcircle[black, fill=green]) Max Pooling Layer 1, (\protect\solidcircle[black, fill=blue]) Convolutional Layer 3, (\protect\solidcircle[black, fill=darkpink]) Convolutional Layer 4, (\protect\solidcircle[black, fill=henna]) Max Pooling Layer 2, (\protect\solidcircle[black, fill=skyblue]) Convolutional Layer 5, (\protect\solidcircle[black, fill=purple]) Fully Connected Layer}\label{Fig:Big-NMI}
\end{figure*}

\begin{figure*}[h!]
  \centering
  \begin{subfigure}[]{0.49\textwidth}
    \centering
    \includegraphics[width=\linewidth]{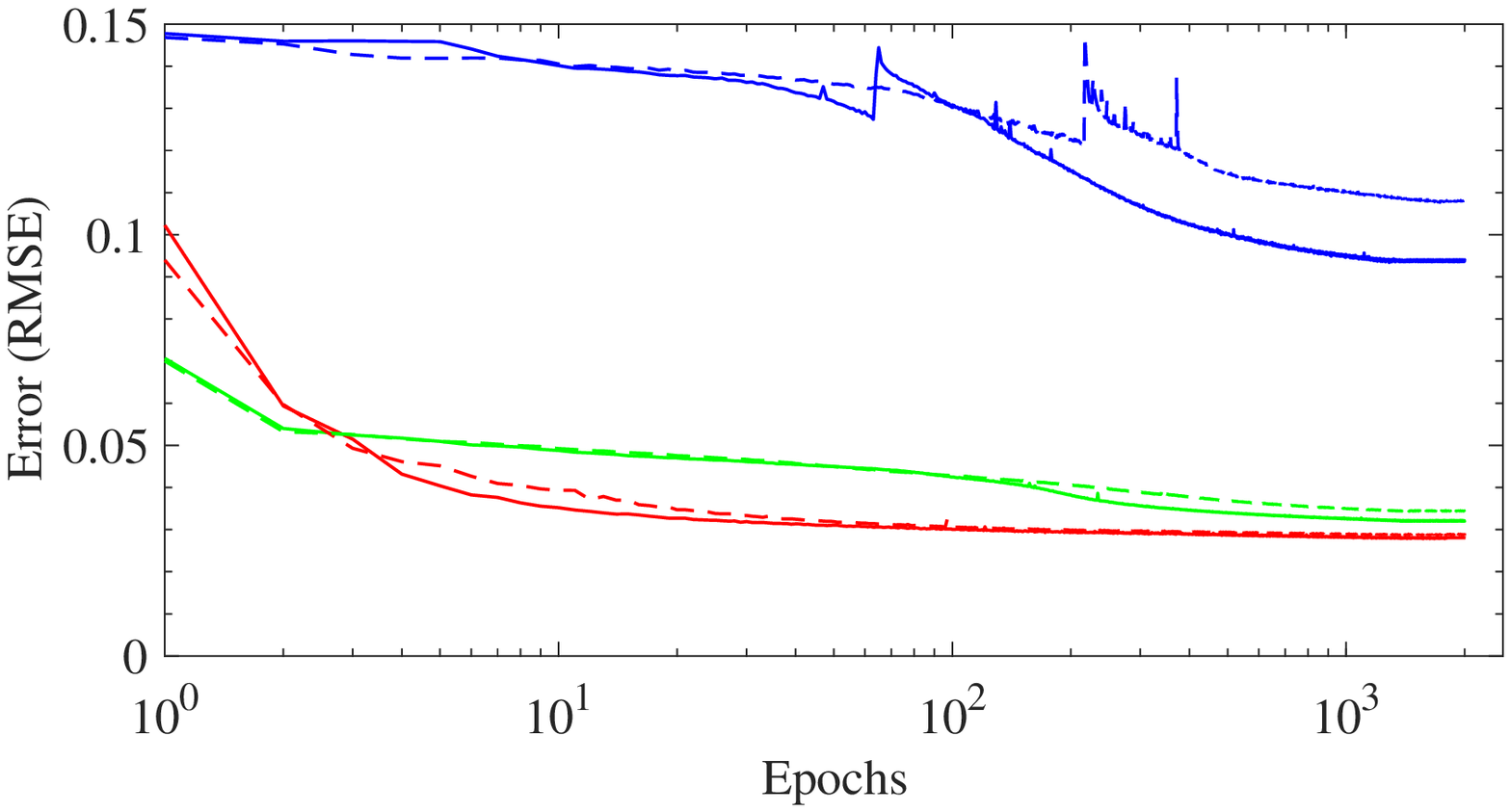}
    \caption{}\label{Fig:TrainingLoss}
  \end{subfigure}
  \hfill
  \begin{subfigure}[]{0.49\textwidth}
    \centering
    \includegraphics[width=\linewidth]{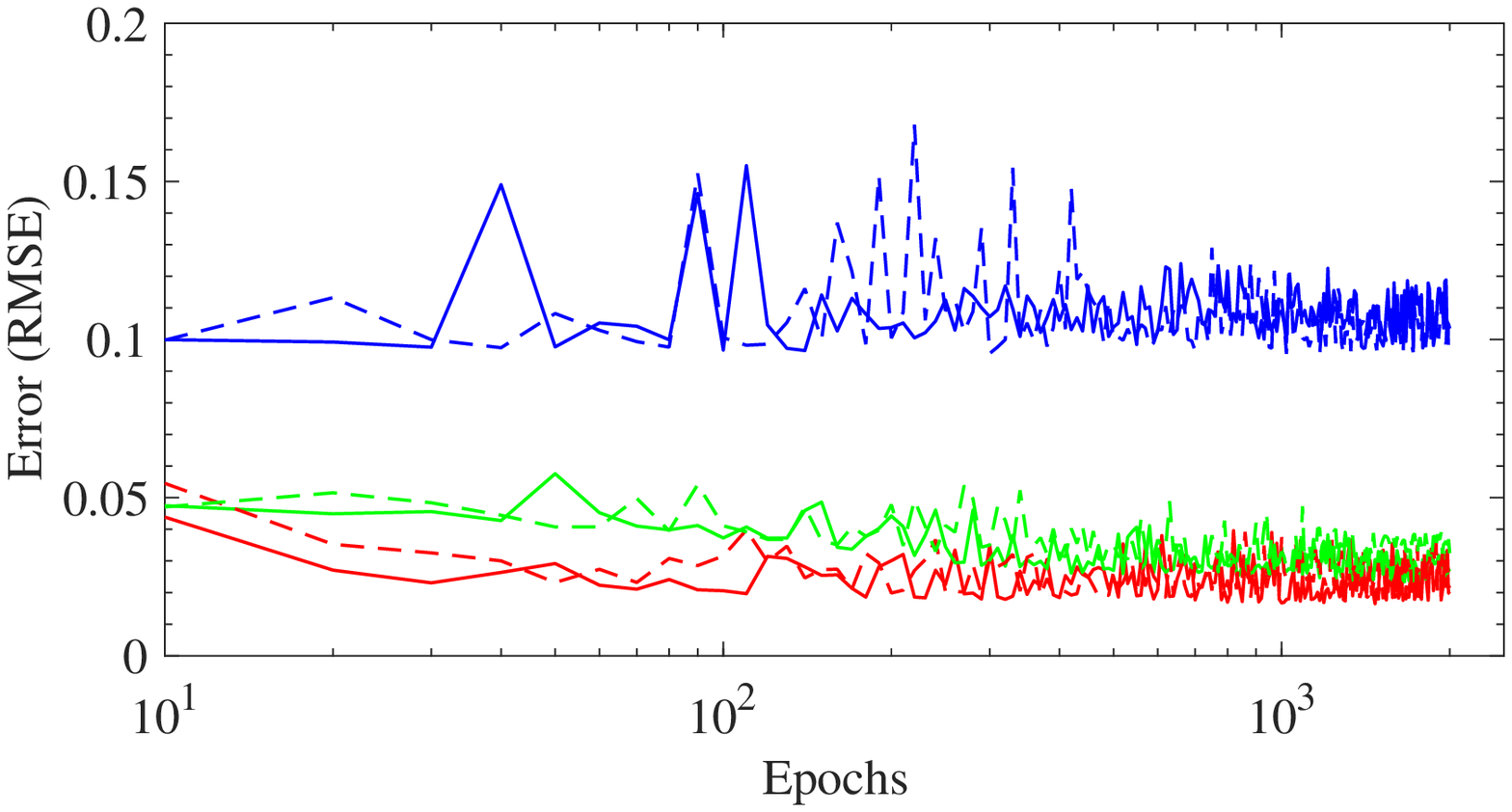}
    \caption{}\label{Fig:ValidationLoss}
  \end{subfigure}
  \caption{(a) Training and (b) Validation of proposed CNN models across 2000 epochs. Legend: (\protect\bluesolidline) $CNN_{eig}^9$, (\protect\bluedashedline) $CNN_{eig}^7$, (\protect\greensolidline) $CNN_{gaf}^9$, (\protect\greendashedline) $CNN_{gaf}^7$, (\protect\redsolidline) $CNN_{cov}^9$, (\protect\reddashedline) $CNN_{cov}^7$}\label{Fig:TrainingValidation}
\end{figure*}

For the experimental purpose, the normalized mutual information for each CNN model was calculated between each layer's output and the model's input i.e. $NMI(X;L^i)$ and between each layer's output and the model's output i.e. $NMI(L^i;Y)$. Here $L^i$ represent the $i^{th}$ layer's output, $X \subseteq \mathds{X}$ and $Y \subseteq \mathds{Y}$ represent the input and output of particular CNN model. For calculating the normalized mutual information, $X$, $Y$, and $L^i$ were binned into 5000 equal intervals. Then these discretized $X$, $Y$ and $L^i$ were used to calculate their joint distributions and hence, normalized mutual information $NMI(X;L^i)$ and $NMI(L^i;Y)$. These calculations were performed repeatedly for 20 randomly initialized CNN models for each CNN architecture trained with 75\% of randomly selected training samples. The variations in normalized mutual information for CNN models trained using the dataset $DS-I$ after preprocessing using GAF, Covariance and Eigen Vector methods are shown in Figures \ref{Fig:Small-NMI} and \ref{Fig:Big-NMI}. It can be clearly seen from these figures that normalized mutual information grows as the training progresses and all the layers starting from different initial state try to obtain the relevant information. The information gain was quite large from initial state till approximately $300^{th}$ epoch and after that not much information gain has happened. So, it is evident from the figures that the networks are training well and can be used even after approximately $300^{th}$ epoch because after that the layers are just optimizing themselves and not much information gain is happening. Also, the layers of randomized networks form clusters and behave similarly. So, it is justified to take the average of randomized networks and plot the average training and validation error across 2000 epochs as shown in Figure \ref{Fig:TrainingValidation}. The training and validation error shown in the figure was calculated from normalized actual and predicted output of the particular CNN model.

Validation was performed at every $10^{th}$ epoch. Figure \ref{Fig:TrainingValidation}, shows how the choice of input feature descriptor effects the training and performance of CNN models. It can be clearly observed that the CNN models trained with Eigen Vectors, i.e., $CNN^7_{eig}$ and $CNN^9_{eig}$, have high training and validation error as compared to other CNN models trained with covariance and GAF features. The CNN models trained with covariance feature descriptors outperformed the other models with minimum training and validation error, but out of $CNN_{cov}^7$ and $CNN_{cov}^9$, which one is better, it is very hard to conclude from the figure as their training and validation error are overlapping. Also, it can be observed that the CNN models $CNN^7_{cov}$ and $CNN^9_{cov}$ converged before other models at approximately $300^{th}$ epoch. These all observations are in accordance to the discussion in section \ref{Subsec:TimeSeriesImageEncoder}, where it has been explained that the GAF features have some information loss which effect the performance of CNN models trained with GAF. Similarly, the eigen vectors descriptors contain only the direction of variance and loose most of the relevant information as compared to the covariance features descriptors.

\begin{figure*}[h!]
  \centering
  \begin{subfigure}[t]{0.32\textwidth}
    \centering
    \includegraphics[width=\textwidth]{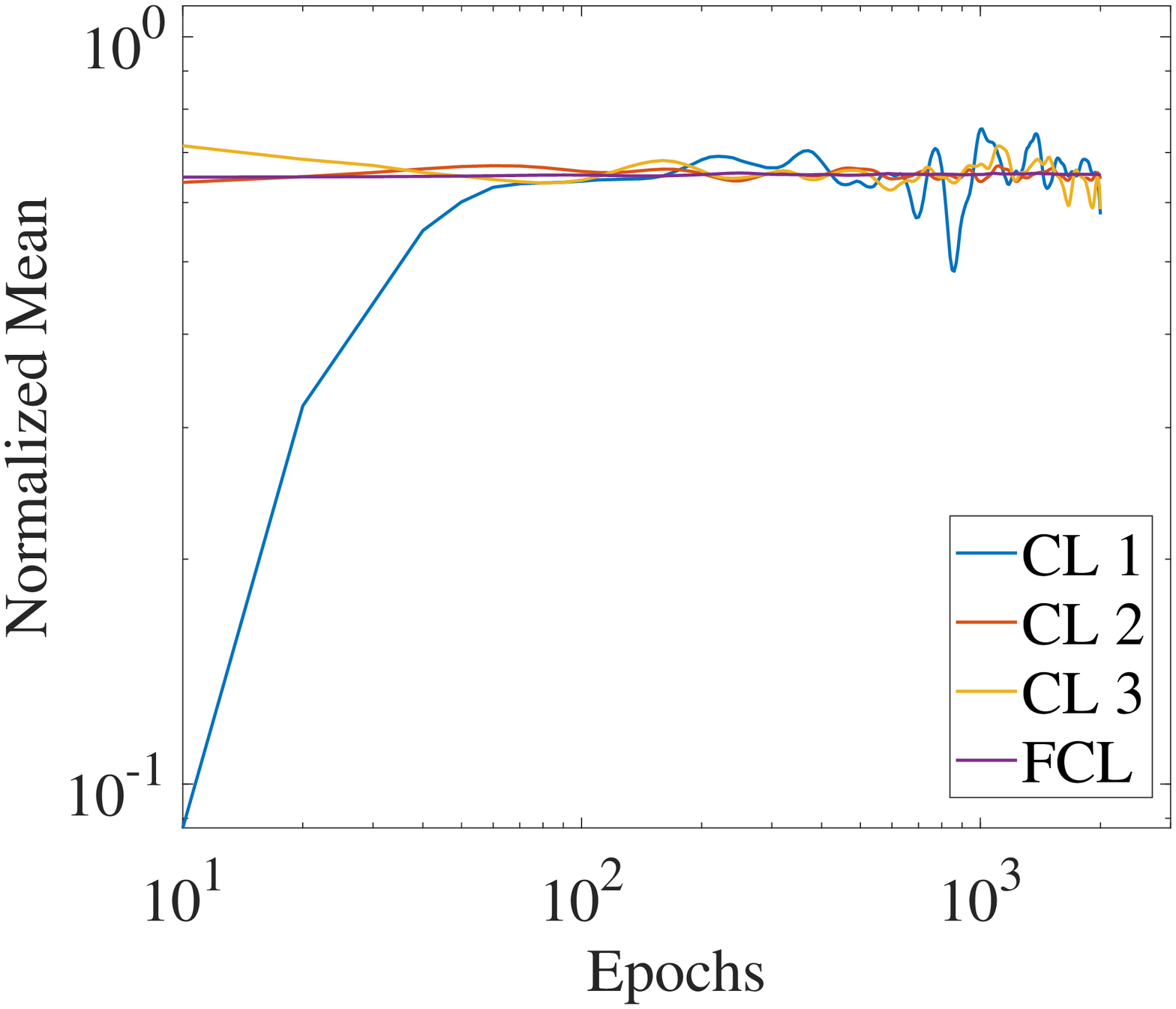}
    \caption{Normalized mean for layers of $CNN_{cov}^7$}
  \end{subfigure}
  \hfill
  \begin{subfigure}[t]{0.32\textwidth}
    \centering
    \includegraphics[width=\textwidth]{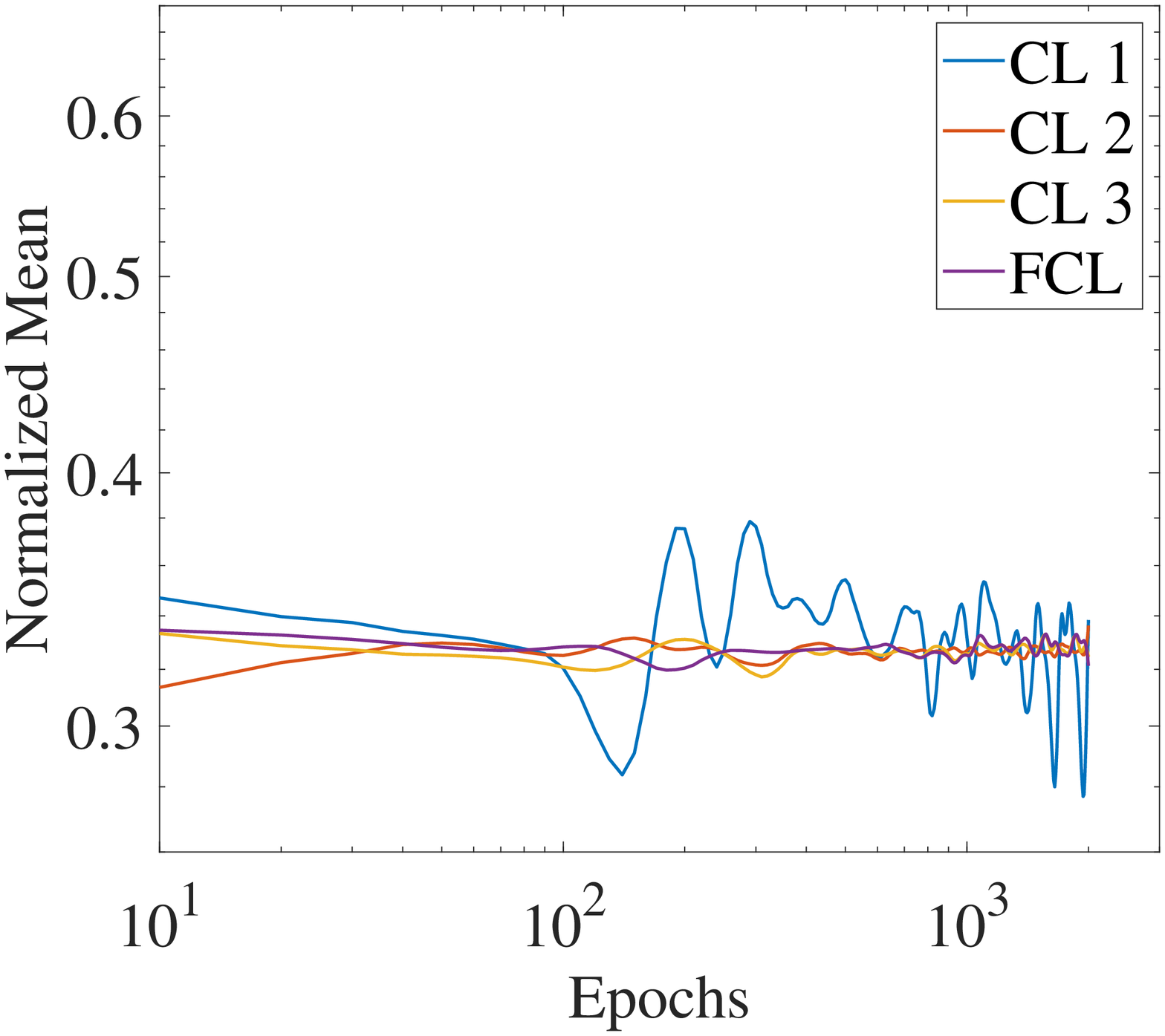}
    \caption{Normalized mean for layers of $CNN_{eig}^7$}
  \end{subfigure}
  \hfill
  \begin{subfigure}[t]{0.32\textwidth}
    \centering
    \includegraphics[width=\textwidth]{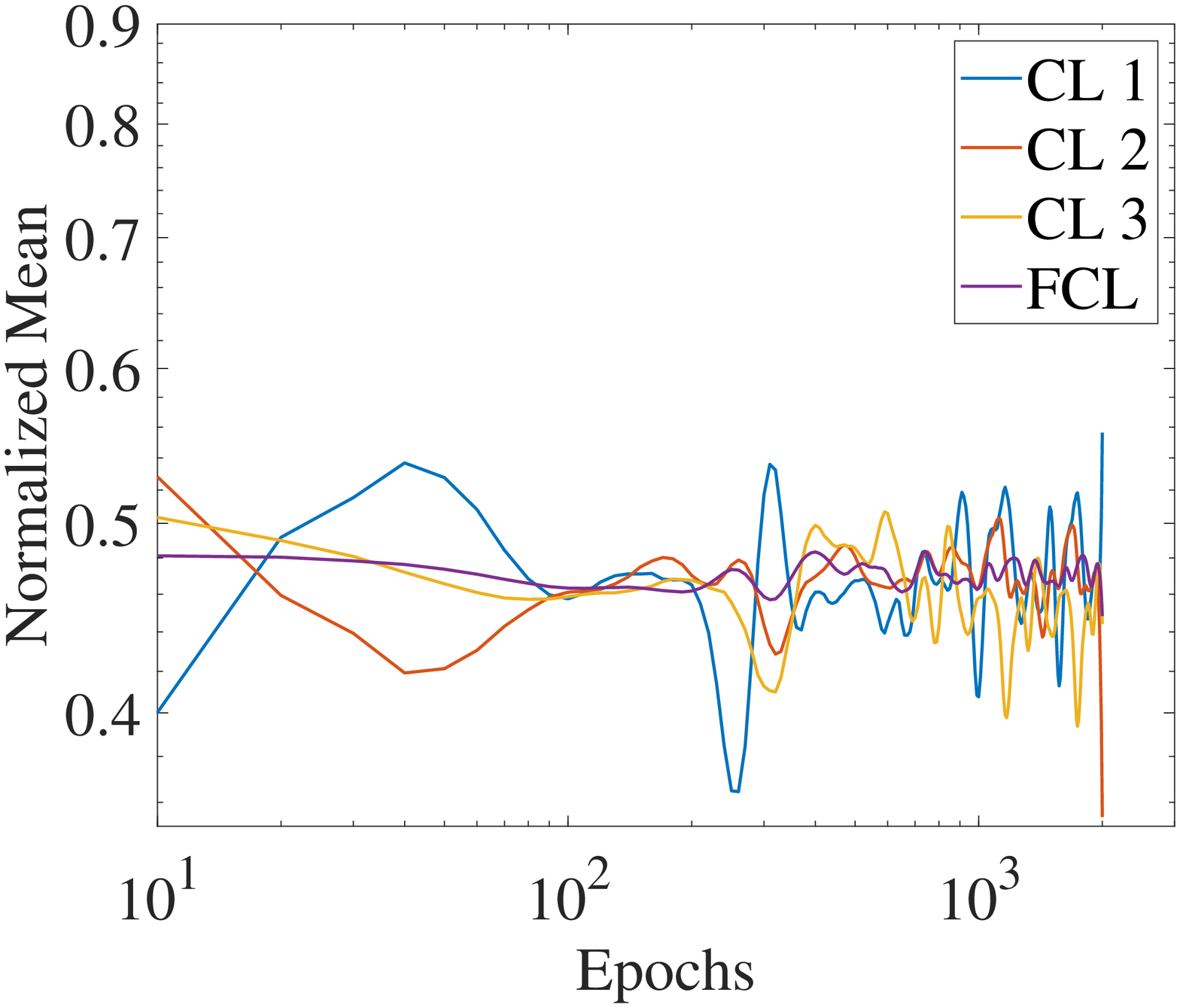}
    \caption{Normalized mean for layers of $CNN_{gaf}^7$}
  \end{subfigure}
  \\
  \vspace{2mm}

  \begin{subfigure}[t]{0.32\textwidth}
    \centering
    \includegraphics[width=\textwidth]{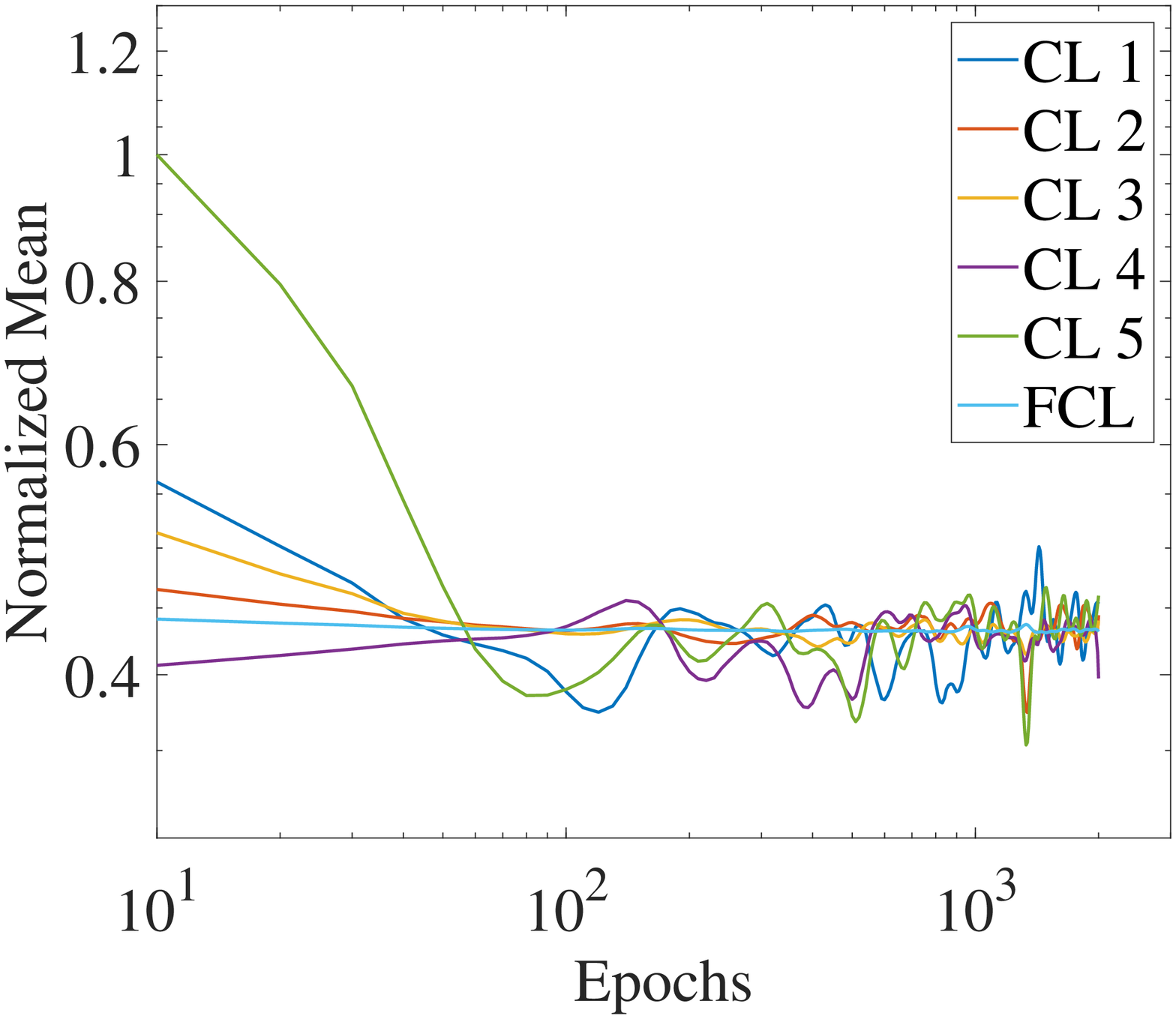}
    \caption{Normalized mean for layers of $CNN_{cov}^9$}
  \end{subfigure}
  \hfill
  \begin{subfigure}[t]{0.32\textwidth}
    \centering
    \includegraphics[width=\textwidth]{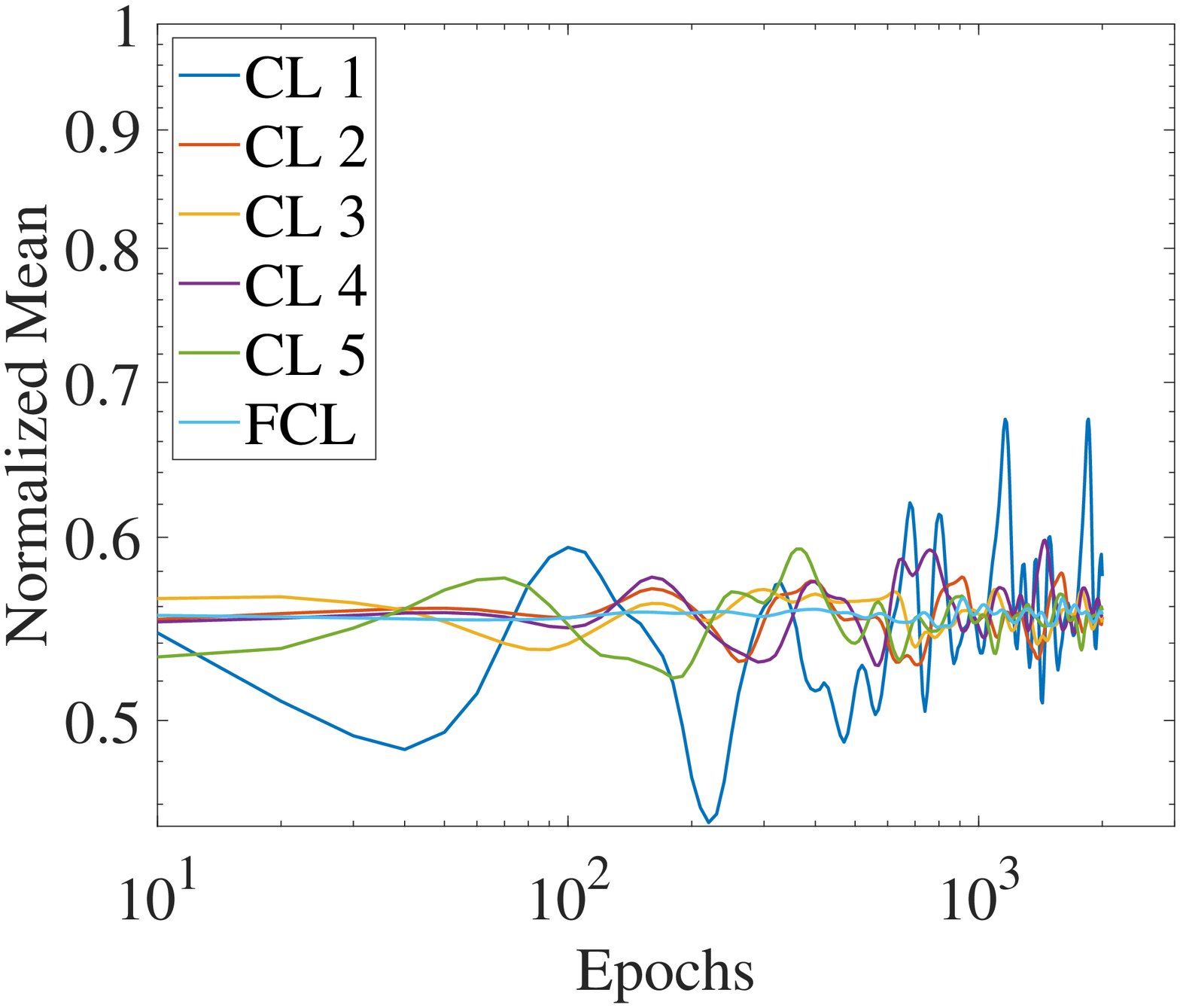}
    \caption{Normalized mean for layers of $CNN_{eig}^9$}
  \end{subfigure}
  \hfill
  \begin{subfigure}[t]{0.32\textwidth}
    \centering
    \includegraphics[width=\textwidth]{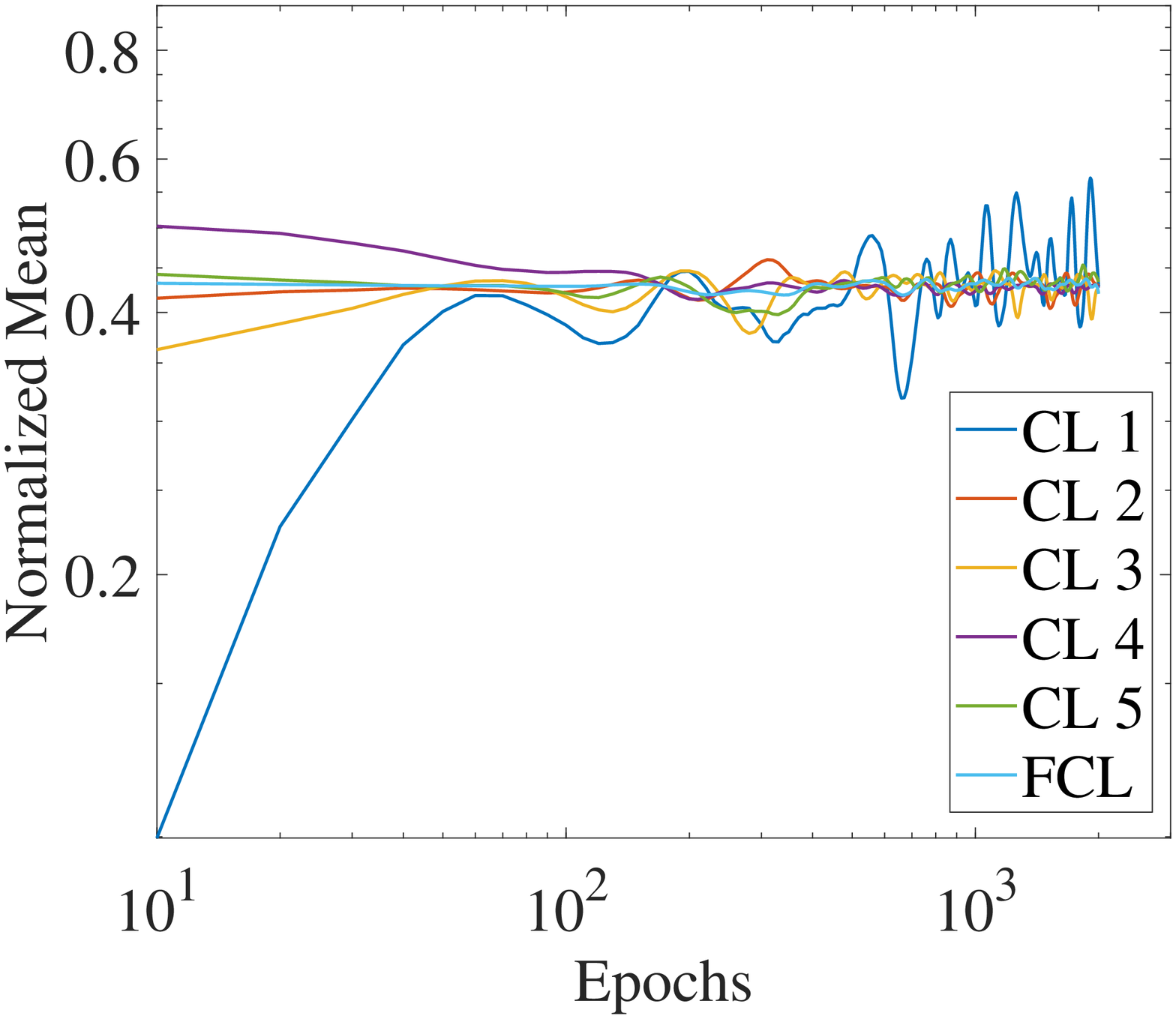}
    \caption{Normalized mean for layers of $CNN_{gaf}^9$}
  \end{subfigure}
  \caption{Normalized Mean of gradient weights of each layer of proposed CNN models. In legend CL and FCL stands for Convolutional and Fully Connected Layer, respectively}\label{Fig:LayersMean}
\end{figure*}

\begin{figure*}[h!]
  \centering
  \begin{subfigure}{0.32\textwidth}
    \centering
    \includegraphics[width=\textwidth]{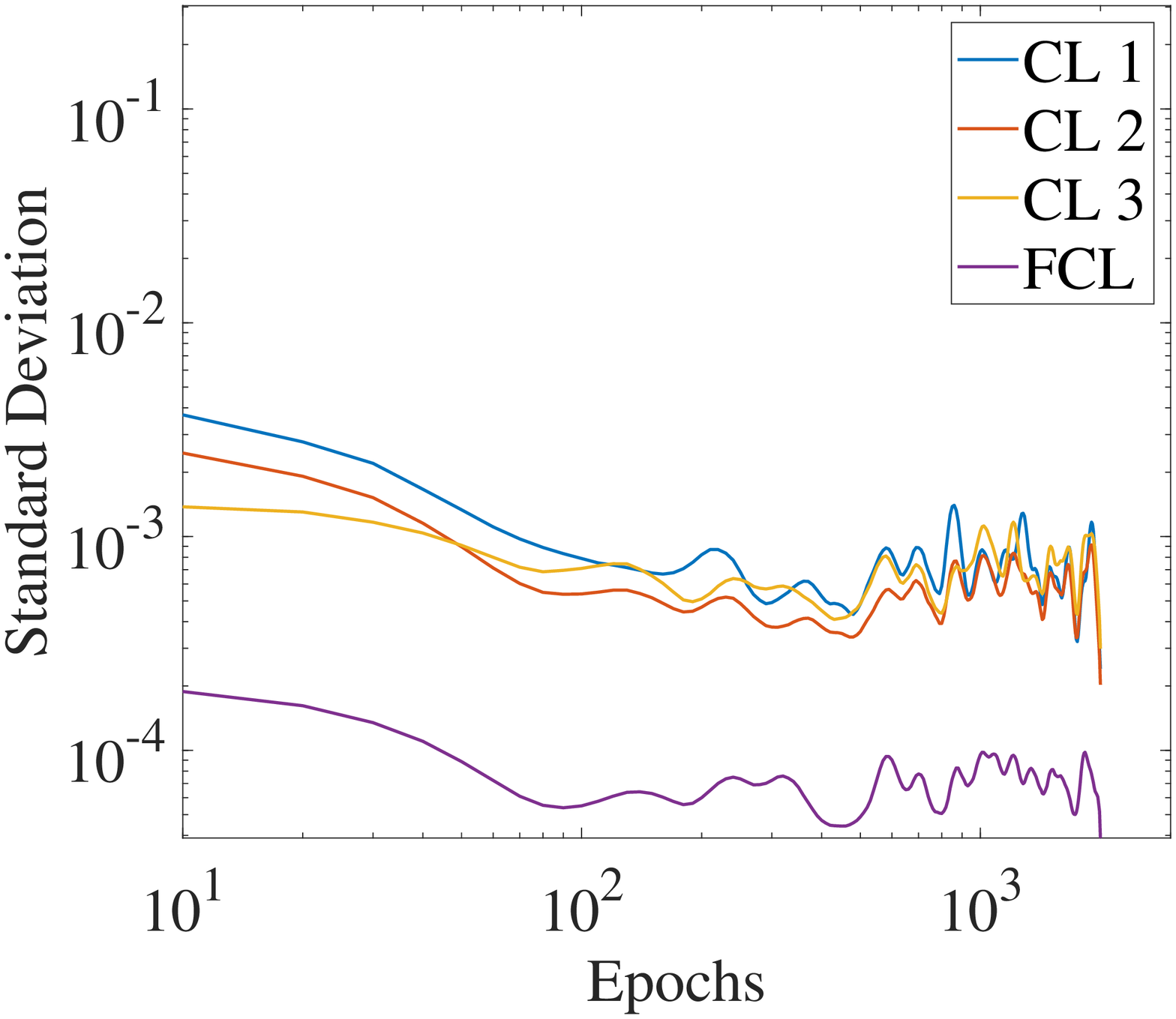}
    \caption{Standard deviation for layers of $CNN_{cov}^7$}
  \end{subfigure}
  \hfill
  \begin{subfigure}{0.32\textwidth}
    \centering
    \includegraphics[width=\textwidth]{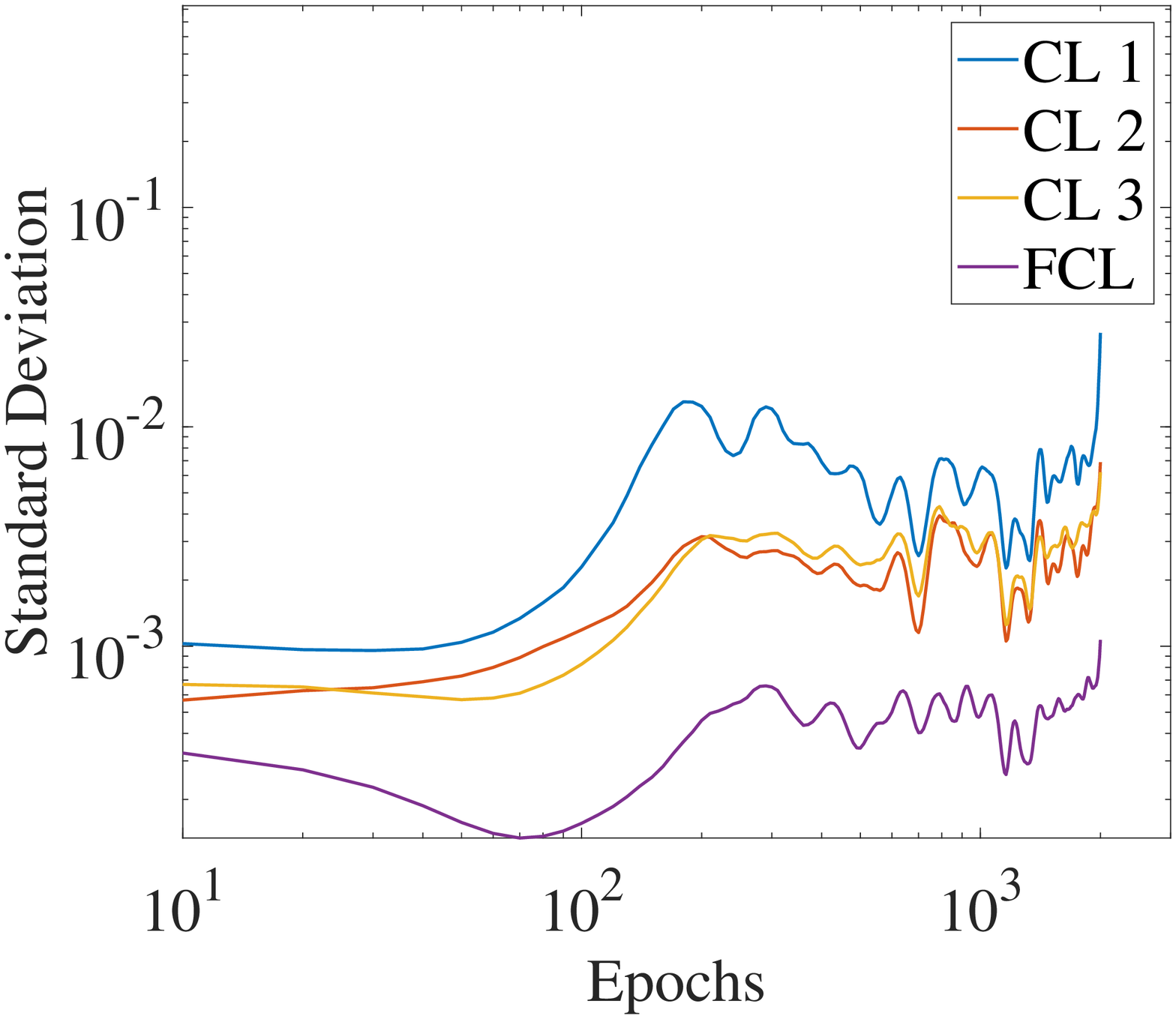}
    \caption{Standard deviation for layers of $CNN_{eig}^7$}
  \end{subfigure}
  \hfill
  \begin{subfigure}{0.32\textwidth}
    \centering
    \includegraphics[width=\textwidth]{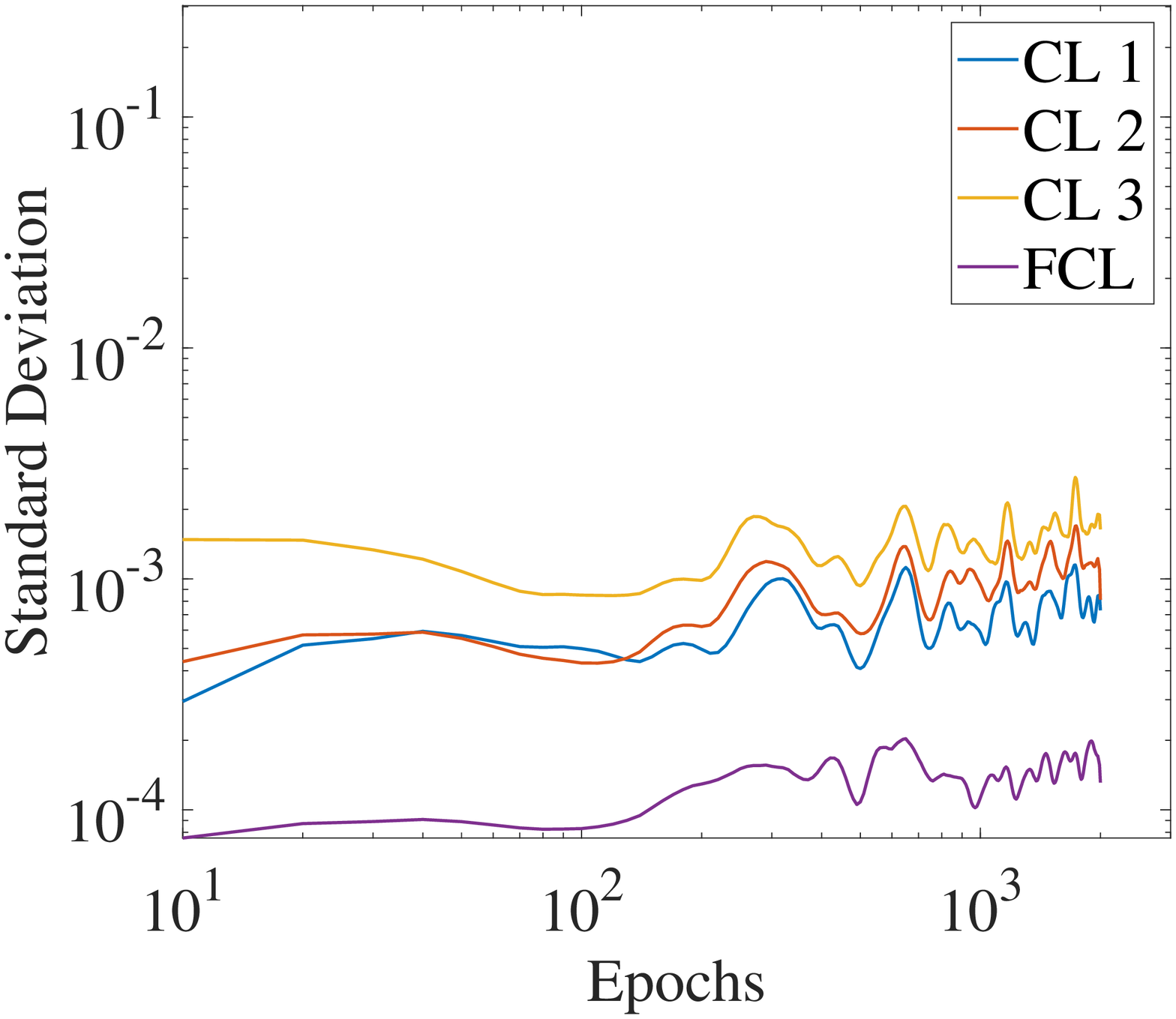}
    \caption{Standard deviation for layers of $CNN_{gaf}^7$}
  \end{subfigure}
  \\
  \vspace{2mm}

  \begin{subfigure}{0.32\textwidth}
    \centering
    \includegraphics[width=\textwidth]{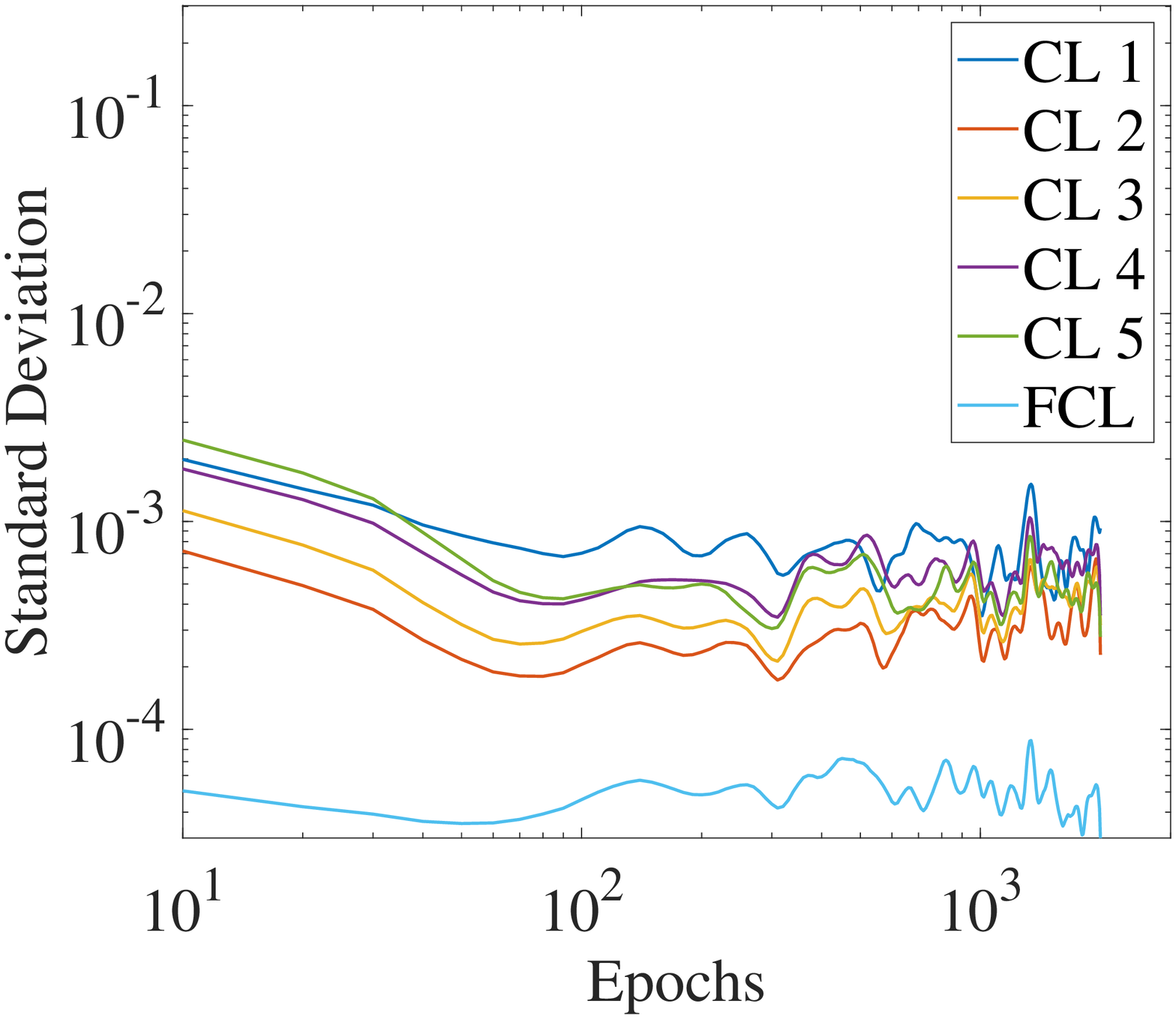}
    \caption{Standard deviation for layers of $CNN_{cov}^9$}
  \end{subfigure}
  \hfill
  \begin{subfigure}{0.32\textwidth}
    \centering
    \includegraphics[width=\textwidth]{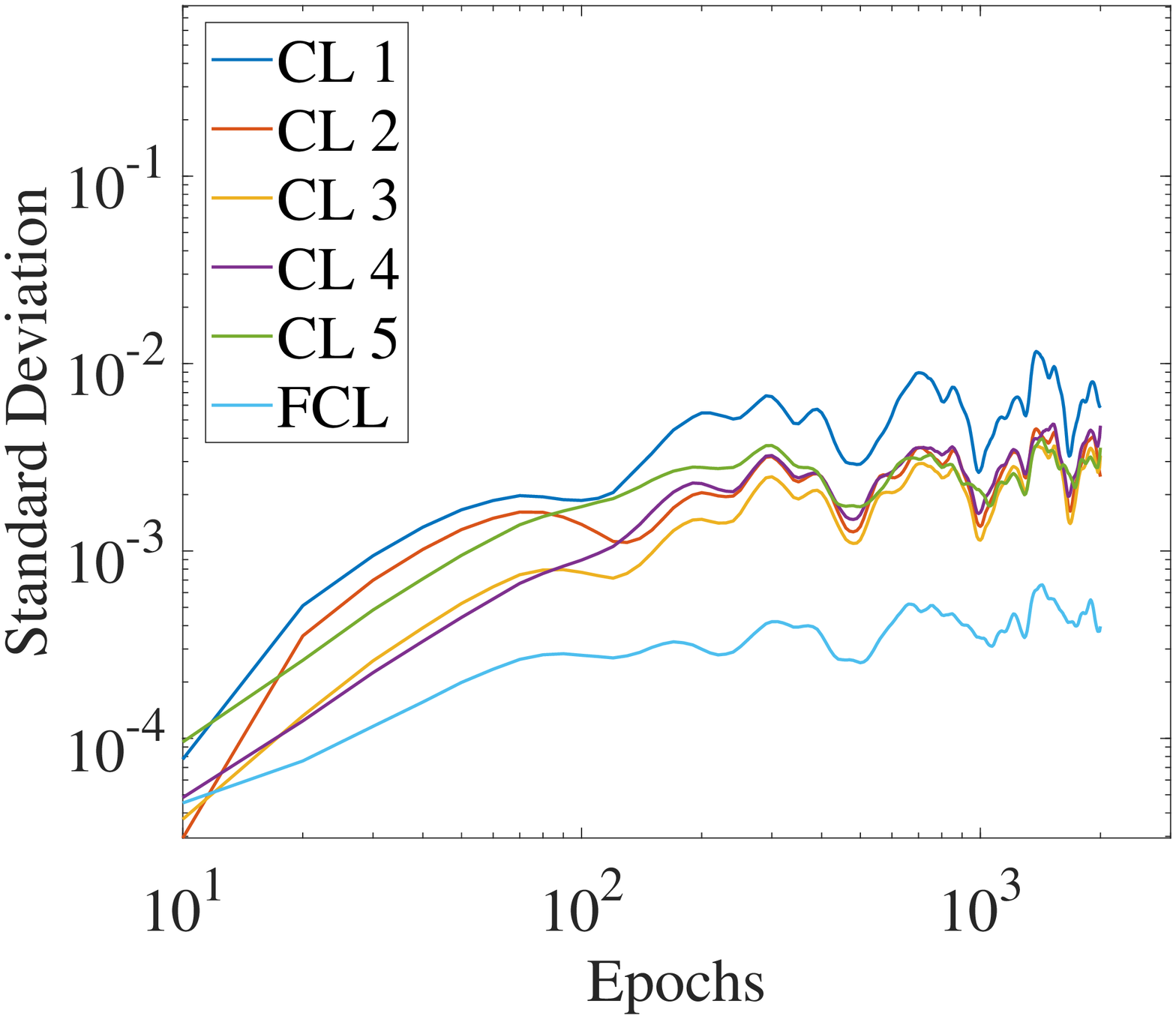}
    \caption{Standard deviation for layers of $CNN_{eig}^9$}
  \end{subfigure}
  \hfill
  \begin{subfigure}{0.32\textwidth}
    \centering
    \includegraphics[width=\textwidth]{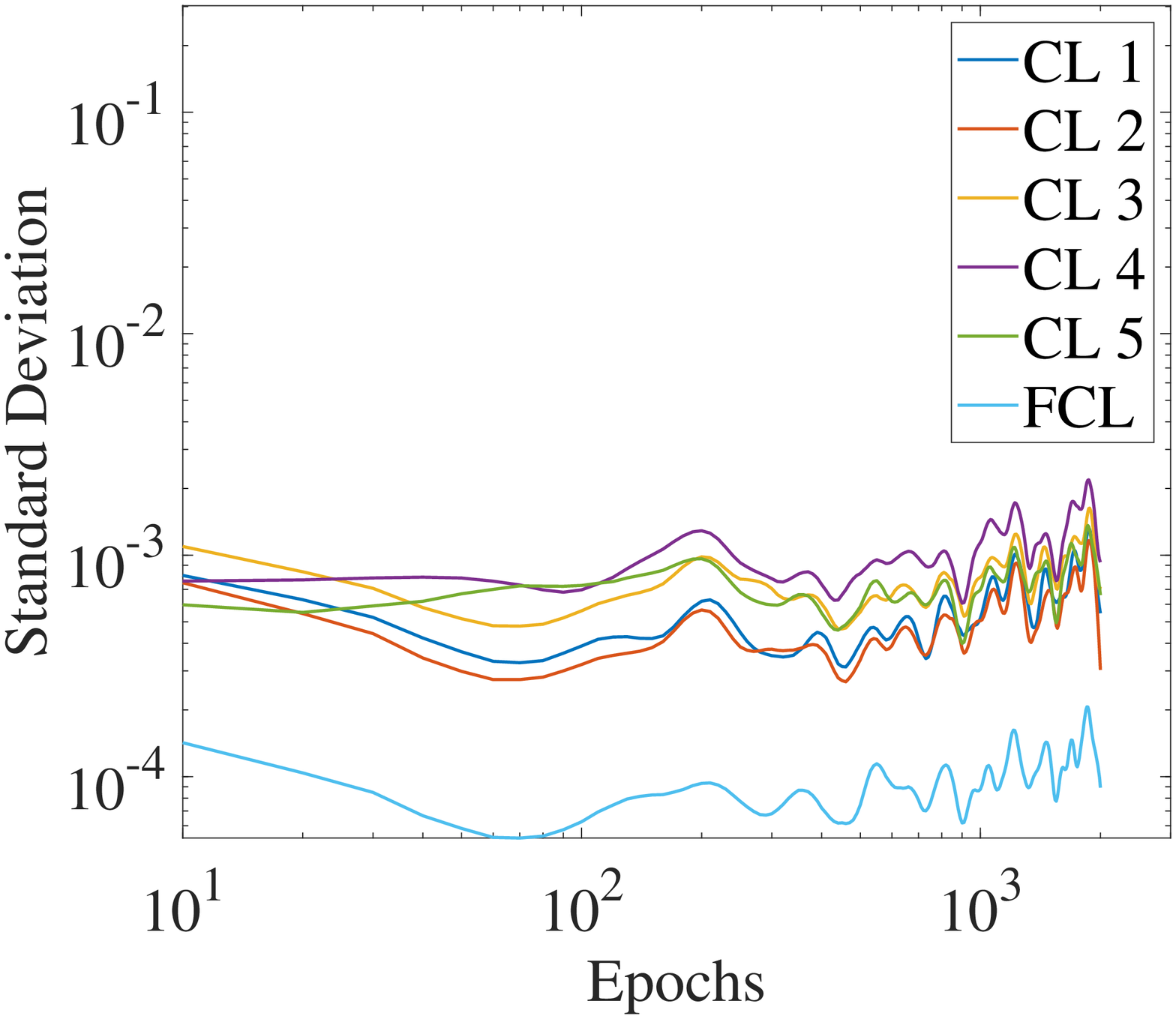}
    \caption{Standard deviation for layers of $CNN_{gaf}^9$}
  \end{subfigure}
  \caption{Standard deviation of gradient weights of each layer of proposed CNN models. In legend CL and FCL represents Convolutional and Fully Connected Layer, respectively}\label{Fig:LayersStd}
\end{figure*}

For a better understanding of the behaviour of layers over training, plots of normalized mean and standard deviation of stochastic gradients of each convolutional and fully connected layer are shown in Figure \ref{Fig:LayersMean} and \ref{Fig:LayersStd}. In the figures, the layers are numbered like CL 1, CL 2 and so on. The lower number represents the layer is near to the input and number increases as moving towards the output. Also, it can be concluded that the normalized mean of each layer of each CNN model is converging around a single value. So after converging, the layers are optimizing themselves which can also be observed in Figures \ref{Fig:Small-NMI} and \ref{Fig:Big-NMI}. Another observation is that mean of gradient weights are larger than the standard deviation of gradient weights which indicates small gradient stochasticity which implies high signal to noise ratio (SNR). Also, it can be observed that the difference between the normalized mean and standard deviation of gradient weights become nearly constant as the training progresses which means that with the training the empirical error saturates. Another observation is that the layers near to the output have less standard deviation of gradient weights. The main reason for that is the layers near to the output already have a great amount of information regarding the output which results in less deviation in gradient weights. So, it can be concluded that more the information a layer has less is the standard deviation in gradient weight of that layer.

\subsection{Testing of the models}\label{SubSec:Testing}
\begin{figure*}[h!]
  \centering
  \begin{subfigure}{0.49\textwidth}
    \centering
    \includegraphics[width=\linewidth]{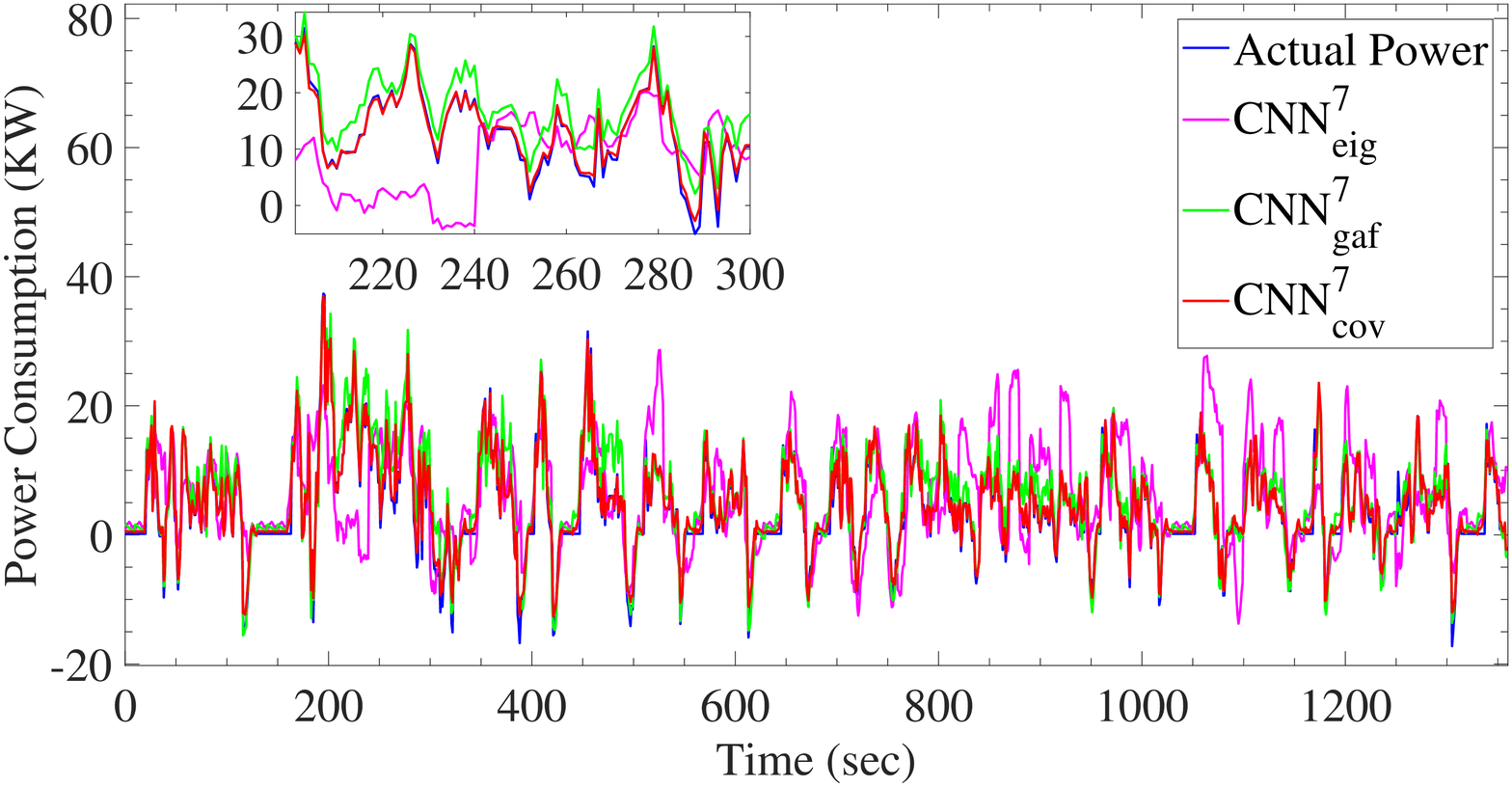}
    \caption{Estimated power consumption by $CNN^7$ for UDDS drive cycle}\label{Fig:Small_UDDS}
  \end{subfigure}
  \hfill
  \begin{subfigure}{0.49\textwidth}
    \centering
    \includegraphics[width=\linewidth]{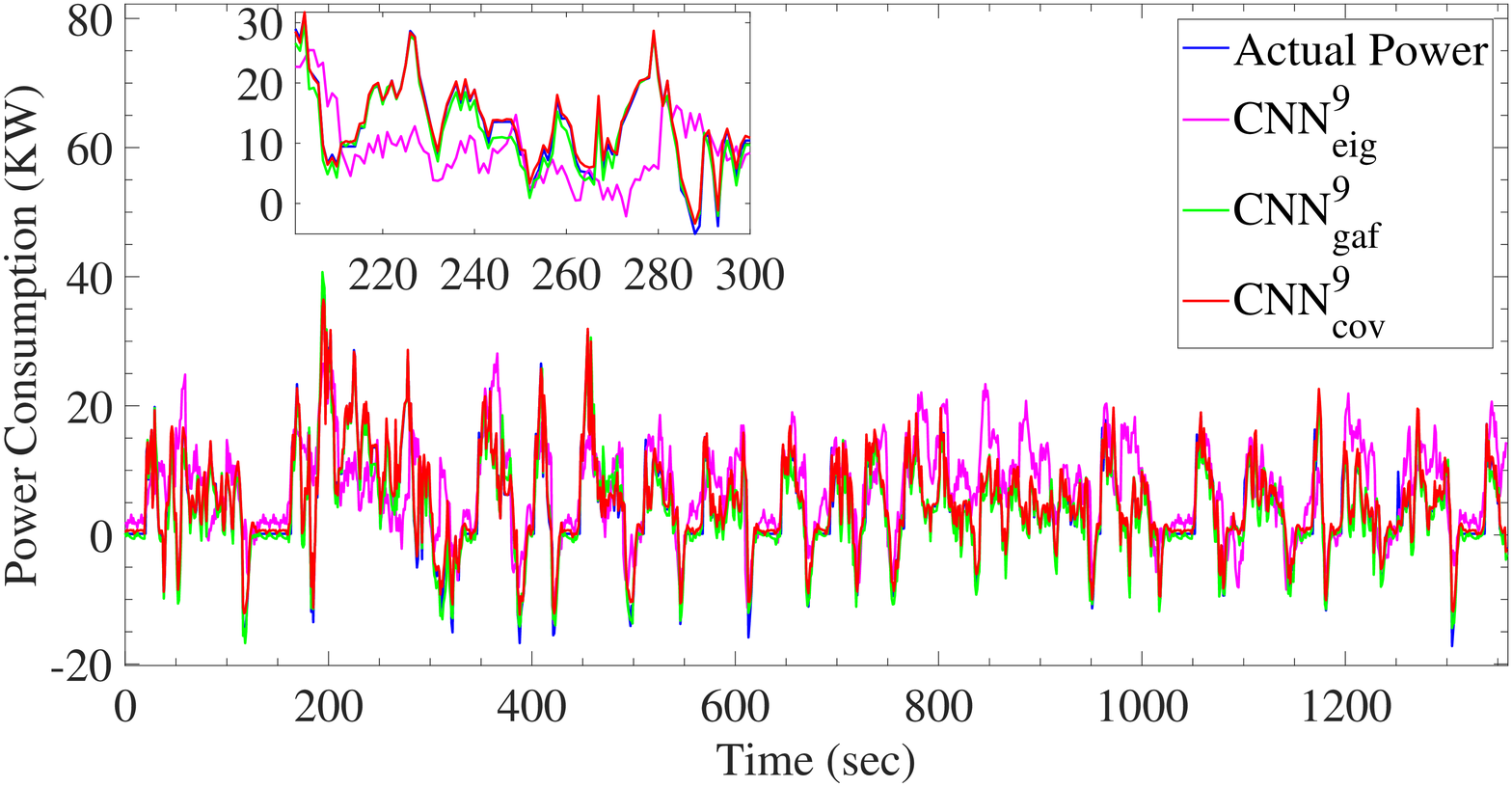}
    \caption{Estimated power consumption by $CNN^9$ for UDDS drive cycle}\label{Fig:Big_UDDS}
  \end{subfigure}
  \begin{subfigure}{0.49\textwidth}
    \centering
    \includegraphics[width=\linewidth]{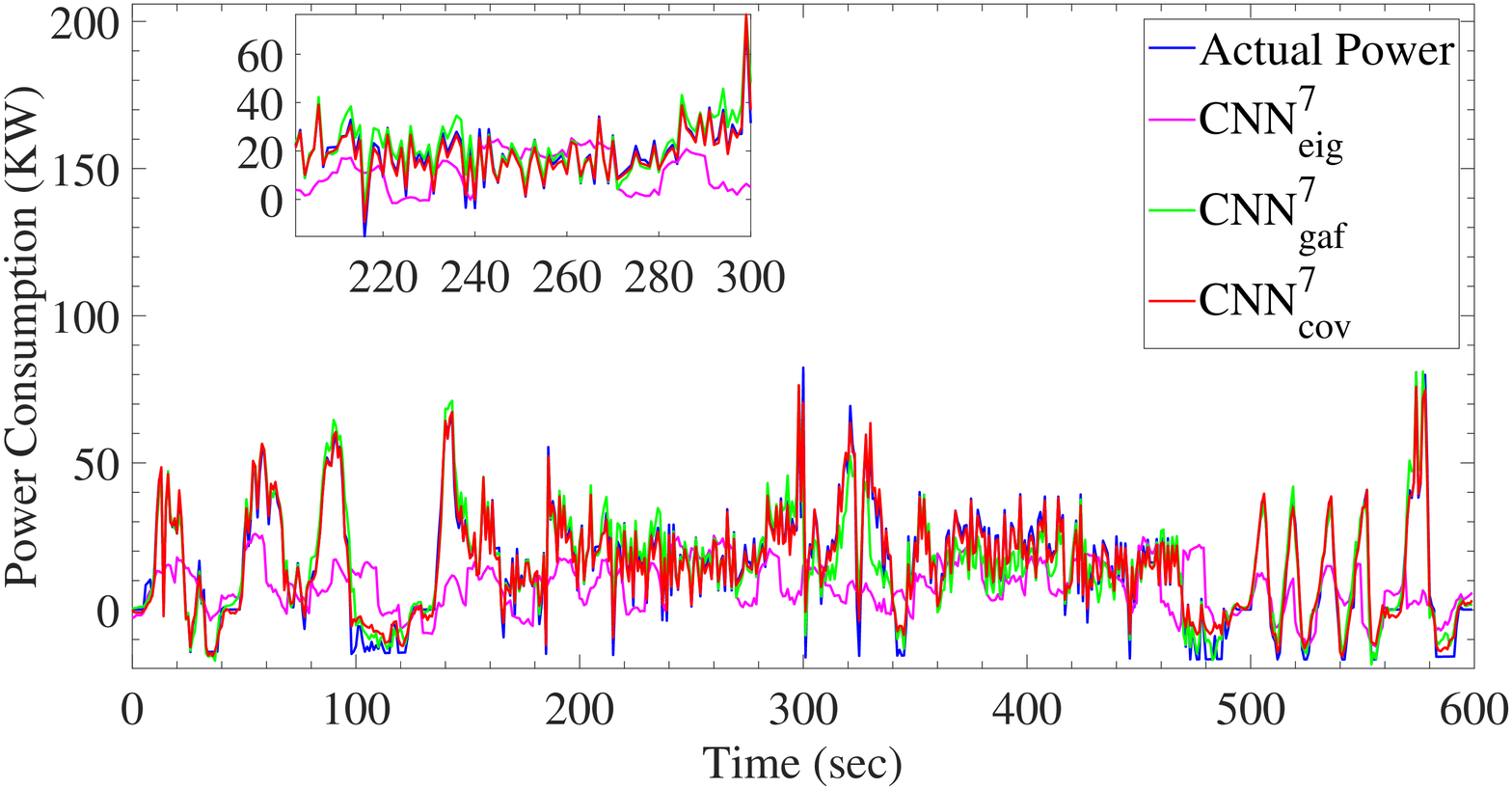}
    \caption{Estimated power consumption by $CNN^7$ for SFTP drive cycle}\label{Fig:Small_US06}
  \end{subfigure}
  \hfill
  \begin{subfigure}{0.49\textwidth}
    \centering
    \includegraphics[width=\linewidth]{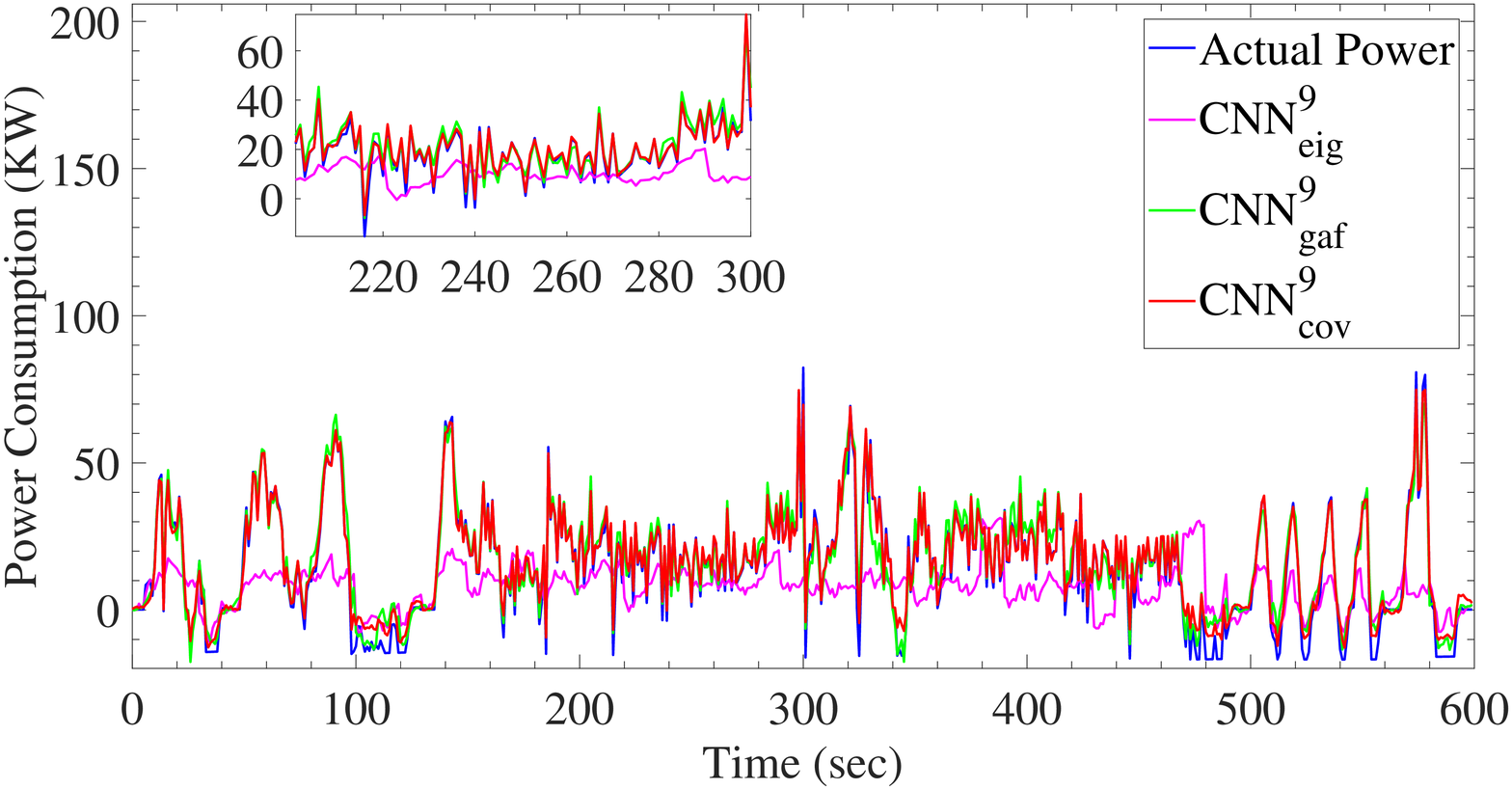}
    \caption{Estimated power consumption by $CNN^9$ for SFTP drive cycle}\label{Fig:Big_US06}
  \end{subfigure}
  \begin{subfigure}{0.49\textwidth}
    \centering
    \includegraphics[width=\linewidth]{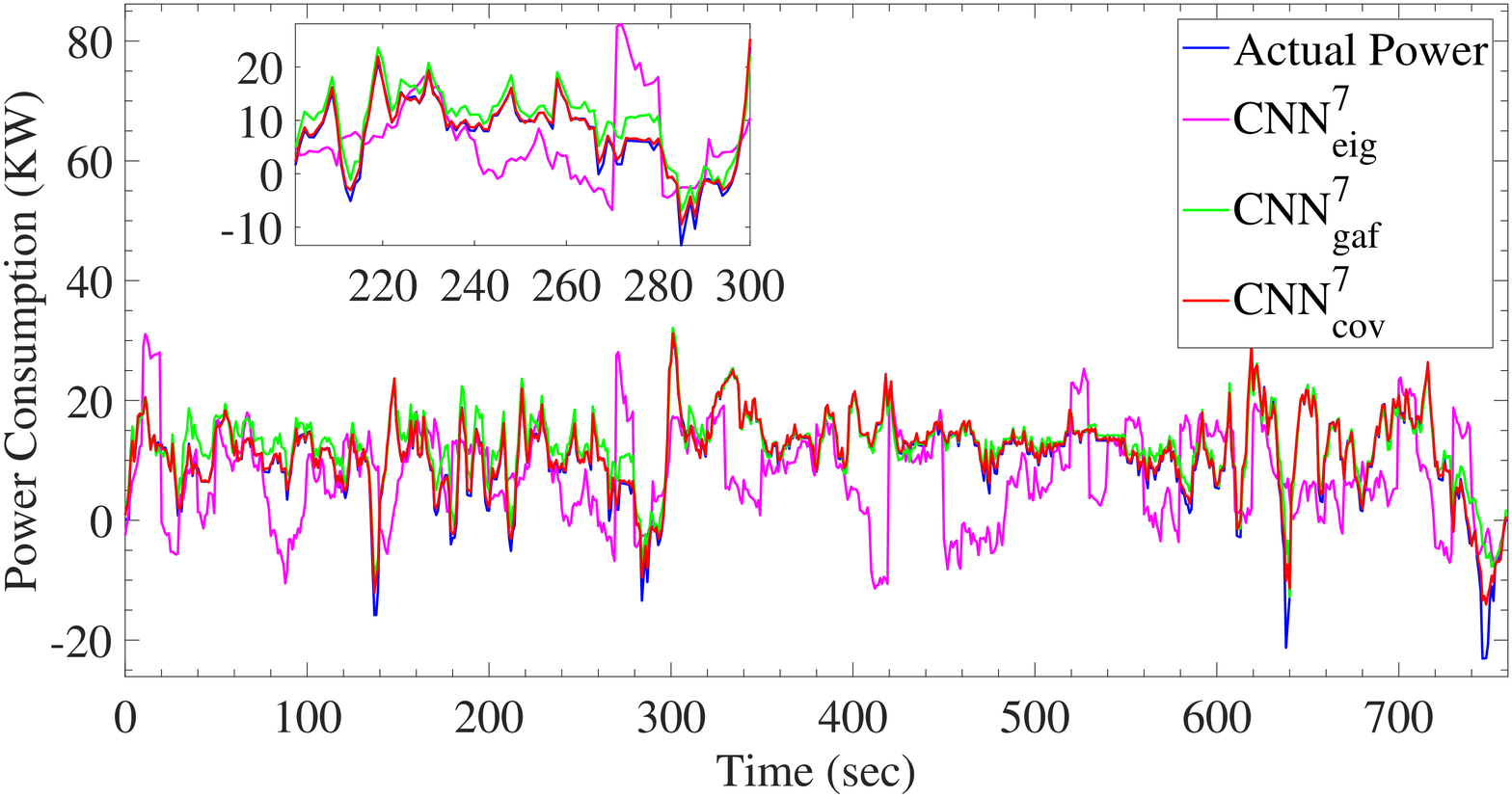}
    \caption{Estimated power consumption by $CNN^7$ for HWFET drive cycle}\label{Fig:Small_HWFET}
  \end{subfigure}
  \hfill
  \begin{subfigure}{0.49\textwidth}
    \centering
    \includegraphics[width=\linewidth]{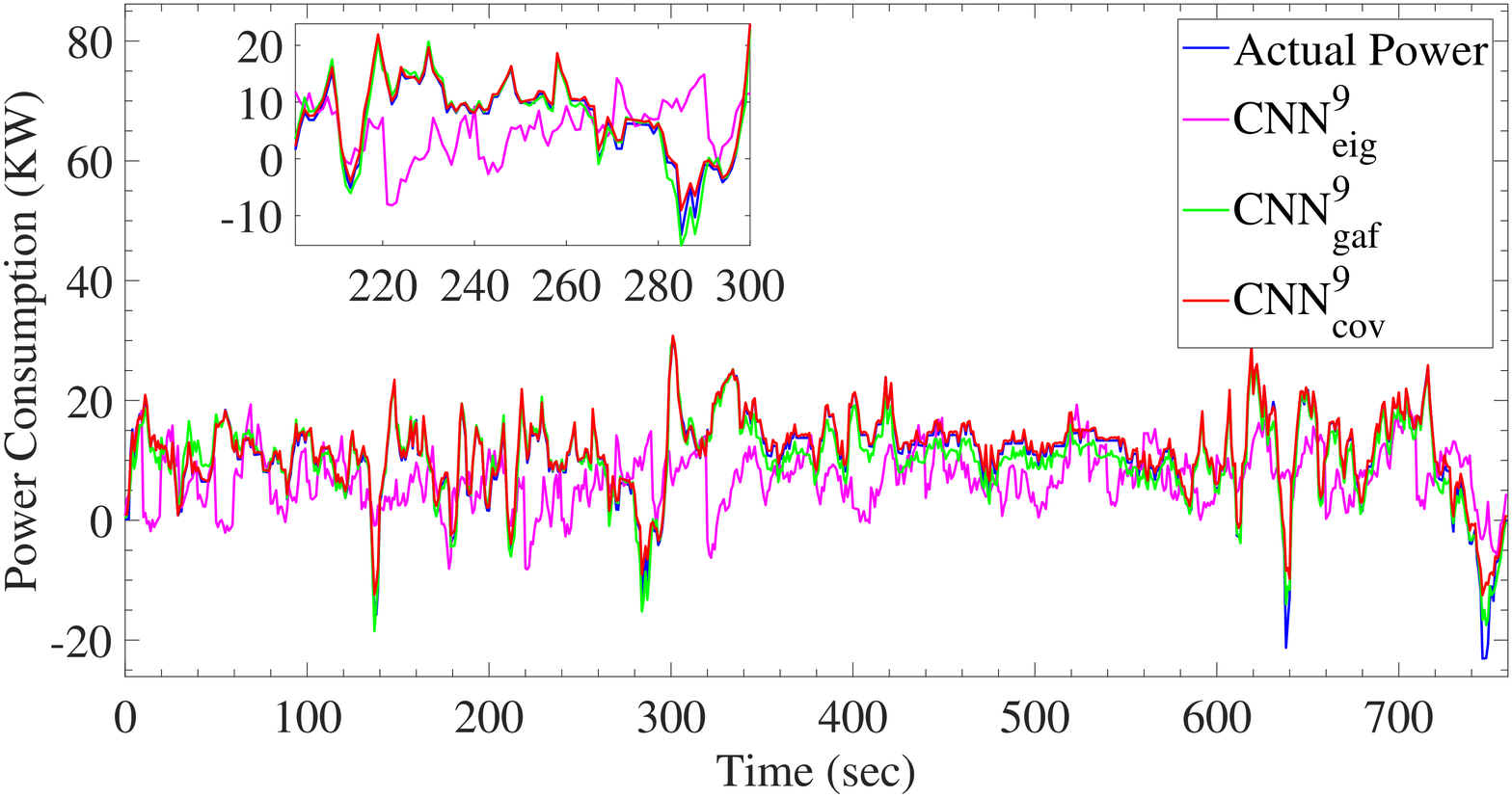}
    \caption{Estimated power consumption by $CNN^9$ for HWFET drive cycle}\label{Fig:Big_HWFET}
  \end{subfigure}
  \begin{subfigure}{0.49\textwidth}
    \centering
    \includegraphics[width=\linewidth]{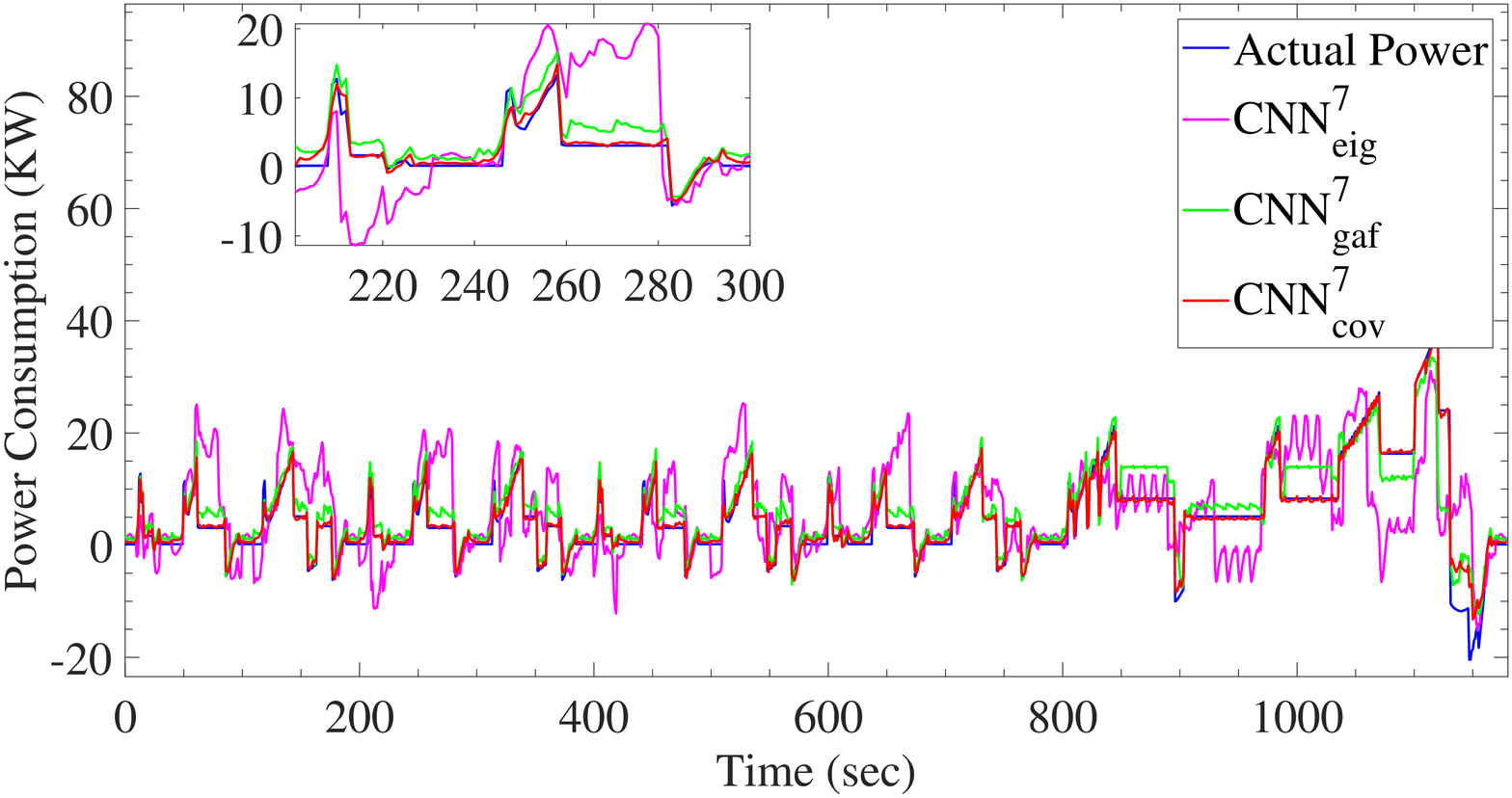}
    \caption{Estimated power consumption by $CNN^7$ for NEDC drive cycle}\label{Fig:Small_NEDC}
  \end{subfigure}
  \hfill
  \begin{subfigure}{0.49\textwidth}
    \centering
    \includegraphics[width=\linewidth]{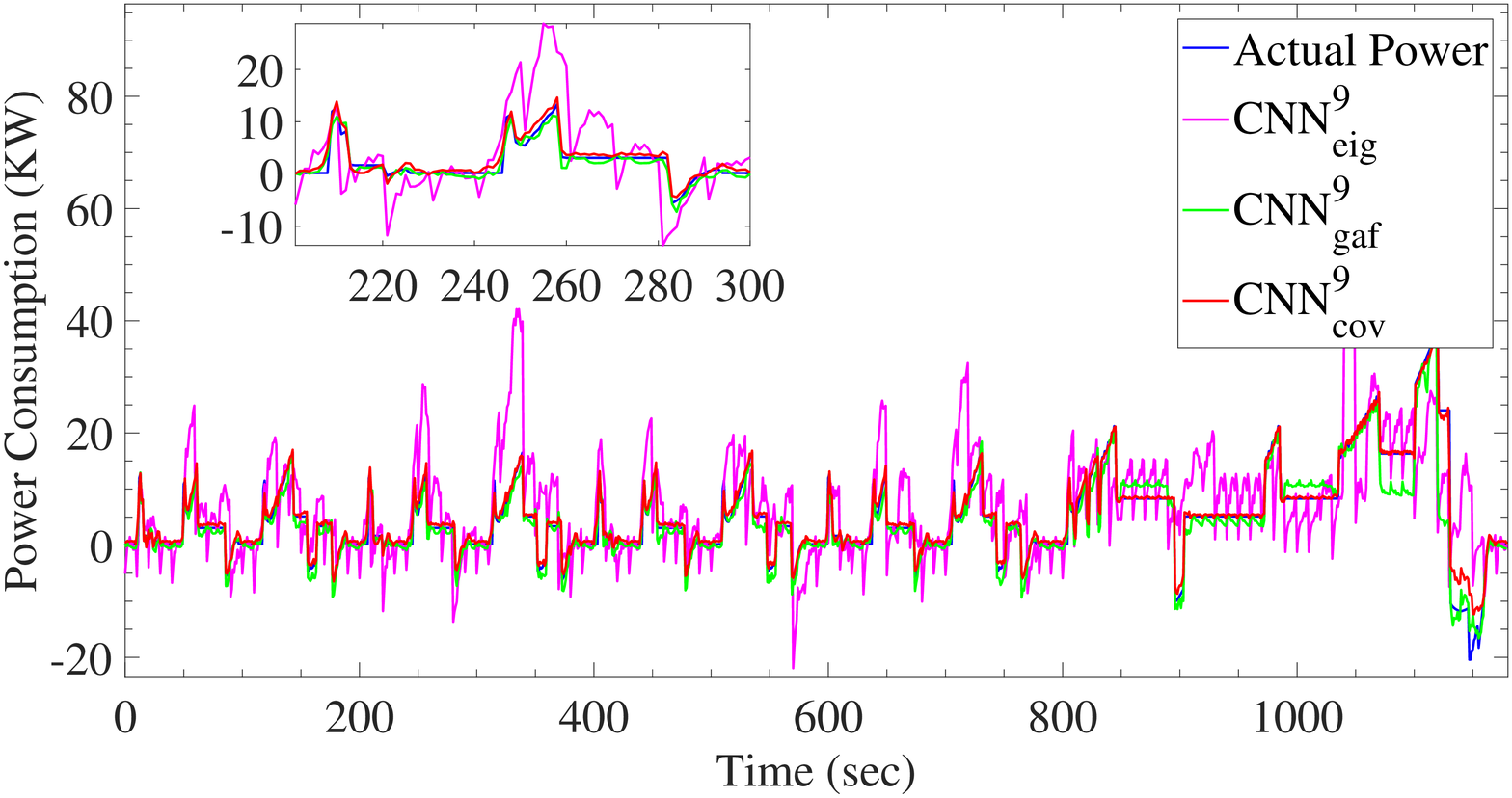}
    \caption{Estimated power consumption by $CNN^9$ for NEDC drive cycle}\label{Fig:Big_NEDC}
  \end{subfigure}
  \caption{Estimated Power Consumption for different driving cycles by proposed CNN Architectures} \label{Fig:DriveCyclePowerConsumption}
\end{figure*}

As discussed previously, the models were tested using the dataset $DS-II$ after preprocessing with the three methods explained in Section \ref{Subsec:TimeSeriesImageEncoder}. Dataset $DS-II$ has number of different drive cycles to test upon. Figure \ref{Fig:DriveCyclePowerConsumption} show the testing results for four such drive cycles namely, Urban Dynamometer Driving Schedule (UDDS), US06 Supplemental Federal Test Procedure (SFTP),  Highway Fuel Economy Test (HWFET) and New European Driving Cycle (NEDC) with each CNN model. It can be clearly seen from the figure that the CNN models trained with eigenvector features i.e. $CNN^7_{eig}$ and $CNN^9_{eig}$, in most of the cases, are no-where near the target value which is in agreement with the conclusion drawn from Figure \ref{Fig:TrainingValidation}. The models trained with GAF and covariance feature are really close to the target power consumption. On close observation, it can be found that CNN model $CNN^7_{cov}$ consistently performed better in all of the cases as compared to others. The main reason for this is the amount of information each feature descriptors holds. It has been explained previously also, that the covariance feature descriptors holds the maximum amount of information as compared to GAF and eigenvectors. To make the above conclusion clear, Table \ref{Tab:EnergyConsumption} has been presented, which shows the percentage energy consumption deviation (calculated using equation \eqref{Eq:EnergyConsumptionDeviation}) for above four drive cycles by each CNN model as compared to actual energy consumption.

\begin{equation}\label{Eq:EnergyConsumptionDeviation}
    E_{dev} = \frac{|\int^T_{t=0} P_{act}(t)dt - \int^T_{t=0} P_{est}(t)dt|}{\int^T_{t=0} P_{act}(t)dt} \times 100
\end{equation}

where $E_{dev}$ represent the percentage energy consumption deviation, $P_{act}(t)$ and $P_{est}(t)$ is the actual and estimated instantaneous power consumption at time $t$, respectively.

\begin{table}[h!]
    \centering
    \caption{Total energy consumption deviation (in percentage) for different drive cycles and proposed CNN models}
    \label{Tab:EnergyConsumption}
    \resizebox{0.6\linewidth}{!}{
    \begin{tabular}{|l|l|l|l|l|l|l|}
        \hline
                & $CNN^9_{cov}$ & $CNN^9_{eig}$ & $CNN^9_{gaf}$ & $CNN^7_{cov}$ & $CNN^7_{eig}$ & $CNN^7_{gaf}$  \\ \hline
        UDDS    & 13.18         & 59.56         & 13.90         & 7.04          & 48.09         & 28.92          \\ \hline
        SFTP    & 10.75         & 48.58         & 10.68         & 2.93          & 36.22         & 13.47          \\ \hline
        HWFET   & 06.31         & 35.86         & 07.97         & 6.61          & 32.20         & 12.18          \\ \hline
        NEDC    & 11.39         & 55.90         & 17.50         & 6.04          & 28.94         & 21.53          \\ \hline

    \end{tabular}
    }
\end{table}


From Table \ref{Tab:EnergyConsumption}, it can be clearly concluded that $CNN_{cov}^7$ performed consistently better as compared to other CNN models with lowest energy consumption deviation. There are some exceptions like in case of HWFET drive cycle $CNN_{cov}^9$ performed marginally better than $CNN_{cov}^7$. So, to justify and generalize the above conclusion cross validation has been performed for all the proposed CNN models.

\subsection{Cross Validation}\label{SubSec:CrossValidation}
To show the generalization ability and to measure the robustness of the proposed methodology, cross validation has been performed for all the CNN models. For this, a cross validation technique named $k$-fold cross validation has been used. The dataset $DS-I$ was partitioned into $k$ equally sized partitions. Then, $70\%$ of these $k$ partitions was selected and used for training the CNN models and remaining $30\%$ was used for validation. This process was repeated $k$ times (the folds), such that each of the $k$ partitions used at least once as part of validation set. Following are the different metrics that have been used as performance indicators.

\begin{enumerate}[i)]
    \item \emph{Root Mean Square Error (RMSE):} RMSE is a very popular and standardized formula to measure the error rate and hence the performance of a system. It can be calculated using the below equation:

        \begin{equation}\label{Eq:RMSE}
            RMSE = \sqrt{\frac{\sum^n_{i=1}(P_{act}^i - P_{est}^i)^2}{n}}
        \end{equation}

        where $P_{act}$ is the actual power consumption, $P_{est}$ is the estimated power consumption by the CNN model and $n$ is the total number of instances.

    \item \emph{Mean Absolute Error (MAE):} MAE is also a standardized measure which gives the idea of absolute deviation of estimated value with respect to actual value. Following is the equation which was used to calculate mean absolute error between actual and estimated power consumption:

        \begin{equation}\label{Eq:MAE}
            MAE = \frac{\sum^n_{i=1}|(P_{act}^i - P_{est}^i)|}{n}
        \end{equation}

        Similar to RMSE, the symbols $P_{act}$, $P_{est}$ and $n$ represent actual power consumption, estimated power consumption and total number of instances, respectively.

    \item \emph{Correlation (Corr):} Correlation represents the statistical relationship between actual and estimated value. It can be calculated using the below equation:

        \begin{equation}\label{Eq:Correlation}
            Corr = \frac{\sum^n_{i=1}(P_{act}^i - \overline{P_{act}}) (P_{est}^i - \overline{P_{est}})} {\sqrt{\sum^n_{i=1}(P_{act}^i - \overline{P_{act}})^2 \sum^n_{i=1}(P_{est}^i - \overline{P_{est}})^2}}
        \end{equation}

        where $\overline{P_{act}}$ and $\overline{P_{est}}$ represent the mean of actual power consumption and mean of estimated power consumption and rest of the symbols are same as RMSE or MAE. The correlation lies in the range of [-1,1]. If correlation value for two variables $x$ and $y$ is negative, it means that when $x$ increases $y$ decreases and vice versa. If correlation is 0, it means the two variables $x$ and $y$ are not related whereas if correlation is positive, it means the two variables are linearly related to each other and have similar behaviour which means if one increases other also increases and vice versa. So, correlation between actual and predicted variable should be close to 1 for any algorithm to be considered good.
\end{enumerate}

\begin{figure*}[h!]
  \centering
  \begin{subfigure}[t]{0.32\linewidth}
    \centering
    \includegraphics[width=\linewidth]{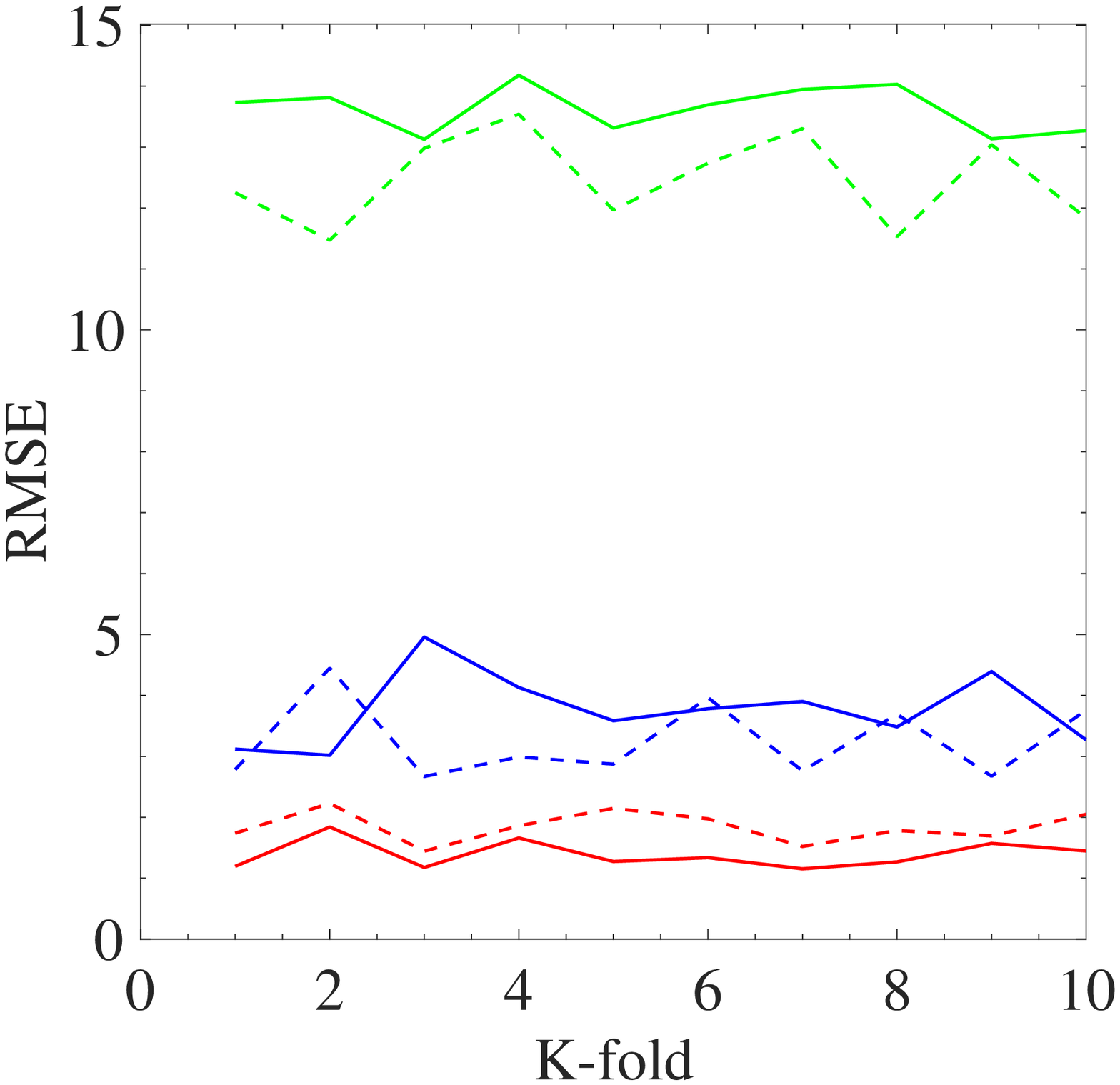}
    \caption{RMSE}
  \end{subfigure}
  \hfill
  \begin{subfigure}[t]{0.32\linewidth}
    \centering
    \includegraphics[width=\linewidth]{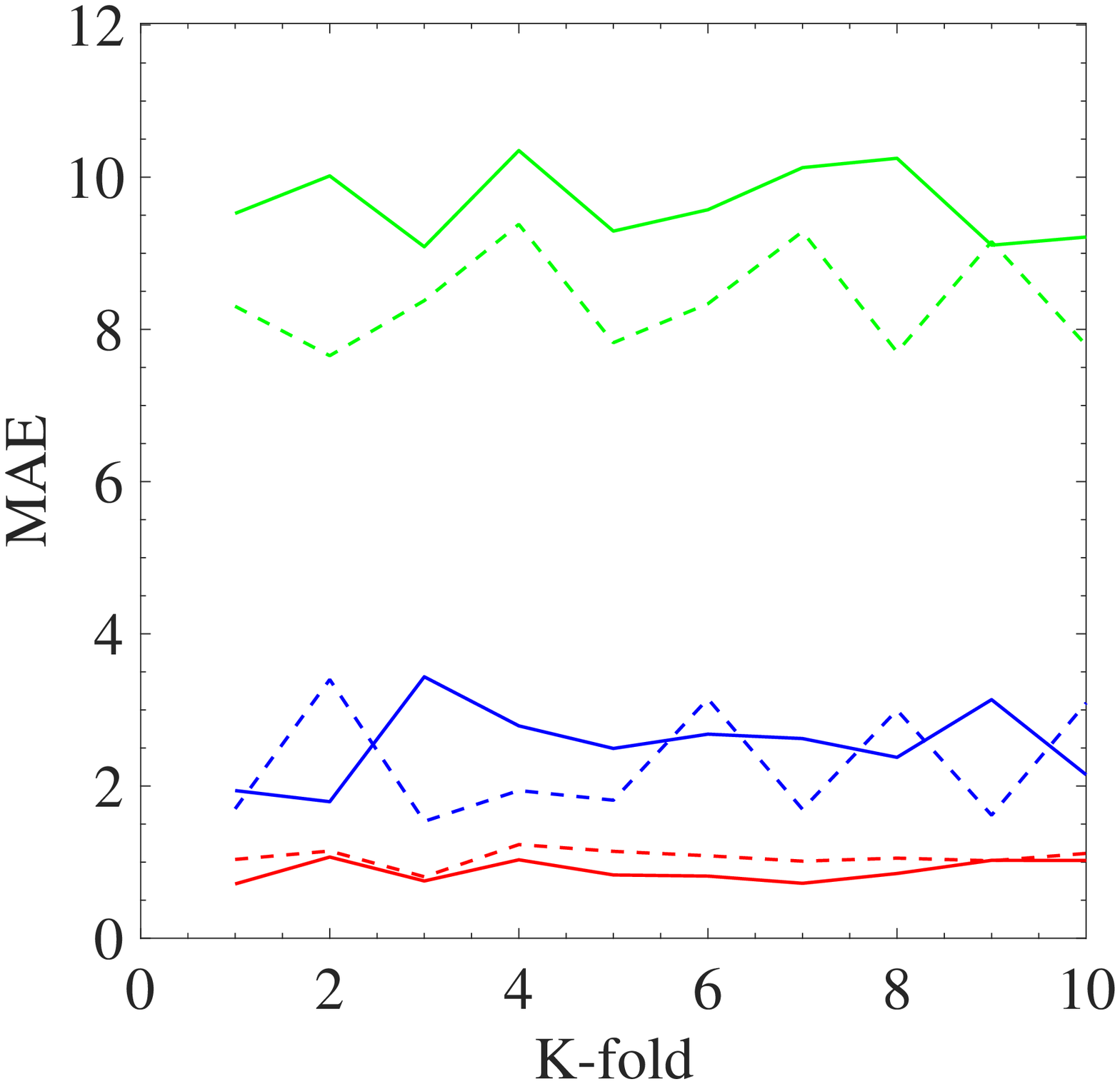}
    \caption{MAE}
  \end{subfigure}
  \hfill
  \begin{subfigure}[t]{0.32\linewidth}
    \centering
    \includegraphics[width=\linewidth]{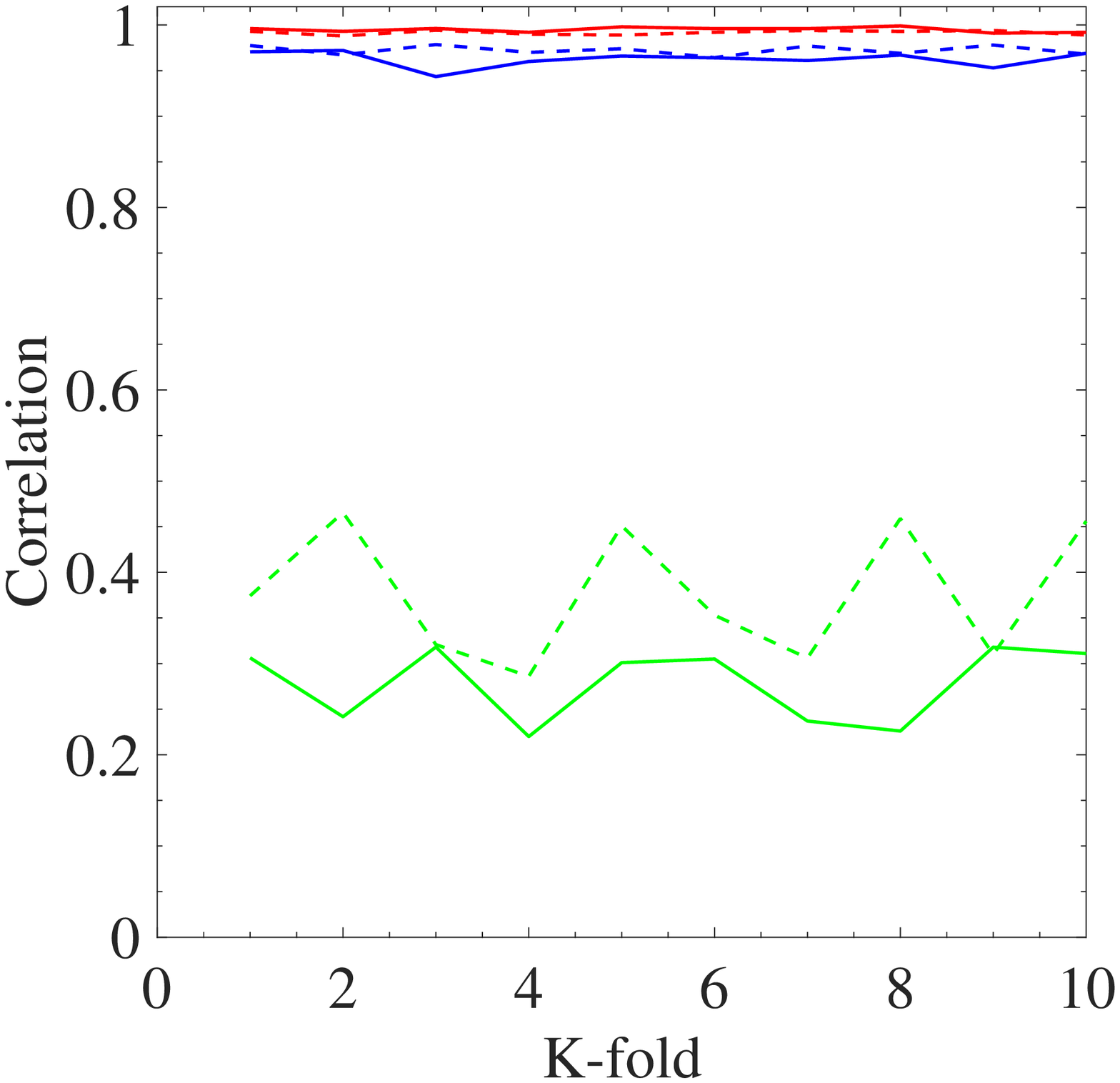}
    \caption{Correlation}
  \end{subfigure}
  \caption{10-fold cross validation of the different CNN models using RMSE, MAE and Correlation as performance indicators. Legend: (\protect\redsolidline) $CNN_{cov}^7$, (\protect\reddashedline) $CNN_{cov}^9$, (\protect\bluesolidline) $CNN_{gaf}^7$, (\protect\bluedashedline) $CNN_{gaf}^9$, (\protect\greensolidline) $CNN_{eig}^7$, (\protect\greendashedline) $CNN_{eig}^9$}\label{Fig:CrossValidationMetrics}
\end{figure*}

\begin{table}[h!]
    \centering
    \caption{Statistical analysis of the computed results by two sample $t$-test with equal variance}
    \label{Tab:pValueforCNNModels}
    \resizebox{\linewidth}{!}{
    \begin{tabular}{|c|c|c|c|c|c|c|}
    \hline
                    &$CNN^9_{gaf}$ & $CNN^7_{gaf}$ & $CNN^9_{eig}$ & $CNN^7_{eig}$ & $CNN^9_{cov}$ & $CNN^7_{cov}$   \\ \hline
                    &\multicolumn{6}{c|}{Results using RMSE values from 10-fold cross validation}                    \\ \hline
    Mean 						     & 3.26 & 3.76  & 12.46 & 13.62 & 1.84 & 1.39                \\ \hline
    Variance						 & 0.41 & 0.37  & 0.56  & 0.15  & 0.07 & 0.05                \\ \hline
    Observations					 & 10   & 10    & 10    & 10    & 10   & 10                  \\ \hline
    Pooled Variance					 & 0.26 & 0.23  & 0.34  & 0.11  & 0.07 & -                   \\ \hline
    Hypothetical mean difference	 & 0    & 0     & 0     & 0     & 0    & -                   \\ \hline
    Degree of freedom				 & 18   & 18    & 18    & 18    & 18   & -                   \\ \hline
    $t$-stat						 & 8.19 & 10.97 & 42.28 & 81.61 & 3.91 & -                   \\ \hline
    $P(T \leq t)$ one tail			 & 8.70 $\times 10^{-8}$  & 1.05 $\times 10^{-9}$  & 0  & 0  & 5.17 $\times 10^{-4}$ & -                \\ \hline

    t–critical one tail		         & 1.734 & 1.734 & 1.734 & 1.734 & 1.734 & -                \\ \hline
                    &\multicolumn{6}{c|}{Results using MAE values from 10-fold cross validation}                    \\ \hline
    Mean 						     & 2.29  & 2.54  & 8.38  & 9.65  & 1.06  & 0.88                \\ \hline
    Variance						 & 0.57  & 0.26  & 0.45  & 0.24  & 0.01  & 0.02                \\ \hline
    Observations					 & 10    & 10    & 10    & 10    & 10    & 10                  \\ \hline
    Pooled Variance					 & 0.33  & 0.15  & 0.26  & 0.14  & 0.02  & -                   \\ \hline
    Hypothetical mean difference	 & 0     & 0     & 0     & 0     & 0     & -                   \\ \hline
    Degree of freedom				 & 18    & 18    & 18    & 18    & 18    & -                   \\ \hline
    $t$-stat						     & 5.49  & 9.41  & 32.79 & 51.61 & 3.03  & -                   \\ \hline
    $P(T \leq t)$ one tail			 & 1.63 $\times 10^{-5}$ & 1.13 $\times 10^{-8}$ & 0 & 0 & 3.56 $\times 10^{-3}$ & -                \\ \hline

    t–critical one tail 		     & 1.734 & 1.734 & 1.734 & 1.734 & 1.734 & -                \\ \hline
\end{tabular}}
\end{table}

In this work, value of $k$ has been taken as 10. So, the results for 10-fold cross validation using RMSE, MAE and correlation as the performance indicators has been shown in Figure \ref{Fig:CrossValidationMetrics}. It can be observed that $CNN_{cov}^7$ has minimum RMSE and MAE and maximum correlation as compared to the other CNN models. To further validate this conclusion, statistical analysis has also been performed using two sample $t$-test between the results of $CNN_{cov}^7$ and other CNN models i.e. $CNN^7_{cov}$ with $CNN^9_{cov}$, $CNN^7_{cov}$ with $CNN^9_{eig}$ and so on. An analysis has been conducted with the assumption that the populations have equal variances at the significance level of $\alpha=0.05$. For the test, null hypothesis has been taken as that the difference of population means is zero. Under this null hypothesis, the pooled $t$-test has been performed with the assumption of equal variance and hence t-statistics values, shown in Table \ref{Tab:pValueforCNNModels}, are obtained. It can be observed from the table that the $t$-stat values are greater than the t-critical values. Also, the p-values corresponding to each pair are less than the significance level of $\alpha=0.05$. Thus, the population means differ significantly, which leads to the rejection of the null hypothesis. Further, the mean of the RMSE and MAE values in case of $CNN^7_{cov}$ is less than other CNN models and hence, the results obtained from $CNN^7_{cov}$ are better than the other CNN models and this difference is statistically significant.

\subsection{Analysis}\label{SubSec:Analysis}
A number of experiments were performed by training a number of CNN models with different number of layers (such as 5, 6, 7, 8, 9, 10 and 11) and varying the input feature descriptors, namely covariance, eigenvectors and GAF. From the results discussed above, it can be observed that different number of layers in CNN model and different input feature descriptors have great impact on the performance of the proposed methodology. In brief, it can be concluded that the CNN models trained with covariance have performed really well as compared to CNN models trained with eigenvectors and GAF. The main reason for that is a lot of information loss while calculating GAF and eigenvectors as compared to covariance. Also, it can be concluded that as the number of layers increases the CNN models converge faster, for instance, the CNN models with 5, 7, 9 and 11 layers when trained using dataset preprocessed with covariance method converged at approximately 420, 380, 330 and 290, respectively. Although, increasing the layers help the models to converge early but it also increases the computational cost. Also, it is well known fact that the CNN models with more number of layers require more training data to achieve the same level of accuracy as the CNN models with less number of layers. So, it is important to find the minimum possible number of layers with acceptable performance. In this work, it has been observed that CNN model with 7 layers and trained with covariance feature descriptors, represented as $CNN^7_{cov}$, performed consistently better than other CNN models and it has also been statistically validated in the previous subsection.

\section{Comparative Analysis}\label{Sec:ComparativeAnalysis}
To benchmark the results and to show the efficiency of the proposed approach, the computed results are compared with five of the existing approaches. From discussions in section \ref{SubSec:Testing}, \ref{SubSec:CrossValidation} and \ref{SubSec:Analysis}, it was observed that $CNN_{cov}^7$ performed better than other CNN models. Also, CNN models $CNN^9_{cov}$ and $CNN_{gaf}^9$ have comparable performance. So, in this section comparison results of $CNN_{cov}^7$, $CNN^9_{cov}$ and $CNN_{gaf}^9$ with below discussed five state-of-the-art techniques has been shown.

\begin{enumerate}[i)]
    \item To implement the multivariate model for power consumption estimation of EV the equation \eqref{Eq:GalvinModel} proposed by Galvin \cite{GALVIN2017234} was used.

        \begin{equation}\label{Eq:GalvinModel}
            P = rV + sV^2 + tV^3 + uVA
        \end{equation}

        where $P$, $V$ and $A$ represent the power demand, speed and acceleration, respectively and $r$, $s$, $t$ and $u$ are regression coefficients. The values for these coefficients for NissanSV as given in \cite{GALVIN2017234} are $r = 479.1$, $s = -18.93$, $t = 0.7876$ and $u = 1507$. According to the dataset used in this paper the variable $V$ correspond to $v_{sp}$ and $A$ correspond to change in speed per unit time. So the equation \eqref{Eq:GalvinModel} becomes

        \begin{equation}\label{Eq:GalvinNissanSVModel}
        \begin{split}
            P(t) = 479.1 v_{sp}(t) - 18.93 v_{sp}(t)^2 + 0.7876 v_{sp}(t)^3 \\ + 1507 v_{sp}(t)\Big(\frac{v_{sp}(t) - v_{sp}(t-1)}{(t)-(t-1)}\Big)
        \end{split}
        \end{equation}

        where $P(t)$, $v_{sp}(t)$ represent the power demand and speed of the vehicle at time $t$, respectively.

    \item The model proposed by Yang et al. \cite{YANG201441} was also implemented for comparison with the CNN models. Yang et al. \cite{YANG201441} proposed the equations \eqref{Eq:YangPowerConsumption} and \eqref{Eq:YangRegenrativePower} for estimating the power consumption of EV when motor runs in normal and regenerative mode, respectively.

        \begin{equation}\label{Eq:YangPowerConsumption}
            P = \frac{v}{\eta_{te}\eta_e}\bigg(\delta m \frac{dv}{dt} + mg(f+i) + \frac{\rho C_D A}{2}v^2\bigg) + P_{accessory}
        \end{equation}

        \begin{equation}\label{Eq:YangRegenrativePower}
            P_{reg} = kv\eta_{te}\eta_m\bigg(\delta m \frac{dv}{dt} + mg(f+i) + \frac{\rho C_D A}{2}v^2\bigg) + P_{accessory}
        \end{equation}

        where $P$ is the power consumption, $P_{reg}$ is power regenerated, $v$ is the speed (correspond to $v_{sp}$ in $DS-I$ and $DS-II$ dataset), $\eta_{te}$ is the transmission efficiency, $\eta_e$ is driving efficiency, $\delta$ is the coefficient related to weight of EV, $m$ is mass of vehicle, $f$ is rolling resistance coefficient, $i$ is the road grade (correspond to $r_{el}$ in $DS-I$ and $DS-II$ dataset), $\rho$ is air density, $C_D$ is the aerodynamic drag coefficient, $A$ is the frontal area of vehicle, $P_{accessory}$ is the power consumed by accessories, $k$ is the percentage of total energy during braking that can be recovered by the motor and $\eta_m$ is the motor efficiency. Parameter $k$ was defined using the following equation:

        \begin{equation}\label{Eq:ParameterK}
                k =
                        \begin{dcases}
                            0.5*\frac{v}{5}         &  v < 5\text{m/s} \\
                            0.5+0.3\frac{v-5}{20}   &  v \geq 5\text{m/s}
                        \end{dcases}
        \end{equation}

        For implementing the above model for Nissan Leaf 2013, values of $m$, $C_D$ and $A$ were used from Table \ref{Tab:VehicleParameters}. Other than these, values of $\delta$, $\eta_{te}$, $\eta_m$, $\eta_e$, $\rho$, $P_{accessory}$ (assuming no AC or heater is running) and $f$ given in \cite{YANG201441} were 1.1, 0.9, 0.9, 0.8, 1.2, 150 and 0.015, respectively.

    \item A neural network model, as proposed by Alvarez et al. \cite{6861542}, with 14 inputs and 1 output but without hidden layer was trained using the mean and variance of three parameters as inputs. The parameters include speed, acceleration (further divided into positive and negative acceleration) and jerk (further partitioned into Starting Movement Jerk (SMJ), Cruising Track Jerk (CTJ), Starting Brake Jerk (SBJ) and Ending Brake Jerk (EBJ)). The neural network was trained with $DS-I_{tr}$ which is the 70\% of data from $DS-I$ and validated using the rest 30\%, denoted by $DS-I_{val}$.

    \item Similar to the above for comparison purpose a neural network, as developed by Felipe et al. \cite{7313117} which is the extension of the neural network in \cite{6861542}, was trained with 137 inputs, 1 output, and no hidden layer. The input parameters include the mean and variance of road grade, number of lanes etc along with the parameters used in \cite{6861542}. This neural network was also trained with 70\% of data from $DS-I$ and validated with the rest.

    \item The MLR (Multiple Linear Regression) model proposed by De Cauwer et al. \cite{de2017data} was implemented to estimate the energy consumption of EV for a trip divided into number of small segments using the following equation:

        \begin{equation}\label{Eq:MLRModel}
        \begin{split}
        \triangle E & = \sum_{\text{\emph{segments j}}}^{trip}\Big[B_1 \triangle s_j + B_2 \sum_i^n (v_{EV_i} + v_{wi})^2 \triangle s_j \\
                    & \quad + B_3 (CMF_j^+) \triangle s_j + B_4 (CMF_j^-) \triangle s_j + B_5 \triangle Hpos_j \\
                    & \quad + B_6 \triangle Hneg_j + B_7 Aux_{Tj} \triangle t_j + \varepsilon \Big]
        \end{split}
        \end{equation}

        with:
        \begin{equation*}
        CMF_j = \frac{\sum_{i=2}^n \mid v_{EV_i}^2 - v_{EV_{i-1}}^2 \mid }{\triangle s}
        \end{equation*}

        where $B_i$, $\triangle E$, $v_{EV_i}$, $v_{wi}$, $\triangle s$, $\triangle s_i$, $Aux_T$, $\triangle t$, $\triangle Hpos_j$, $\triangle Hneg_j$, $\varepsilon$, $n$ are regression coefficients, energy, vehicle speed at time $t_i$, wind speed at time $t_i$, distance, driven distance between $t_{i-1}$ and $t_i$, temperature scaling, time, positive elevation changes, negative elevation changes, error term and number of data points in segment $j$, respectively. For implementing the above model the drive cycles were divided into small segments each of 10 sec duration. Then for each segment equation \eqref{Eq:MLRModel} was applied by using the values from dataset $DS-I$ and $DS-II$. For instance $v_{sp}$ for $v_{EV}$, average $v_{sp}$ of segment $j$ multiplied by time for $\triangle s_j$ etc. The values of regression coefficients $B_1$ to $B_7$ for Nissan Leaf was provided in \cite{de2017data}, so those values were used for comparison with the proposed CNN model.
\end{enumerate}


\begin{table}[h!]
    \caption{Comparison using different performance metrics}
    \label{Tab:MeanEnergyConsumptionComparison}
    \resizebox{\linewidth}{!}{

    \begin{tabular}{|c|c|c|c|c|c|c|c|c|c|}
        \hline
        \multirow{2}{*}{Approach}          & \multicolumn{4}{c|}{$DS-I_{val}$} &  \multicolumn{4}{c|}{$DS-II$} & Average Prediction Time \\ \cline{2-9}
        & Mean $E_{dev}$   & RMSE       & MAE     & Corr    & Mean $E_{dev}$  & RMSE       & MAE      & Corr    &  / drive cycle (in sec) \\ \hline

        De Cauwer et al. \cite{de2017data} & 7.25            & 4.22       & 1.70    & 0.953    & 6.09      & 1.85         & 0.95      & 0.982     & 1.58 $\times 10^{-3}$                 \\ \hline

        Yang et al. \cite{YANG201441}      & 8.78            & 6.19       & 3.13    & 0.935    & 8.13      & 3.45         & 2.36               & 0.977     & 1.97 $\times 10^{-3}$           \\ \hline

        Galvin \cite{GALVIN2017234}        & 13.63            & 8.54       & 3.87    & 0.763    & 11.56      & 2.33         & 1.11               & 0.981      &  \textbf{3.47} $\mathbf{\times} \mathbf{10^{-4}}$          \\ \hline

        Alvarez et al. \cite{6861542}      & 12.37            & NA         & NA      & NA      & 10.21      & NA           & NA                 & NA        &   1.14 $\times 10^{-2}$         \\ \hline

        Felipe et al. \cite{7313117}       & 7.41            & NA         & NA      & NA      & 7.34      & NA           & NA                 & NA        &   3.06 $\times 10^{-2}$         \\ \hline

        The Proposed Models  &  &  &  &  &  &  &  &   &            \\ \hline

        $CNN^7_{cov}$ & \textbf{5.21}   & \textbf{1.39}  & \textbf{0.88}    & \textbf{0.995}  & \textbf{5.09}  & \textbf{1.35}    & \textbf{0.76}      &  \textbf{0.997}   &  1.76    \\ \hline

        $CNN^9_{cov}$ & 9.43             & 1.84       & 1.06    & 0.993    & 8.86       & 1.46            & 0.85               & 0.996         &  2.26                                 \\ \hline

        $CNN^9_{gaf}$ & 13.38            & 3.26       & 2.29    & 0.972    & 12.07       & 3.01            & 1.99              & 0.978       &  2.28                                   \\ \hline

    \multicolumn{10}{|l|}{\footnotesize{NA represent not applicable and bold values are the best ones}}     \\ \hline
    \end{tabular}
    }

\end{table}

The values of different performance metrics of above discussed five techniques and proposed models $CNN_{cov}^7$, $CNN_{cov}^9$ and $CNN_{gaf}^9$ has been presented in Table \ref{Tab:MeanEnergyConsumptionComparison} and it can be observed that $CNN^7_{cov}$ outperforms the other existing approaches with a lowest mean $E_{dev}$ of 5.21\% and 5.09\% on dataset $DS-I_{val}$ and $DS-II$, respectively. These results have also been validated by other metrics like Root Mean Square Error (RMSE), Mean Absolute Error (MAE) and Correlation (Corr). Also, it can be observed that the proposed models $CNN_{cov}^7$ and $CNN_{cov}^9$ have lowest RMSE (i.e. 1.39 and 1.84 on $DS-I_{val}$ and 1.35 and 1.46 on $DS-II$) and MAE (i.e. 0.88 and 1.06 on $DS-I_{val}$ and 0.76 and 0.85 on $DS-II$) and highest Corr (i.e. 0.995 and 0.993 on $DS-I_{val}$ and 0.997 and 0.996 on $DS-II$) values, as compared to the existing techniques. This shows that $CNN^9_{cov}$ is the second best model after $CNN^7_{cov}$, in terms of RMSE, MAE and Corr. Values of RMSE, MAE and Corr can not be calculated for \cite{6861542, 7313117} as the techniques presented in these do not give real-time power/energy consumption as output and provide only single value of total energy consumption for the trip. It can be observed that all the approaches performed better on $DS-II$ than on $DS-I_{val}$. The main reason for that is dataset $DS-II$ had readings on constant road grade of 0\% i.e. no change in road elevation whereas $DS-I_{val}$ had readings with road grade varying from -20\% to 20\%. Results for another metric of average prediction time per drive cycle has also been given in Table \ref{Tab:MeanEnergyConsumptionComparison}. The average prediction time per drive cycle is the time, the trained model takes to predict the output for given driven cycle and does not include the training time of the model. It has been calculated on a system with Intel i5 Processor, 8GB RAM on a torch-lua platform. It can be seen that the proposed models $CNN_{cov}^7$, $CNN_{cov}^9$ and $CNN_{gaf}^9$ take more inference time as compared to existing techniques. This is due to the fact that the proposed models have more number of layers in the architecture which increases the amount of computation required. This is also evident from the prediction time of $CNN^7_{cov}$ and $CNN^9_{cov}$, where $CNN^9_{cov}$ takes 2.26 sec as compared to $CNN^7_{cov}$ which takes 1.76 sec. As $CNN^7_{cov}$ has less number of layers so takes less time to predict the output. However, the current aim is to compare the accuracies of the architecture with the previous techniques. Once the model is verified in terms of accuracy, it will be converted to a TensorFlow Lite format to be suitable for execution on Google Coral boards or Odroid boards with Movidius sticks, hence providing a real time performance with a very low inference time.

\begin{figure*}[h!]
  \centering
  \begin{subfigure}[]{0.49\textwidth}
    \centering
    \includegraphics[width=\textwidth]{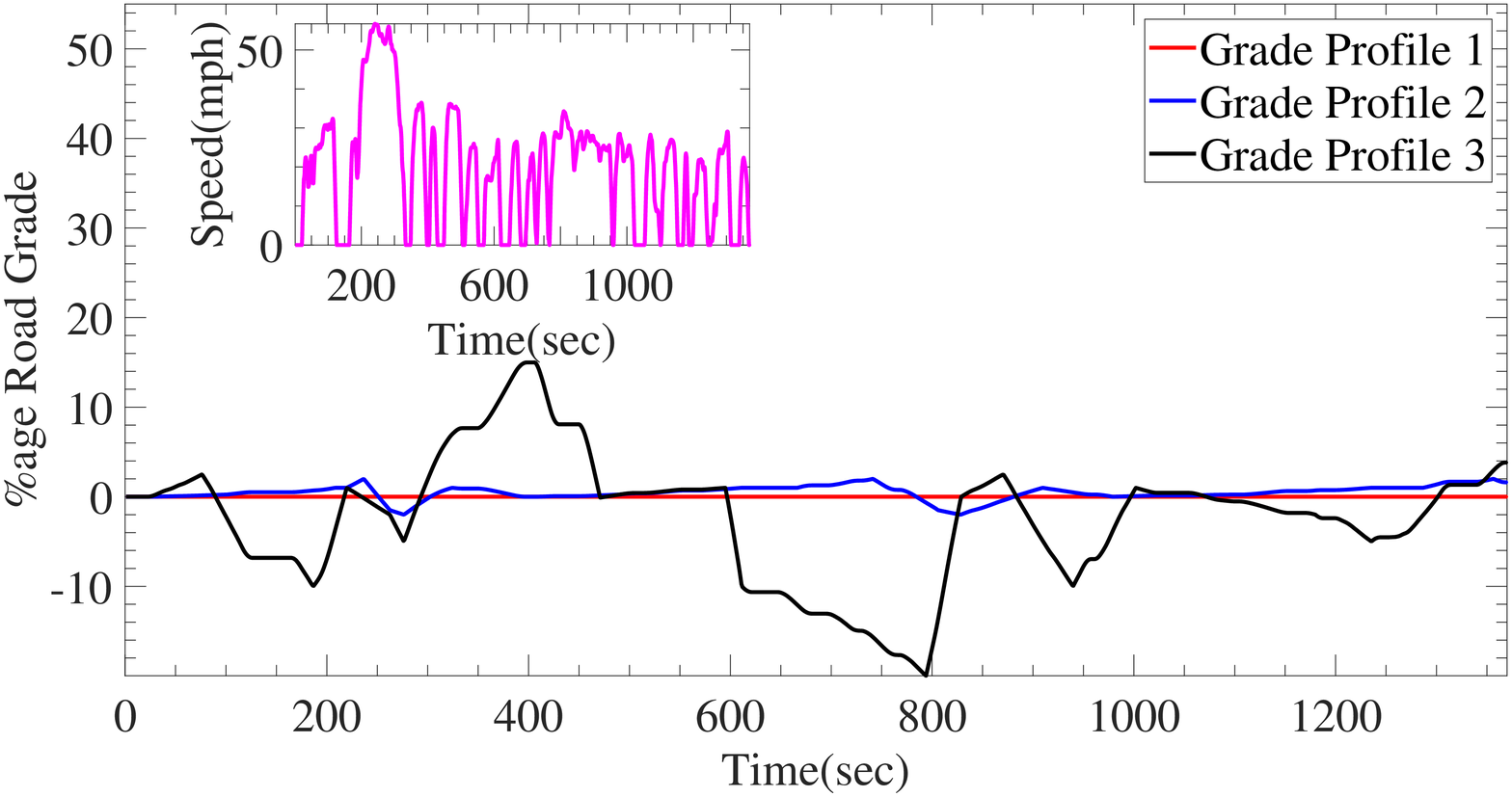}
    \caption{}
  \end{subfigure}
  \hfill
  \begin{subfigure}[]{0.49\textwidth}
    \centering
    \includegraphics[width=\linewidth]{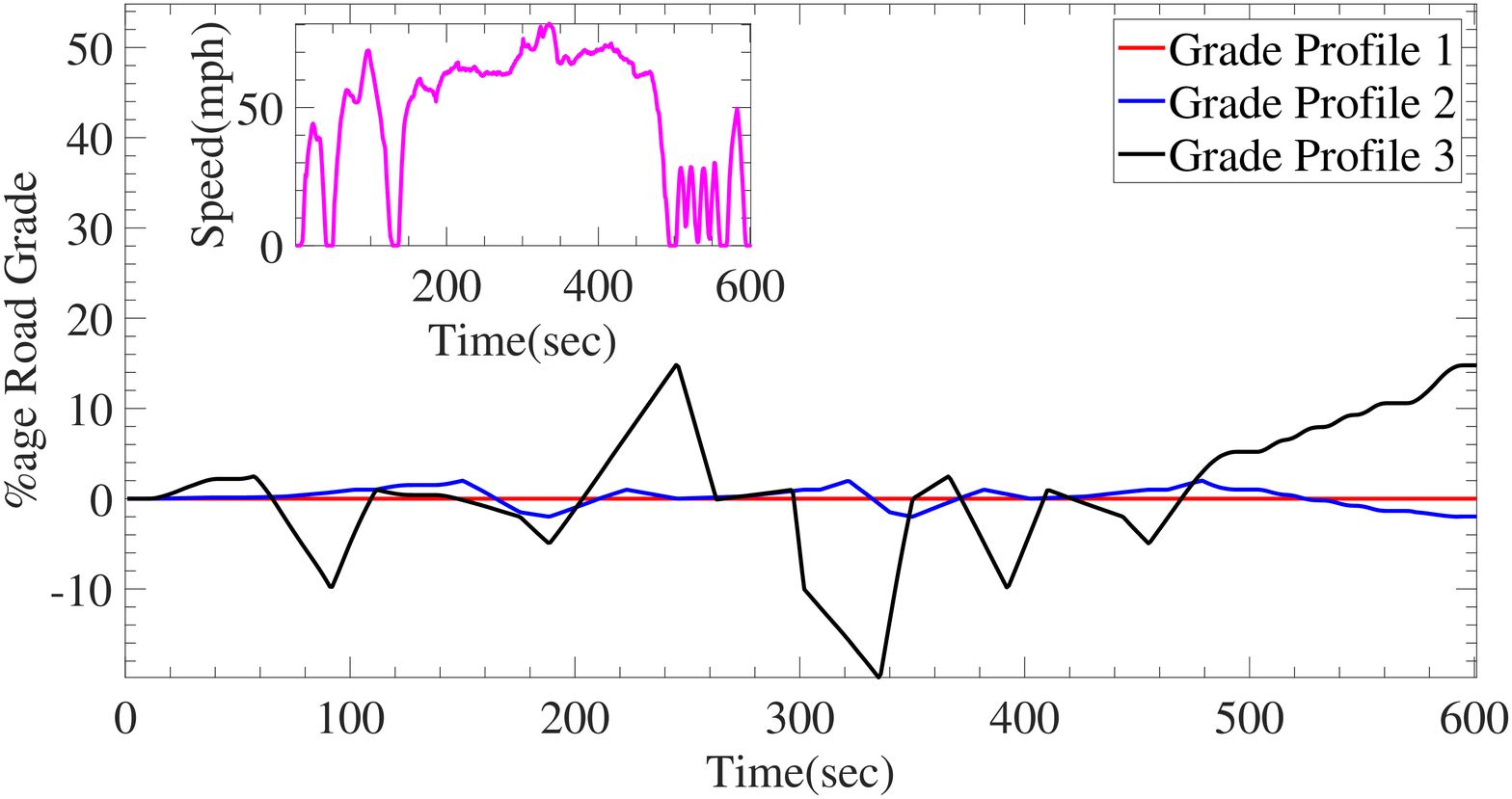}
    \caption{}
  \end{subfigure}
  \begin{subfigure}[]{0.49\textwidth}
    \centering
    \includegraphics[width=\linewidth]{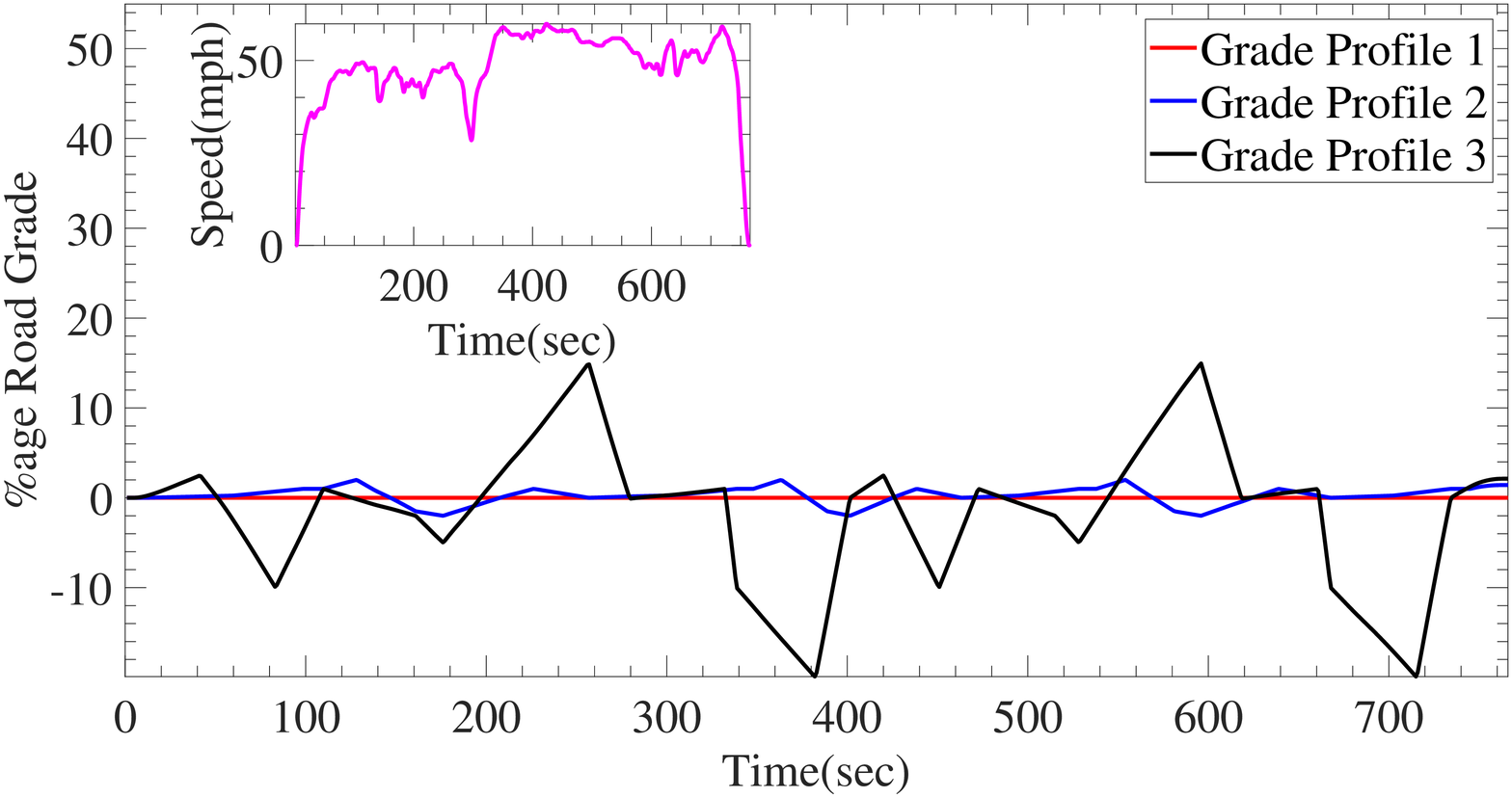}
    \caption{}
  \end{subfigure}
  \hfill
  \begin{subfigure}[]{0.49\textwidth}
    \centering
    \includegraphics[width=\linewidth]{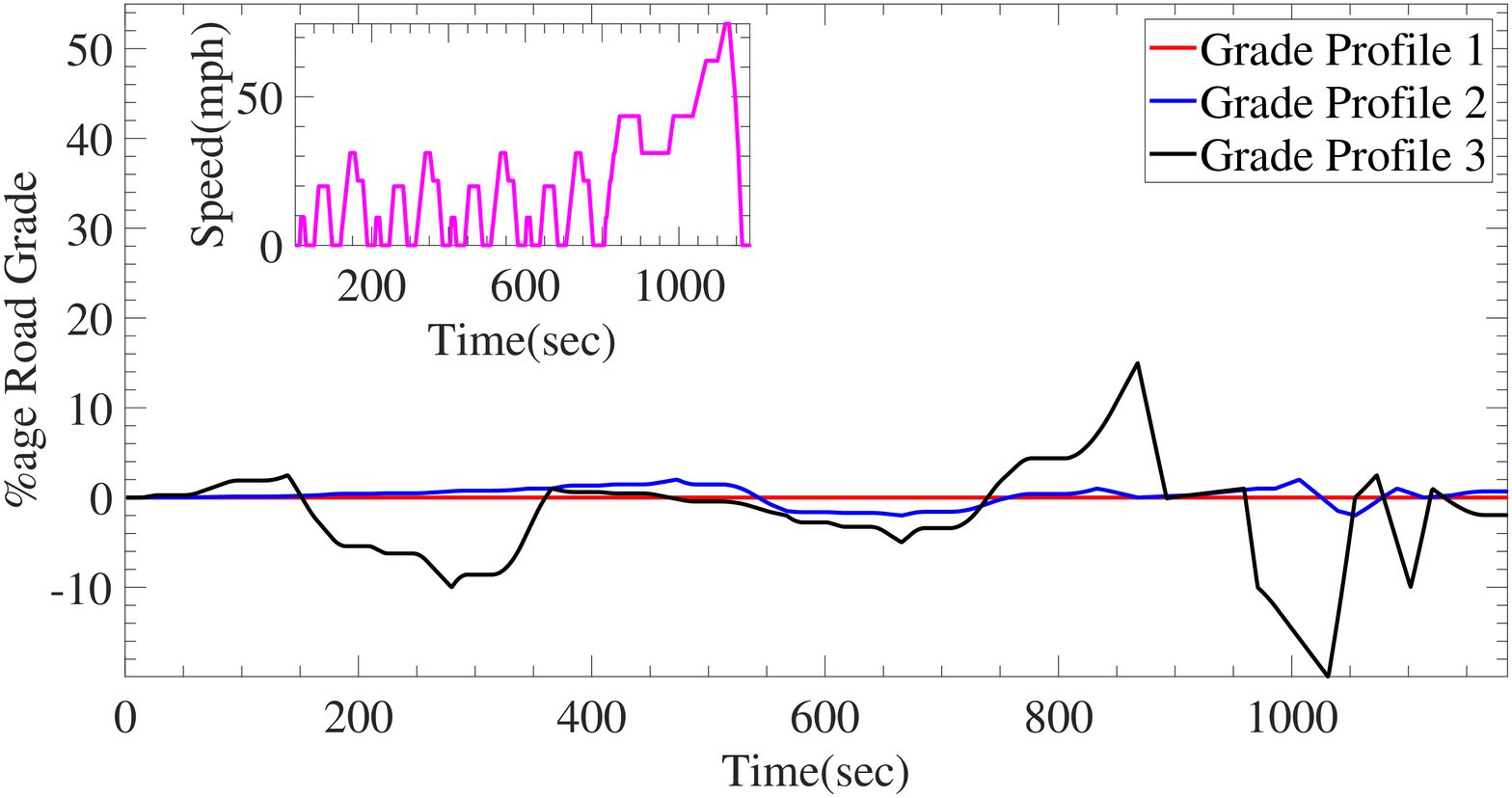}
    \caption{}
  \end{subfigure}
  \caption{Grade Profiles for different drive cycles (shown in sub-axis), (a) UDDS, (b) SFTP, (c) HWFET and (d) NEDC} \label{Fig:DriveCycleGrade}
\end{figure*}

\begin{figure*}[h!]
  \centering
  \begin{subfigure}[]{0.3\textwidth}
    \centering
    \includegraphics[width=\textwidth]{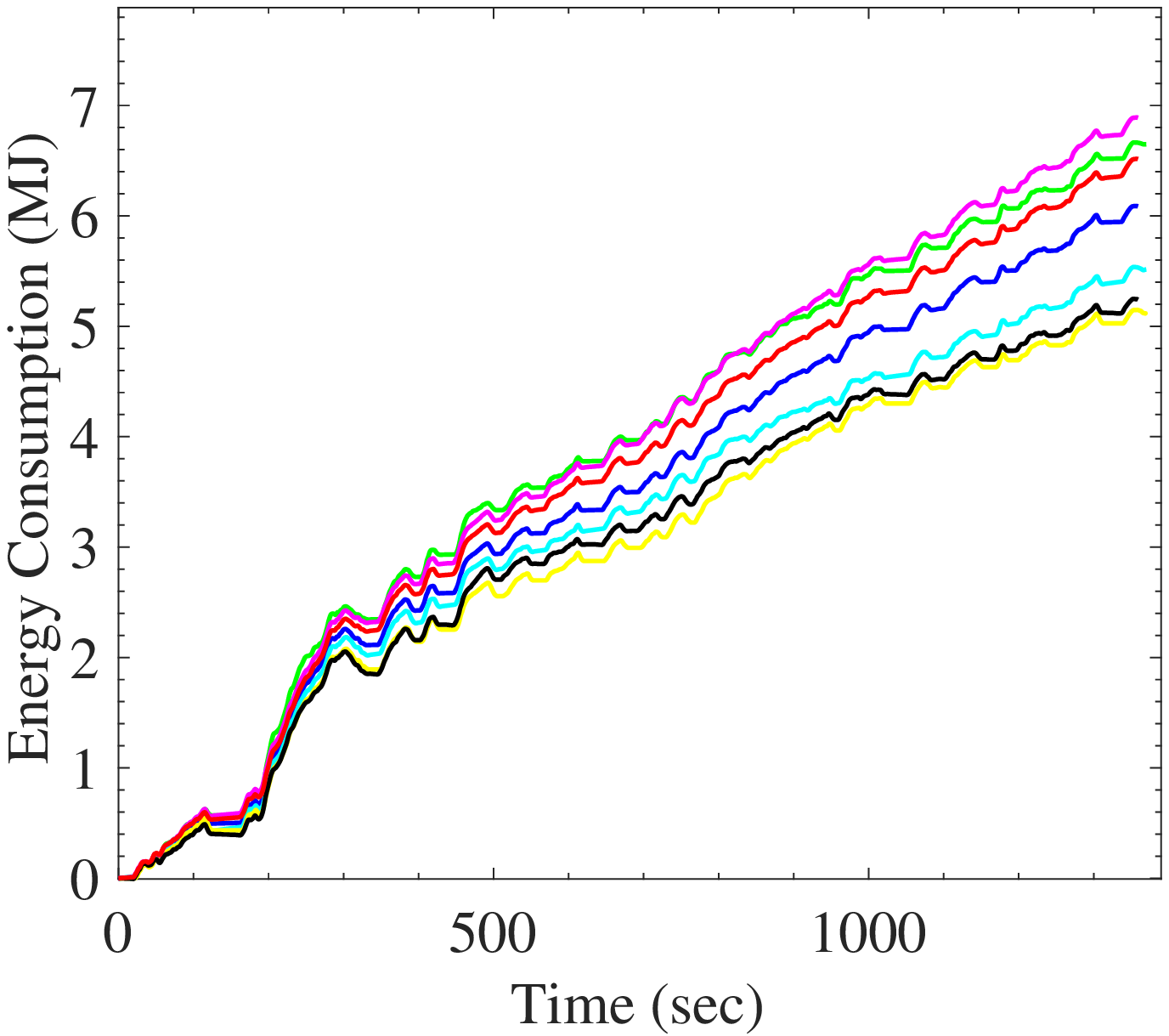}
    \caption{Results for UDDS with grade profile 1}
  \end{subfigure}
  \hfill
  \begin{subfigure}[]{0.3\textwidth}
    \centering
    \includegraphics[width=\linewidth]{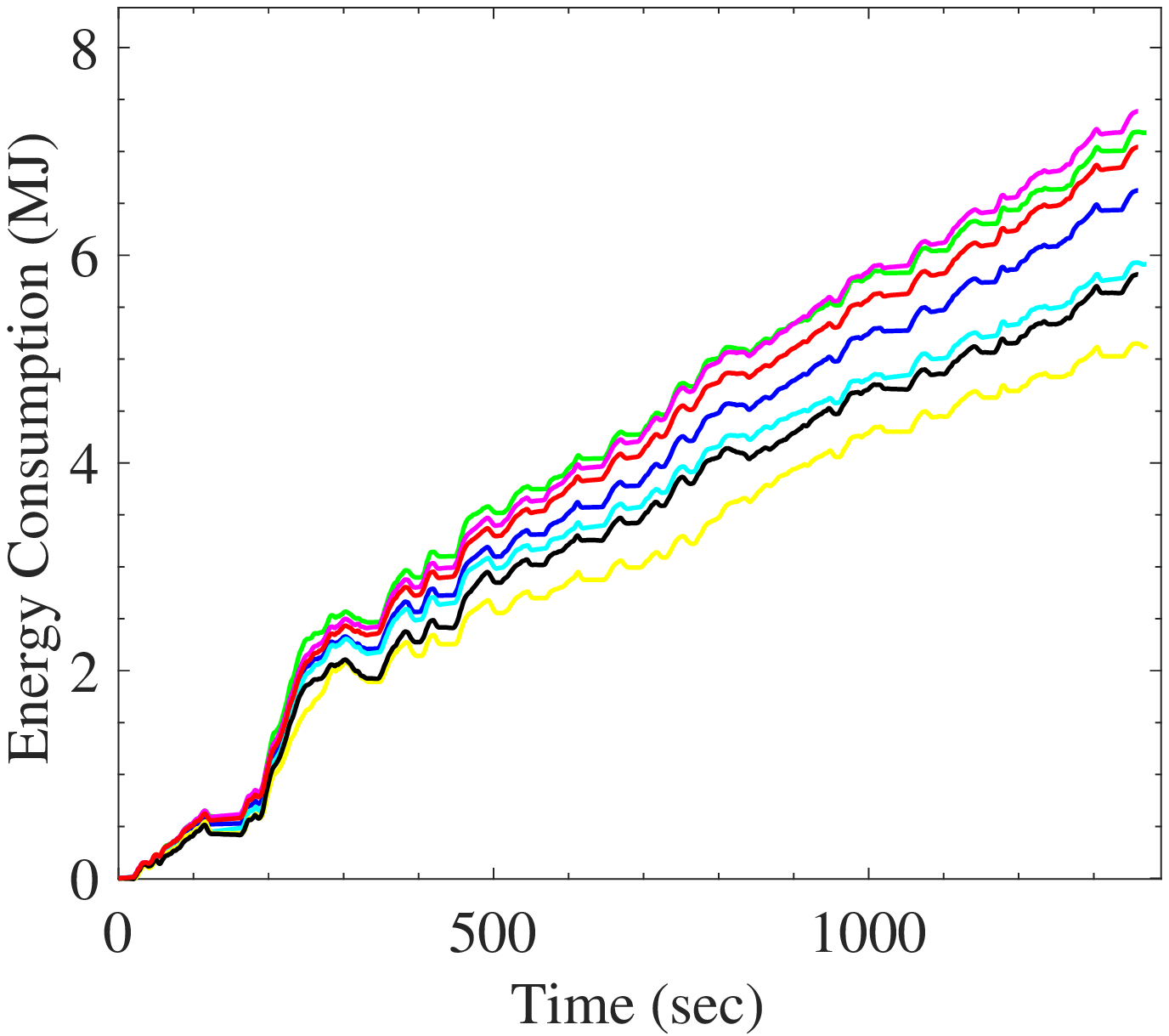}
    \caption{Results for UDDS with grade profile 2}
  \end{subfigure}
  \hfill
  \begin{subfigure}[]{0.3\textwidth}
    \centering
    \includegraphics[width=\linewidth]{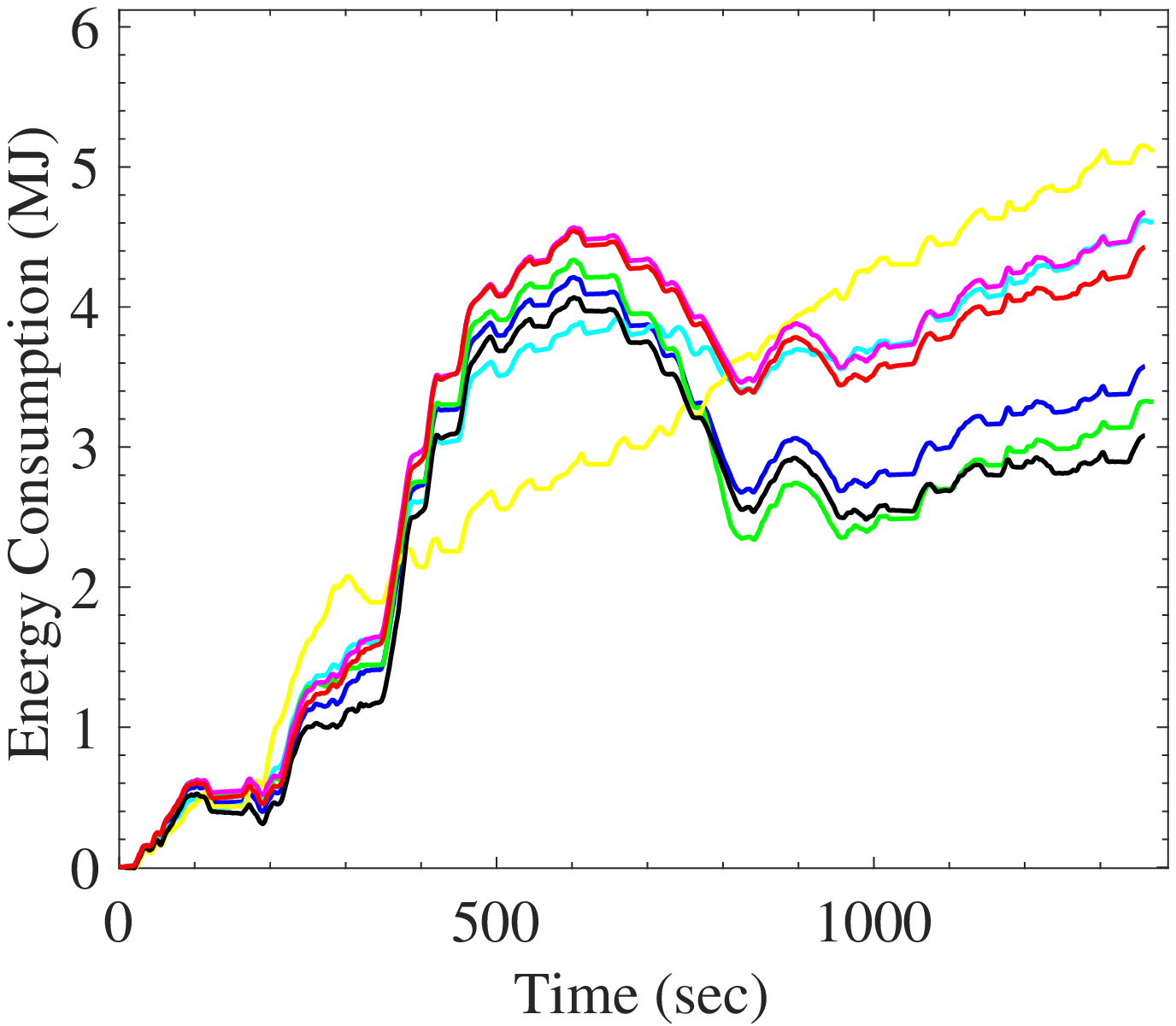}
    \caption{Results for UDDS with grade profile 3}
  \end{subfigure}
  \begin{subfigure}[]{0.3\textwidth}
    \centering
    \includegraphics[width=\linewidth]{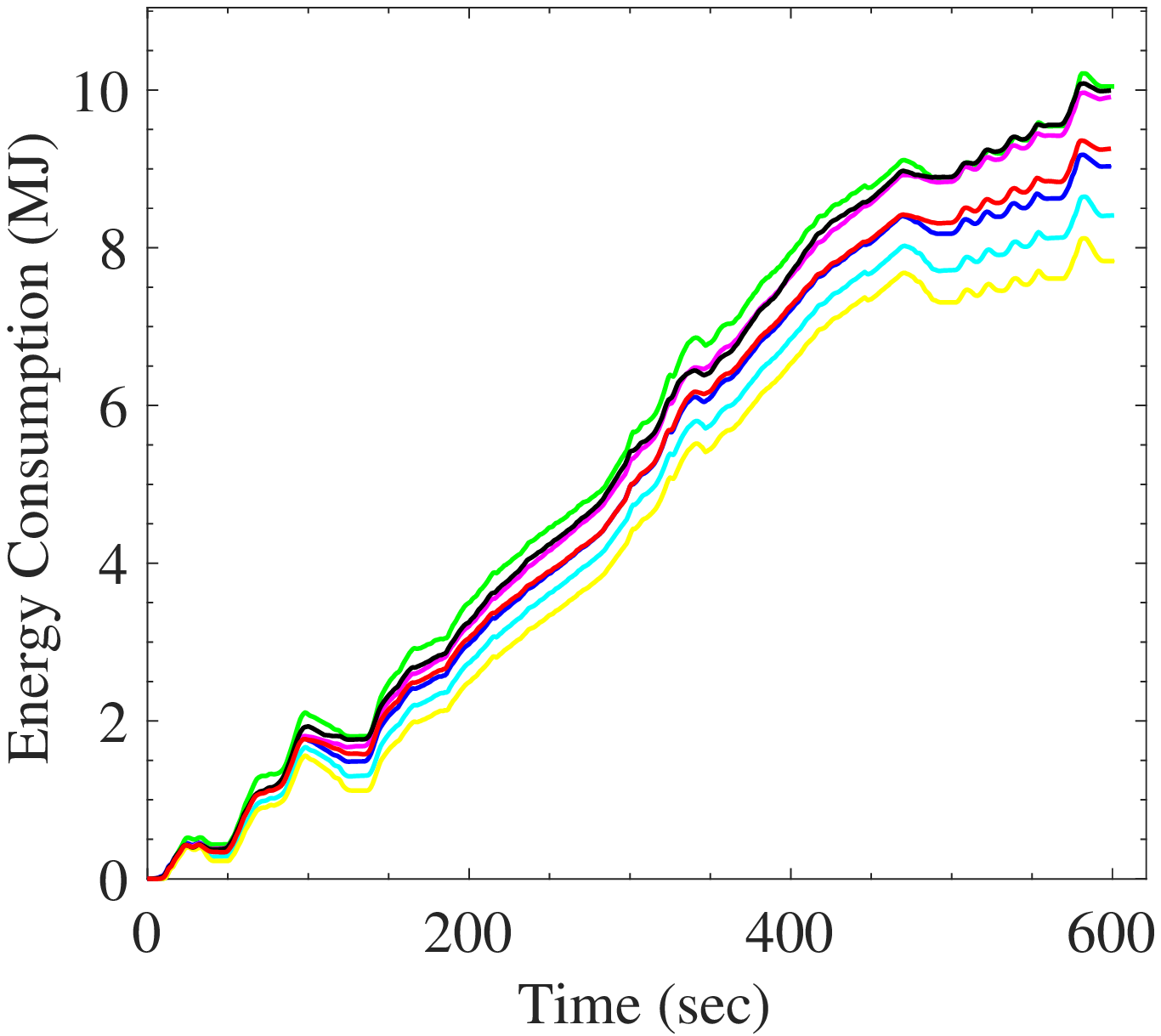}
    \caption{Results for SFTP with grade profile 1}
  \end{subfigure}
  \hfill
  \begin{subfigure}[]{0.3\textwidth}
    \centering
    \includegraphics[width=\linewidth]{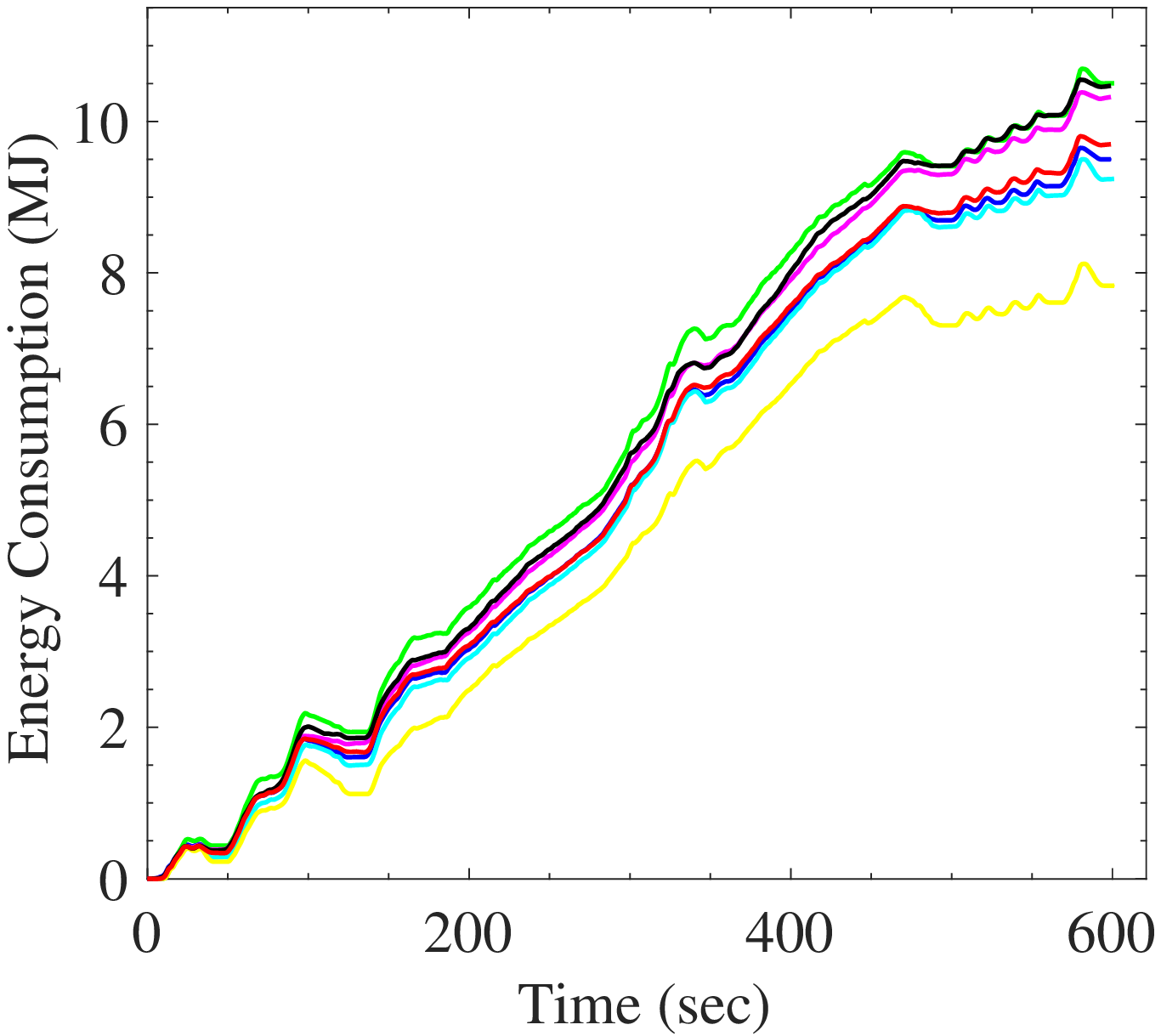}
    \caption{Results for SFTP with grade profile 2}
  \end{subfigure}
  \hfill
  \begin{subfigure}[]{0.3\textwidth}
    \centering
    \includegraphics[width=\linewidth]{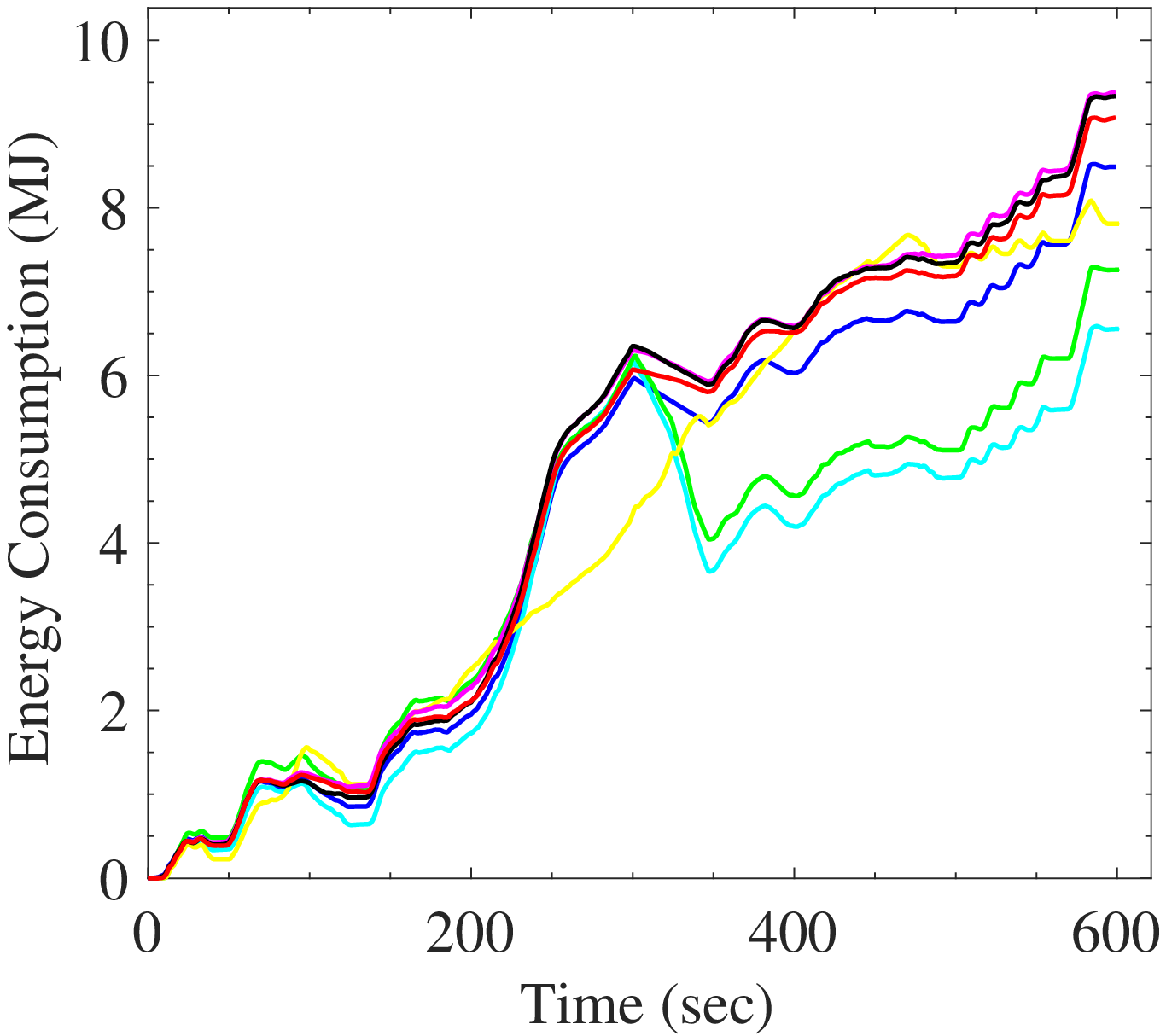}
    \caption{Results for SFTP with grade profile 3}
  \end{subfigure}
  \begin{subfigure}[]{0.3\textwidth}
    \centering
    \includegraphics[width=\linewidth]{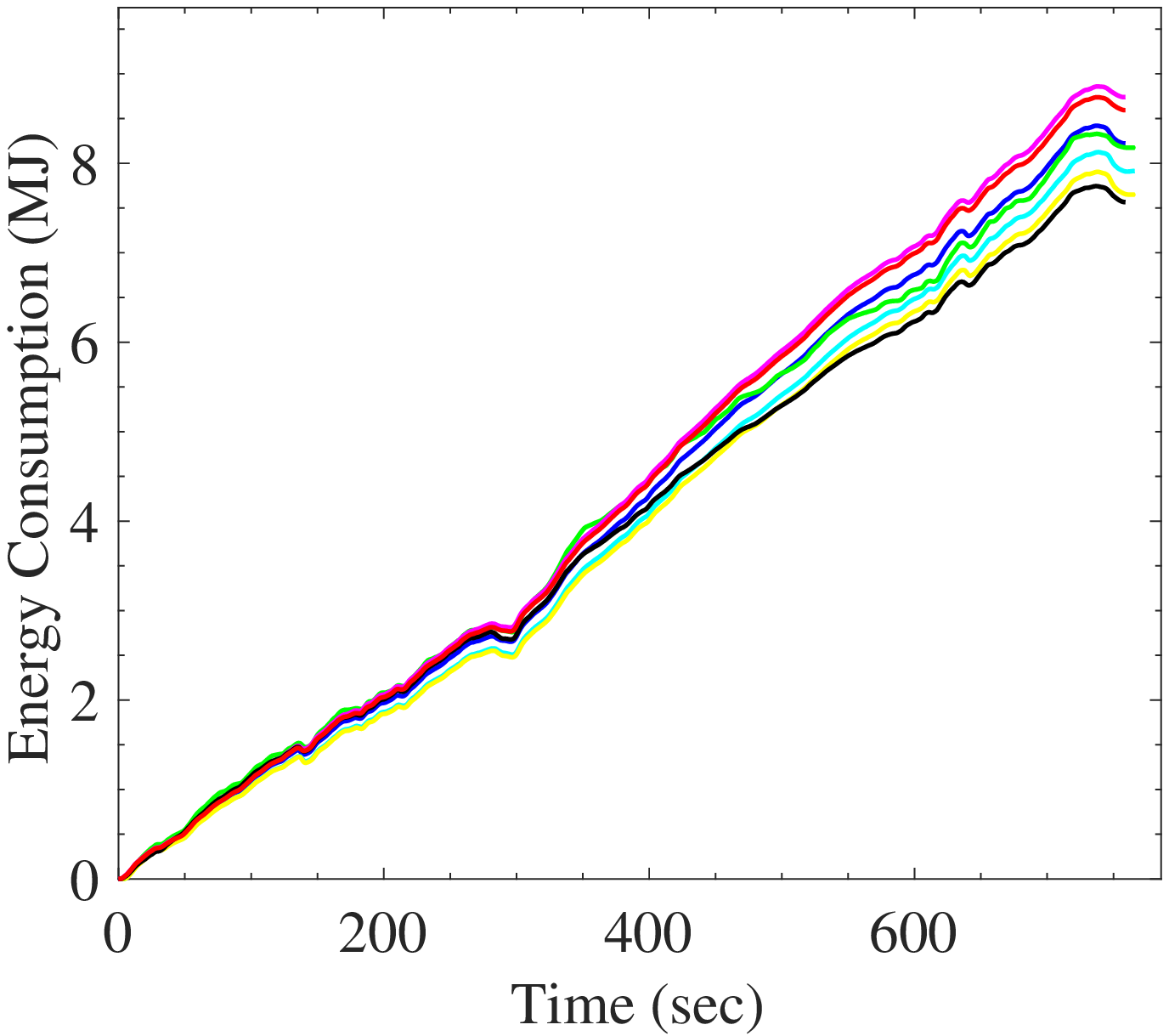}
    \caption{Results for HWFET with grade profile 1}
  \end{subfigure}
  \hfill
  \begin{subfigure}[]{0.3\textwidth}
    \centering
    \includegraphics[width=\linewidth]{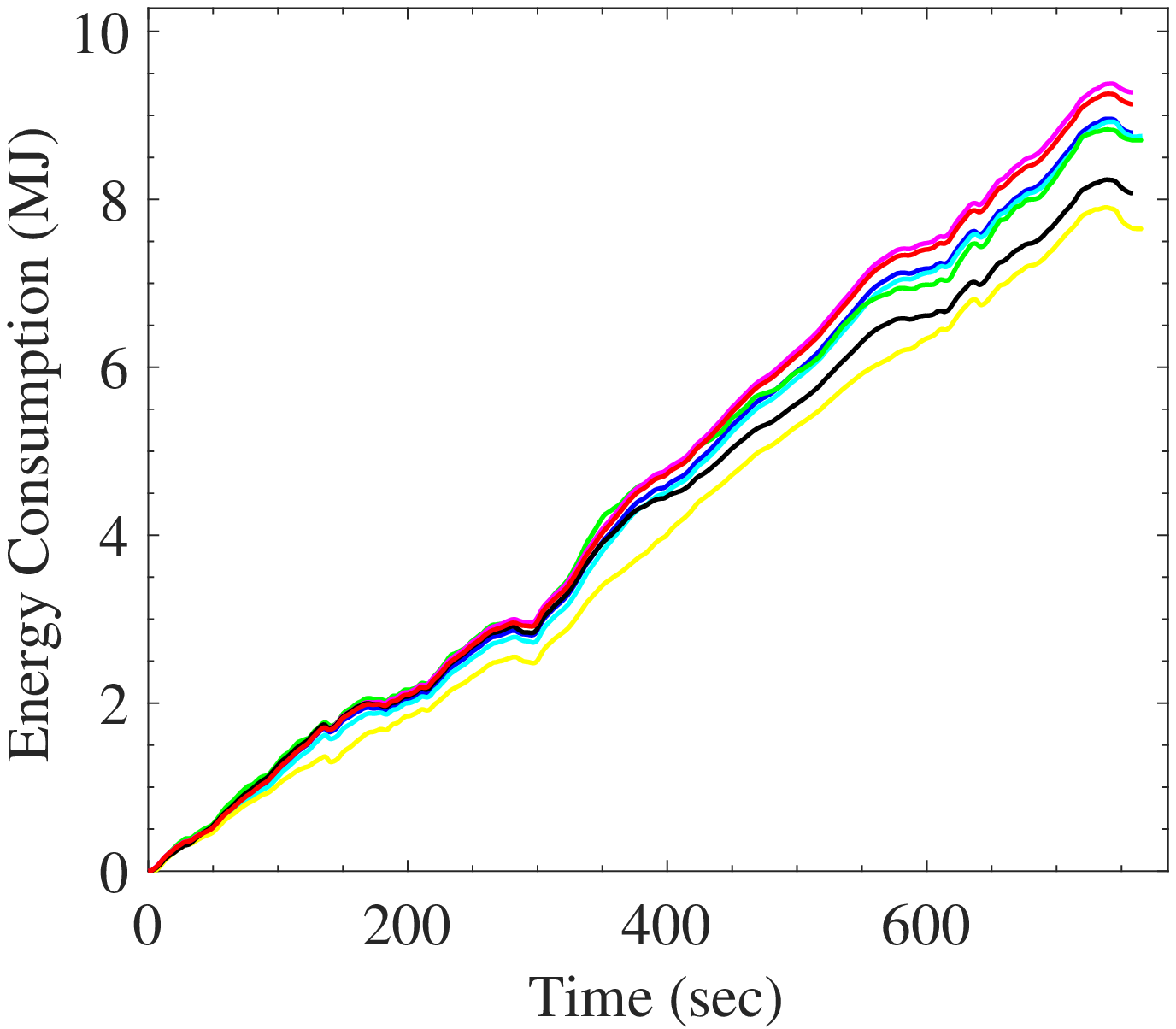}
    \caption{Results for HWFET with grade profile 2}
  \end{subfigure}
  \hfill
  \begin{subfigure}[]{0.3\textwidth}
    \centering
    \includegraphics[width=\linewidth]{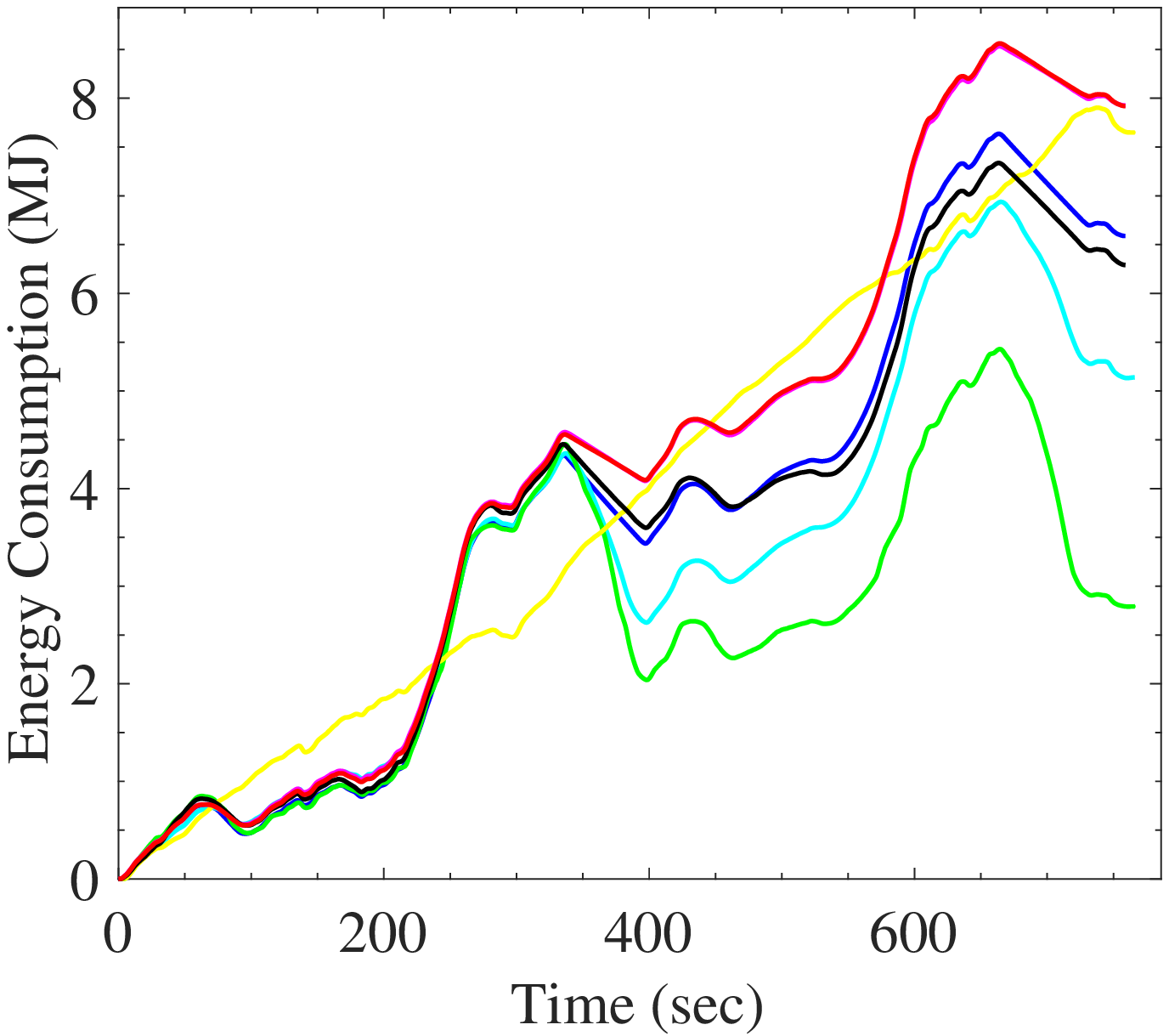}
    \caption{Results for HWFET with grade profile 3}
  \end{subfigure}
  \begin{subfigure}[]{0.3\textwidth}
    \centering
    \includegraphics[width=\linewidth]{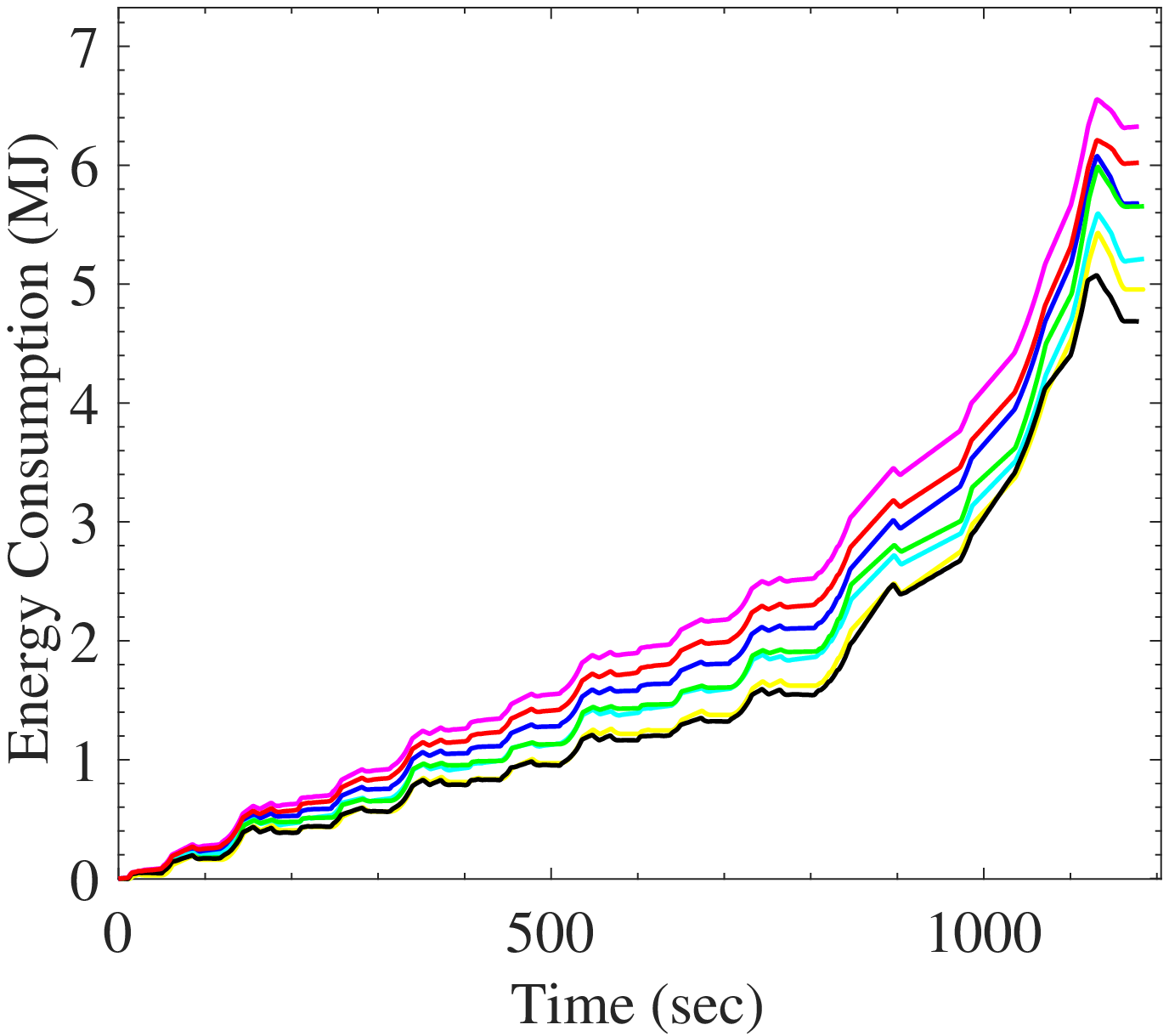}
    \caption{Results for NEDC with grade profile 1}
  \end{subfigure}
  \hfill
  \begin{subfigure}[]{0.3\textwidth}
    \centering
    \includegraphics[width=\linewidth]{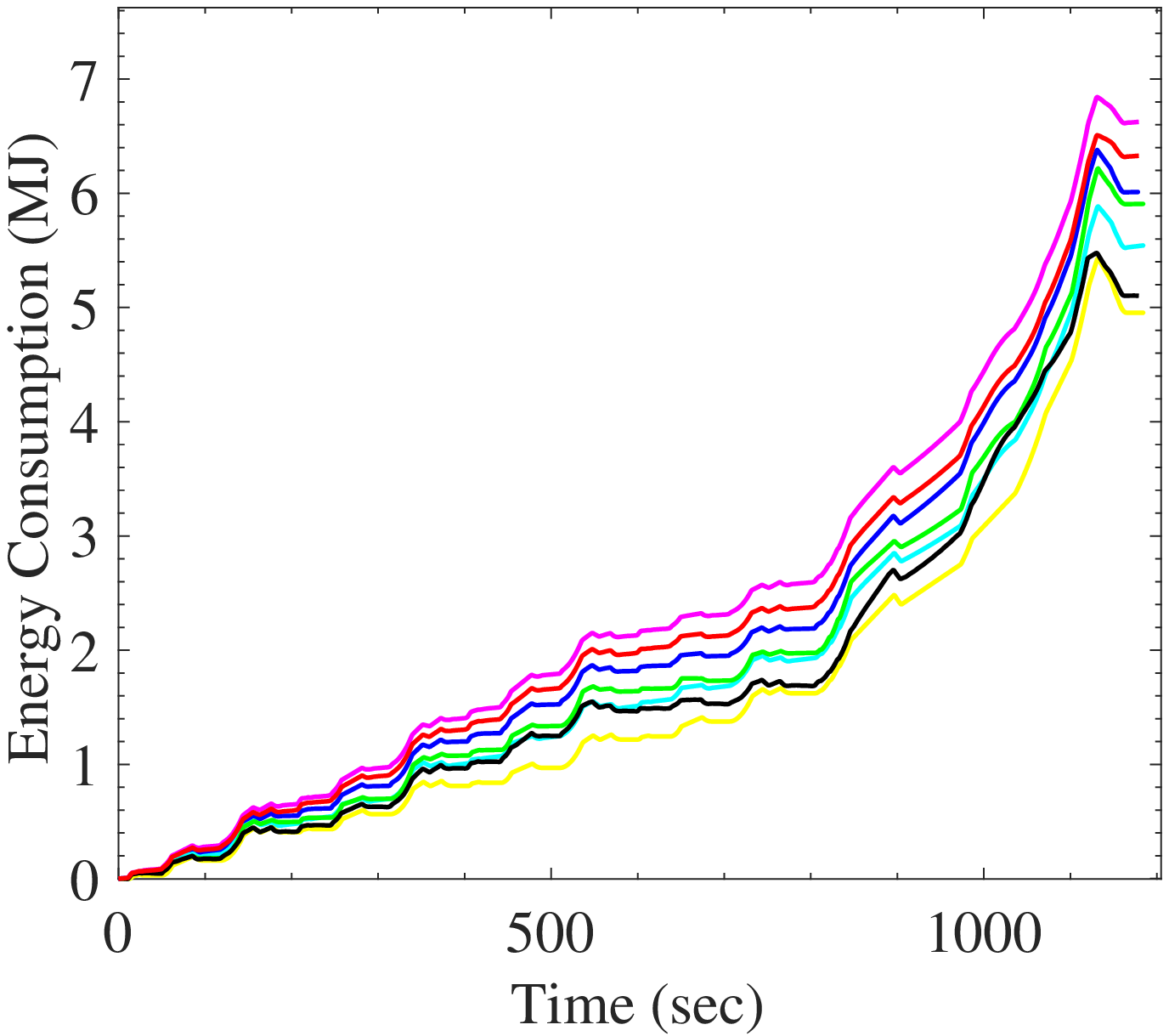}
    \caption{Results for NEDC with grade profile 2}
  \end{subfigure}
  \hfill
  \begin{subfigure}[]{0.3\textwidth}
    \centering
    \includegraphics[width=\linewidth]{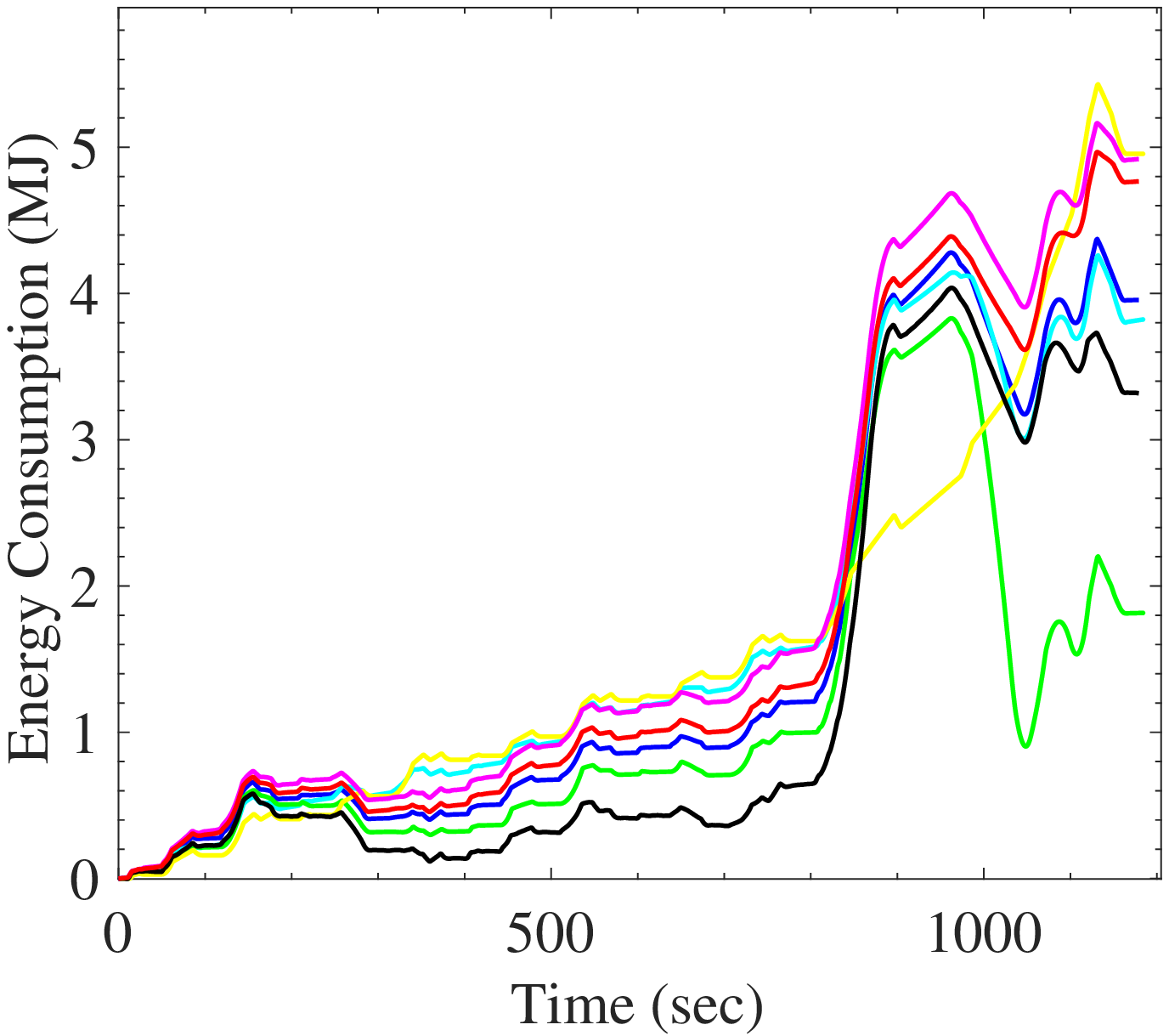}
    \caption{Results for NEDC with grade profile 3}
  \end{subfigure}
  \caption{Energy consumption estimation comparison of proposed and existing techniques for different driving cycles and road grade profiles. Legend: (\protect\bluesolidline) Actual / Target, The existing approaches (\protect\cyansolidline) De Cauwer et al. \cite{de2017data}, (\protect\greensolidline) Yang et al. \cite{YANG201441} and (\protect\yellowsolidline) Galvin \cite{GALVIN2017234}, The proposed models (\protect\redsolidline) $CNN^7_{cov}$, (\protect\magentasolidline) $CNN^9_{cov}$ and (\protect\blacksolidline) $CNN^9_{gaf}$} \label{Fig:EnergyConsumptionComparison}
\end{figure*}

A comparison of energy consumption, estimated using the proposed models $CNN_{cov}^7$, $CNN_{cov}^9$ and $CNN_{gaf}^9$ and above discussed state-of-the-art approaches except, approaches presented in \cite{6861542, 7313117}, has been presented in Figure \ref{Fig:EnergyConsumptionComparison}. The energy consumption was calculated by integrating the power over the time period. As \cite{6861542, 7313117} do not provide real-time power/energy consumption as output so, it was not possible to plot them. In Figure \ref{Fig:EnergyConsumptionComparison}, each column represent the different road grade profile and each row represent the different drive cycle i.e., from top to bottom, rows represent UDDS, SFTP, HWFET and NEDC drive cycles, respectively and from left to right, columns represent Grade Profile 1 (constant grade at $0\%$ i.e., no change in elevation), Grade Profile 2 (varies from -2\% to 2\%) and Grade Profile 3 (varies from -20\% to 15\%), as shown in Figure \ref{Fig:DriveCycleGrade}. It can be clearly observed from the Figure \ref{Fig:EnergyConsumptionComparison} that in most of the cases $CNN^7_{cov}$ has performed better than all of the existing approaches \cite{GALVIN2017234, YANG201441, de2017data} and other CNN models of $CNN_{cov}^9$ and $CNN_{gaf}^9$ in terms of accurate estimates. The difference in estimates can be seen more clearly when road grade varies in large range like in Grade Profile 3 i.e., in the rightmost column of Figure \ref{Fig:EnergyConsumptionComparison}. Also, it can be concluded that the estimates given by proposed CNN models follow the same behaviour or trend as the actual energy consumption whereas the existing techniques deviate from the actual value with large deviation.


As Galvin \cite{GALVIN2017234} did not consider the effect of road grade so it is justified for his model to deviate from actual energy consumption when road grade effect gets introduced. Yang et al. \cite{YANG201441} considered the effect of road grade but they tested their model only for small tilt angles i.e. $0^\circ$, $1^\circ$, $2^\circ$ and $3^\circ$ so when the tilt angle changes with large values like in Grade Profile 3, where road grade varies from -20\% (i.e. $-11.30^\circ$) to 15\% (i.e. $8.53^\circ$), their model fail to estimate the actual energy consumption accurately. The MLR model developed by De Cauwer et al. \cite{de2017data} has performed really well and in some cases even performed better than the $CNN_{cov}^7$ but it also fails to accurately estimate the energy consumption when road grade changes with high values. The main reason for that is the non-linear relationship of the influencing parameters which the MLR model was not able to estimate accurately as compared to the $CNN_{cov}^7$. The NN architecture presented in \cite{6861542, 7313117} have no hidden layer and has one input layer with 14 and 137 inputs, respectively and 1 output corresponding to the inputs. As the networks were shallow so they were also not able to accurately represent the non-linear relationship between the influencing factors. Also, the NN architectures provide only one output of total energy consumption over the whole trip so they can not be used to provide real-time information to the drivers.

From the above discussions, a number of observations can be concluded which are summarized as follow:
\begin{enumerate}[(i)]
    \item The results show that although the CNN model with more layers i.e., $CNN^9$ converge faster than a model with fewer layers i.e., $CNN^7$ but more number of layers does not increase the estimation accuracy.
    \item Road gradient is an important parameter and effects the energy consumption of EV greatly that is why it can be seen that when road grade changes with large values most of the existing techniques fail to accurately estimate the energy consumption.
    \item CNN models trained with covariance feature descriptors i.e. $CNN^7_{cov}$ and $CNN^9_{cov}$ give very good results than other CNN models trained with GAF or Eigen feature descriptors so the choice of input features also affect the performance of CNN architectures.
\end{enumerate}


\section{Conclusion}\label{Sec:Conclusion}
A Deep Convolutional Neural Networks (D-CNN) based solution has been developed for estimation of energy consumption of EVs considering three external parameters namely, road elevation, tractive effort and speed of the vehicle. Unlike previous methods that require either manufacturer data which is not readily available or real-world data which require special sensors to be deployed on EVs, the proposed approach require only three parameters which can easily be obtained. A number of CNN models with different architectures were trained using simulated data after preprocessing, in which the simulated time series data was converted to images. The simulated data was generated from a simulation model of Nissan Leaf 2013, developed in FASTSim. The CNN models were tested using the experimental data obtained from Argonne National Laboratory of US. It has been observed that one of the CNN models with seven layers represented with $CNN^7_{cov}$ has performed really well with average percentage energy consumption deviation of 5.21\% and 5.09\% on dataset $DS-I_{val}$ and $DS-II$, respectively.

In the future, the model will consider the effect of other parameters like environmental temperature, traffic and auxiliary loads etc. The model will also be ported to a TensorFlow Lite format such that it can be used with Google Coral boards or Odroid boards with Movidius sticks etc. and can provide the optimal driving parameters (such as speed, the route to be taken etc.) to the driver in real time. Also, it would be interesting to explore the performance of other advanced deep learning approaches.


\end{document}